\documentclass[twocolumn,superscriptaddress,showpacs,amssymb,amsmath,amsfonts,aps]{revtex4}
\input{epsf}
\usepackage{epsfig}
\usepackage{latexsym}
\usepackage{amssymb}
\usepackage{graphicx}
\usepackage{amsmath} 
\usepackage[dvips]{color}

\begin{document}
\date{\today}

\title {\large {\bf Electroproduction of $p \pi^{+} \pi^{-}$ off
protons   at
0.2 $<$ $\it{Q^{2}}$ $<$ 0.6\enskip {\rm GeV$^{2}$} and
1.3 $<$ $\it{W}$ $<$ 1.57\enskip {\rm GeV} with CLAS} }
\newcommand*{\MOSCOW}{Moscow State University, Skobeltsyn Nuclear Physics
Institute and Physics Department, 119899 Moscow, Russia} \affiliation{\MOSCOW}
\newcommand*{\JLAB}{Thomas Jefferson National Accelerator Facility, Newport News, Virginia 23606} \affiliation{\JLAB}
\newcommand*{\ASU}{Arizona State University, Tempe, Arizona 85287-1504} \affiliation{\ASU}
\newcommand*{\CMU}{Carnegie Mellon University, Pittsburgh, Pennsylvania 15213} \affiliation{\CMU}
\newcommand*{\CUA}{Catholic University of America, Washington, D.C. 20064} \affiliation{\CUA}
\newcommand*{\CNU}{Christopher Newport University, Newport News, Virginia 23606} \affiliation{\CNU}
\newcommand*{\SACLAY}{CEA-Saclay, Service de Physique Nucl\'eaire, F91191 Gif-sur-Yvette,Cedex, France} \affiliation{\SACLAY}
\newcommand*{\WM}{College of William and Mary, Williamsburg, Virginia 23187-8795} \affiliation{\WM}
\newcommand*{\ECOSSEE}{Edinburgh University, Edinburgh EH9 3JZ, United Kingdom} \affiliation{\ECOSSEE}
\newcommand*{\FU}{Fairfield University, Fairfield, CT 06824}
\affiliation{\FU}
\newcommand*{\FIU}{Florida International University, Miami, Florida 33199} \affiliation{\FIU}
\newcommand*{\FSU}{Florida State University, Tallahassee, Florida 32306} \affiliation{\FSU}
\newcommand*{\GWU}{The George Washington University, Washington, DC 20052} \affiliation{\GWU}
\newcommand*{\ISU}{Idaho State University, Pocatello, Idaho 83209} \affiliation{\ISU}
\newcommand*{\ORSAY}{Institut de Physique Nucleaire ORSAY, Orsay, France} \affiliation{\ORSAY}
\newcommand*{\INFNGE}{INFN, Sezione di Genova, 16146 Genova, Italy} \affiliation{\INFNGE}
\newcommand*{\INFNFR}{INFN, Laboratori Nazionali di Frascati, Frascati, Italy 00044} \affiliation{\INFNFR}
\newcommand*{\ITEP}{Institute of Theoretical and Experimental Physics, Moscow, 117259, Russia} \affiliation{\ITEP}
\newcommand*{\JMU}{James Madison University, Harrisonburg, Virginia 22807} \affiliation{\JMU}
\newcommand*{\KYUNGPOOK}{Kyungpook National University, Daegu 702-701, South Korea} \affiliation{\KYUNGPOOK}
\newcommand*{\MIT}{Massachusetts Institute of Technology, Cambridge, Massachusetts  02139-4307} \affiliation{\MIT}
\newcommand*{\NSU}{Norfolk State University, Norfolk, Virginia 23504} \affiliation{\NSU}
\newcommand*{\OHIOU}{Ohio University, Athens, Ohio  45701} \affiliation{\OHIOU}
\newcommand*{\ODU}{Old Dominion University, Norfolk, Virginia 23529} \affiliation{\ODU}
\newcommand*{\RPI}{Rensselaer Polytechnic Institute, Troy, New York 12180-3590} \affiliation{\RPI}
\newcommand*{\RICE}{Rice University, Houston, Texas 77005-1892} \affiliation{\RICE}
\newcommand*{\UNIONC}{Union College, Schenectady, NY 12308} \affiliation{\UNIONC}
\newcommand*{\UCLA}{University of California at Los Angeles, Los Angeles, California  90095-1547} \affiliation{\UCLA}
\newcommand*{\UCONN}{University of Connecticut, Storrs, Connecticut 06269} \affiliation{\UCONN}
\newcommand*{\ECOSSEG}{University of Glasgow, Glasgow G12 8QQ, United Kingdom} \affiliation{\ECOSSEG}
\newcommand*{\UMASS}{University of Massachusetts, Amherst, Massachusetts  01003} \affiliation{\UMASS}
\newcommand*{\UNH}{University of New Hampshire, Durham, New Hampshire 03824-3568} \affiliation{\UNH}
\newcommand*{\SCAROLINA}{University of South Carolina, Columbia, South Carolina 29208} \affiliation{\SCAROLINA}
\newcommand*{\PITT}{University of Pittsburgh, Pittsburgh, Pennsylvania 15260} \affiliation{\PITT}
\newcommand*{\URICH}{University of Richmond, Richmond, Virginia 23173} \affiliation{\URICH}
\newcommand*{\CHILI}{Universidad Técnica Federico Santa María, Av. España 1680 Casilla 110-V
Valparaíso, Chile} \affiliation{\CHILI}
\newcommand*{\VIRGINIA}{University of Virginia, Charlottesville, Virginia 22901} \affiliation{\VIRGINIA}
\newcommand*{\VT}{Virginia Polytechnic Institute and State University, Blacksburg, Virginia   24061-0435} \affiliation{\VT}

\newcommand*{\NOWOHIOU}{Ohio University, Athens, Ohio  45701}
\newcommand*{\NOWUNH}{University of New Hampshire, Durham, New Hampshire 03824-3568}
\newcommand*{\NOWMOSCOW}{Moscow State University, General Nuclear Physics Institute, 119899 Moscow, Russia}
\newcommand*{\NOWSCAROLINA}{University of South Carolina, Columbia, South Carolina 29208}
\newcommand*{\YEREVAN }{ Yerevan Physics Institute, Yerevan 375036, Armenia} \affiliation{\YEREVAN }

\newcommand*{\deceased}{Deceased}

\author {G.V.~Fedotov}
\affiliation{\MOSCOW}
\author {V.I.~Mokeev}
\affiliation{\JLAB}
\affiliation{\MOSCOW}
\author {V.D.~Burkert}
\affiliation{\JLAB}
\author {L.~Elouadrhiri}
\affiliation{\JLAB}
\author {E.N.~Golovatch}
\affiliation{\MOSCOW}
\author {B.S.~Ishkhanov}
\affiliation{\MOSCOW}
\author {E.L.~Isupov}
\affiliation{\MOSCOW}
\author {N.V.~Shvedunov}
\affiliation{\MOSCOW}
\author {G.~Adams}
\affiliation{\RPI}
\author {M.J.~Amaryan}
\affiliation{\ODU}
% ...
\collaboration{The CLAS Collaboration} \noaffiliation

\begin{abstract}
This paper reports on the most comprehensive data set obtained on differential and fully integrated
cross sections for the process $e p \rightarrow e' p \pi^{+} \pi^{-} $. The data were 
collected
with the CLAS detector at Jefferson Laboratory. Measurements were
carried out in the so-far unexplored kinematic region of photon virtuality 
0.2 $<$ $Q^{2}$ $<$ 0.6 GeV$^{2}$ and invariant
mass of the final hadron system $W$ from  1.3 to 1.57~GeV.
For the first time, nine independent 1-fold differential cross sections were determined
in each bin of $W$ and $Q^{2}$ covered by the measurements. A phenomenological analysis of the data allowed us
to establish the most significant
mechanisms contributing to the reaction. The non-resonant mechanisms account for
a major part of cross-sections.  However, we find sensitivity 
to s-channel excitations of low-mass
nucleon resonances, especially 
to the $N(1440)P_{11}$ and $N(1520)D_{13}$ states in kinematical dependencies of
the 1-fold differential cross-sections.
\end{abstract}


\maketitle
\normalsize



\section{Introduction}
An extensive research program is currently underway in Hall B at
Jefferson Lab with the CLAS detector, focused on studies of
nucleon resonances ($N^{*}$) in various exclusive channels of
meson electroproduction off protons~\cite{Burk07,Burk05,Lee04}. Part of this effort 
is aimed at the determination
of electrocouplings for an entire spectrum of excited nucleon $N^*$ and
$\Delta^{*}$ states versus the photon virtuality $Q^{2}=-(e-e')^{2}$,
where $e$ and
$e'$ are the 4-momentum vectors of the incoming and scattered electron, respectively.
Comprehensive information on resonance electrocouplings and their
evolution with $Q^2$ is needed to probe the spatial and spin structure of
the resonance transitions. 
This information is needed to enhance our understanding of the
effective strong interaction that is at the core of internal baryon structure
and the decay of baryons. It is also needed to firmly establish 
the connection of effective degrees of freedom such as a) `constituent' quarks 
in the binding potential or b)  `constituent' quark scattering amplitudes to
the elementary quarks and gauge gluons of QCD, 
the theory of the strong interaction.

In the past five years, single pseudoscalar meson electroproduction has been studied in
several exclusive processes, e.g. $p\pi^0$, $n\pi^{+}$, $p\eta$, $K\Lambda$, and $K\Sigma$,
\cite{Joo02,Ungaro06,Joo03,Joo04,Joo05,egiyan06,park08,thomp01,denizli07,
carman03,ambroz07,devita02,biselli03,biselli06,Az08a}, which included differential cross 
sections with complete polar angle and azimuthal angle distributions, as well 
as several polarization observables. From these data sets, resonance transition 
electromagnetic form factors have been determined for 
the $\Delta(1232)P_{33}$, $N(1535)S_{11}$, and $N(1440)P_{11}$, 
covering a wide range in $Q^2$.

Studies of charged double pion electroproduction off protons represent 
an important part at
this effort. Single and double pion production are the two largest contributors
to the total photo- and electroproduction
cross sections off protons
in the resonance region. The final states produced in these two
exclusive channels have considerable hadronic interactions, or so-called 
final state interactions (FSI).
FSI may be determined using the data of
experiments with hadronic
probes \cite{Pe02}. According to these data, the cross section 
for the $\pi N \rightarrow \pi \pi N$ reaction
has the second largest strength of all of the exclusive channels in the 
$\pi N$  interaction.
Considerable FSI between the $\pi N $ and $\pi \pi N$ final
states result in substantial contributions to the amplitudes of both 
single and double pion
electroproduction from the electroproduction amplitudes of the other channel.
Accounting for 
these coupled-channel effects  is essential in order 
to get the amplitude description 
compatible with the constraints imposed by unitarity.
Therefore, for $N^*$  studies both in single and double pion electroproduction,
information is needed
on the mechanisms contributing to each of these channels in order to take 
properly into account the impact from coupled-channel effects on the 
exclusive channel 
cross sections. 
The knowledge of single and double pion 
electroproduction
mechanisms becomes even more important for $N^*$  studies in channels with
smaller cross sections such as $p\eta$ or $K\Lambda$ and $K\Sigma$ production,
as they could be significantly affected in leading order by coupled-channel
effects produced by  their hadronic 
interactions with the 
dominant single and double pion electroproduction channels.
Therefore, comprehensive studies of single and double pion electroproduction 
are of key importance for the entire $N^*$ research program.

The world data on double pion electroproduction in the nucleon
resonance excitation region were rather
scarce before the data from CLAS became available. Fully integrated cross sections for $\pi
\Delta$ isobar channels as a function of the invariant mass of the 
hadronic system $W$ and photon virtualities $Q^2$ were available from DESY \cite{Wa78}.
However, the data are presented in very large kinematic bins,
$\Delta W = 200 - 300$ MeV  and $\Delta Q^{2}= 0.2 - 0.6$ GeV$^{2}$.
Center-of-mass angular distributions for $\pi^{-}$ were
also measured, but were averaged over a very large intervals in $Q^2$
from 0.3 to 1.4 GeV$^{2}$. This makes it virtually impossible to determine the 
nucleon resonance parameters from such measurements.

The first detailed
data on charged double pion electroproduction cross sections in the resonance
region were obtained with CLAS \cite{Ri03,db07}. The data were collected for 
$W$=1.4 - 2.1~GeV
in 25 MeV bins and for $Q^2 = 0.5 - 1.5$~GeV$^{2}$ in 0.3 GeV$^{2}$ wide bins.
The current experiment covers invariant masses of
$\pi^{-} \pi^{+}$,~ $\pi^{+} p$ and~ $\pi^{-} p$ in each ($W$, $Q^{2}$) bin  for
$Q^{2}$ = 0.2 - 0.6~GeV$^{2}$ in 0.05 GeV$^{2}$ wide bins and for $W$ =1.30 to
1.57 GeV
 with 25 MeV
bins. In addition,
angular distributions for $\pi^{-}$,  $\pi^{+}$, and proton, as well as
angular distributions in $\alpha_{\it{i}}$ ($i$=1,2,3) angles (see Sect. \ref{xsect} for
$\alpha_{\it{i}}$  definitions), were measured. These very detailed measurements are
crucial to determine the most significant production mechanisms for this process.

\section{Analysis Tools}

The presence of three hadrons in
the final state presents considerable complications in the phenomenological
analysis. Efforts to apply partial wave analysis (PWA) techniques to double
pion production by electromagnetic probes are limited to photoproduction, where very high statistics
data are available \cite{Be04,Tho05,An05}. A strong reduction in statistics
for individual bins in $Q^2$ makes application of PWA methods in
double pion electroproduction data much more difficult.
Moreover, there is no model-independent way to disentangle resonant and
non-resonant mechanisms in any given partial wave. Therefore, reaction models
are needed to isolate the resonant parts
in the double pion production amplitudes and evaluate the $N^{*}$
electromagnetic transition form factors.

Following the pioneering effort of Ref.\cite{So71}, several approaches have been
developed more recently for the description of double pion photo- and electroproduction in
the resonance region~\cite{Os1,Os2,La96,Ro96,Ro05,Hi03,Fi05}.
These efforts were based on a very limited amount of experimental data:
mostly on $W$ and $Q^{2}$ dependencies for fully integrated cross sections 
and on 
invariant mass distributions for various pairs of the final state hadrons. 
The reaction models used  meson-baryon
degrees of freedom: $N$, $\Delta$, $\pi$, $\sigma$  and $\rho$. 
Effective meson-baryon Lagrangian operators were constructed based on Lorentz
 invariance,
gauge invariance and crossing symmetry. For the description of experimental 
data, a
limited set of non-resonant meson-baryon
diagrams was used with amplitudes calculated from effective Lagrangians together with
contributions from several, mostly low-lying nucleon resonances ($M$ $<$ 1.6 GeV).
A general framework for the implementation of
other  meson-baryon degrees of freedom was proposed in Ref.\cite{Ro96}, 
but so-far has not been fully realized.

The meson-baryon diagrams in reaction models may
account for many partial waves. However, in any
reaction model we need to truncate the infinite set of meson-baryon diagrams, 
keeping just the relevant mechanisms.
Moreover, the choice of a particular effective Lagrangian, describing
meson-baryon interactions, 
may be done only at a phenomenological level.

At the distance scale appropriate for the size of hadrons, the
amplitudes of effective meson-baryon interactions contributing to the
reaction cannot be expanded over a small parameter except for a small
kinematic region near threshold $W$ and $Q^2$ $<$ 0.2 GeV$^2$ accessible for
chiral perturbation theory. This feature makes it impossible to select
contributing diagrams based on a perturbative expansion for the entire
double pion reaction phase space covered by the CLAS measurements. So-far, 
no approach has been developed
based on a fundamental theory that would allow either a description of an effective
Lagrangian or a selection of the contributing meson-baryon mechanisms from basic 
principles. We therefore
have to rely on fits to the now available detailed experimental data sets to 
develop reaction
 models that contain the relevant mechanisms.

The large  reaction phase space coverage of CLAS data opens up qualitatively
new opportunities for the analysis of charged double pion electroproduction. 
The
exclusive channel $e p \rightarrow e' p \pi^{+} \pi^{-}$
offers many observables for the analysis. The hadronic final state can be
projected on nine independent 1-fold differential cross sections in
each $W$ and $Q^{2}$ bin. For the first time, all of these
observables have become experimentally accessible~\cite{Ri03,Fe07}. By studying 
the kinematical dependencies
of the differential cross section and their correlations we are able to establish
the presence and strength of the relevant reaction mechanisms.
A phenomenological
approach \cite{Ri00,Mo01,Mo03,Mo04,Sh07,Az05,Mo06,Mo06-1,Mo07} was developed in collaboration
between Jefferson Lab and Moscow State University, referred to herein as "JM". 
This approach is intended to establish all
significant
mechanisms seen in the observables of charged double pion electroproduction, 
to isolate
the resonant parts of the amplitudes, and to derive the electrocouplings of nucleon
resonance transitions from fits of all measured observables
combined.

 The CLAS data presented in this paper were collected 
with binning resolutions over $W$ and $Q^{2}$ surpassing by almost an
order of magnitude what was achieved in previous measurements before the
experiments with CLAS. These
data will allow us to extract the electrocoupling amplitudes of the 
$N(1440)P_{11}$
and $N(1520)D_{13}$ states. A certain advantage of the double pion channel is that 
the amplitudes of the $N(1440)P_{11}$ resonance do not interfere with the 
high-mass tail
of the $\Delta(1232)P_{33}$ state, as is the case for the amplitudes in
single $\pi$ production.

The behavior of $N^{*}$ electrocouplings at small photon virtualities is
of particular interest. Studies of the magnetic transition form factor
for the  $\Delta(1232)P_{33}$ in Ref.\cite{Sa01} revealed considerable
meson-baryon dressing effects in addition to the 3-quark core contributions. The dressing is
expected to decrease with increasing $Q^{2}$. It is most pronounced at $Q^{2}$
$<$ 1.0~GeV$^{2}$ \cite{BLee08}. While the role of meson-baryon dressing effects has 
been studied based on 
the data on
electrocouplings of the $\Delta(1232)P_{33}$  state, for resonances heavier than 
the $\Delta(1232)P_{33}$, this remains an open question,
and is currently being addressed at Jefferson Lab through extensive theoretical efforts~\cite{Lee06,Lee07,Lee07a,BLee08}. Accounting for these effects is
a necessary step to probe quark and possibly gluonic degrees of freedom in baryons.
The comparison of constituent quark model
predictions \cite{Ca95,Can98,Az07} with the
measured $N^{*}$ electrocouplings \cite{Burk05,Lee04}, as well as the coupled-channels 
analysis of the data on reactions with hadronic probes \cite{BLee08}, 
suggest considerable meson-baryon dressing effects  at $Q^{2}$ $<$ 0.5~GeV$^{2}$ for the $N(1440)P_{11}$ and
$N(1520)D_{13}$ electrocouplings. Therefore, the information on these states at low
photon virtualities from charged double pion electroproduction may further elucidate the
relevant degrees of freedom in resonance excitation at hadronic
distance scales.

The data presented in this paper can also provide information on the 
$p \to \Delta$ axial transition form factor.
Current algebra~\cite{Ebata} relates the contact term in the set of non-resonant Born
terms for  $\pi \Delta$ isobar channels to the axial transition form factors.
These contact terms could be fit to the data within the framework of the JM
approach \cite{Mo06,Mo06-1}.
So-far, the $p \to \Delta$ axial transition form factor has been
determined mostly from neutrino-induced reactions \cite{Ki86,Be78,Al90}.
Axial transition form factors offer a complementary view of
baryon structure, seen in
the axial vector currents, while electroproduction experiments usually
access baryon structure through vector currents. Recently, lattice QCD
results have become available \cite{Ale07}, which make the experimental
study of the nucleon axial structure an important topic of hadronic physics.

\section{Experiment \label{select}}

The measurement was carried out using the CEBAF Large Acceptance
Spectrometer (CLAS) \cite{Mecking} at the Thomas Jefferson National
Accelerator Facility (Jefferson Lab). CLAS 
provides almost
complete angular coverage in the center-of-mass frame. It is well suited for
conducting experiments that require detection of two or more
particles in the final state. Such a detector and the continuous beam
produced by CEBAF provide excellent conditions for measuring the  $e
p \rightarrow e' p' \pi^{+} \pi^{-}$  cross section
by detecting the outgoing electron, proton and at least one pion in coincidence.

\subsection{Apparatus \label{apparatus}}

\begin{figure}[h!]
\begin{center}
\epsfig{file=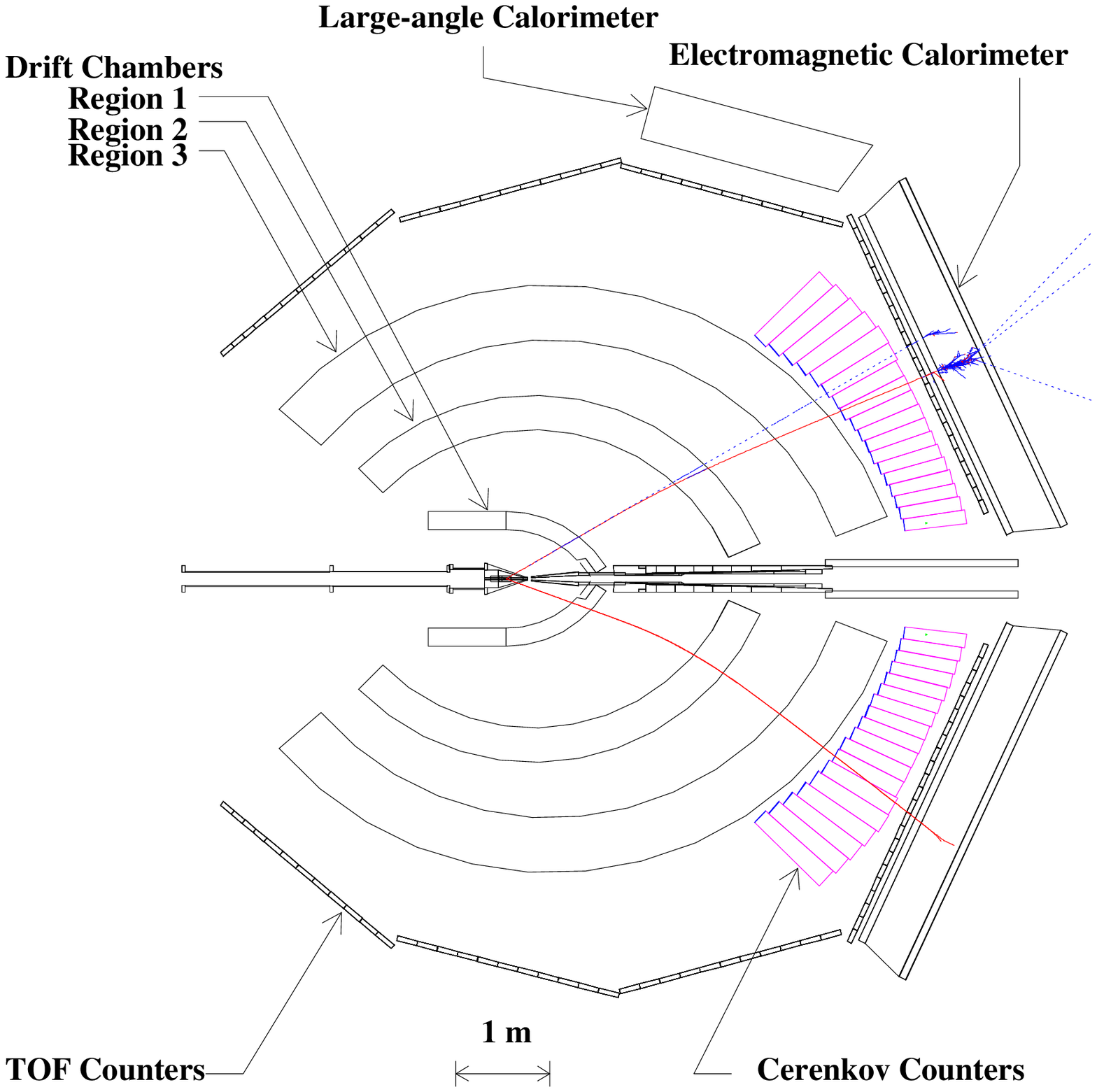,width=8cm}
\epsfig{file=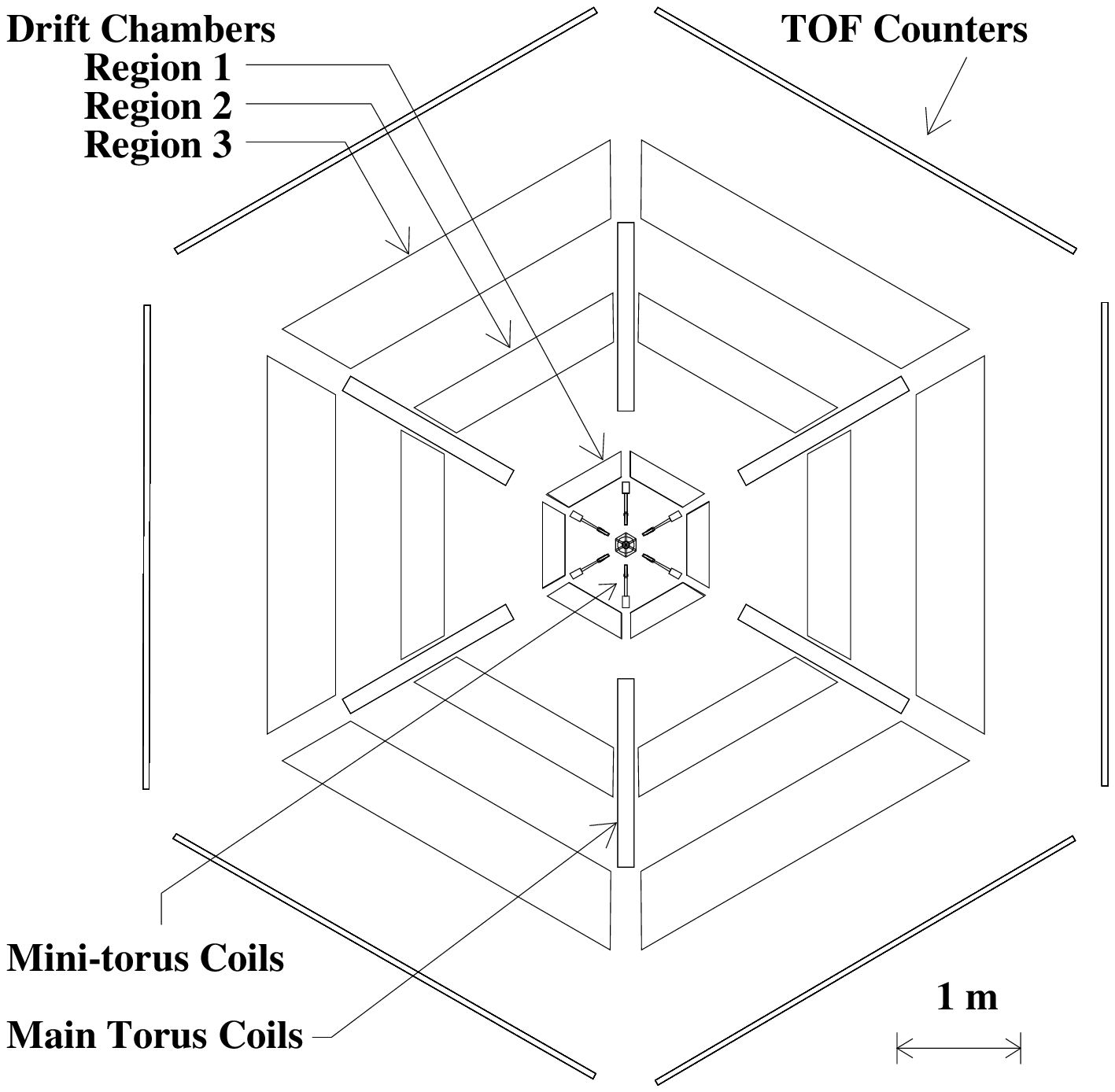,width=8cm}
\end{center}
\vspace{-0.6cm} \caption{\small (color online) Cross-sectional views of the CLAS detector. The top panel shows a cut
along the beam line and through the mid-plane of two opposite sectors.
The bottom panel shows a cut perpendicular to the beam line and through
the nominal target center. Descriptions of the detector elements are
given in the text of Sect.~\ref{apparatus}. \label{detector}}
\end{figure}

The main magnetic field of CLAS is provided by six superconducting
coils, symmetrically arranged around the beam line, which generate an approximately toroidal field in the
azimuthal direction around the beam axis. The gaps between the coil
cryostat are instrumented with six identical detector packages, referred to here as "sectors", as shown in Fig.~\ref{detector}.
Each sector consists of three regions (Region 1, Region 2, Region 3) 
of Drift Chambers (DC) \cite{Mestayer} to determine the trajectories of the charged
particles as they travel from the target outward in magnetic field, a Cherenkov 
Counter \cite{Adamsr} for electron
identification, Scintillator Counters (SC) \cite{Smith} for charged
particle identification using the Time-of-Flight (TOF) method, and 
an Electromagnetic Calorimeter \cite{Amarian} for electron
identification. The
liquid-hydrogen target was located in the center of the detector. To reduce the electromagnetic background resulting
from M{\o}ller scattering off atomic electrons, a second smaller
normal-conducting toroidal magnet (mini-torus) was placed
symmetrically around the target. This additional magnetic field
prevented M{\o}ller electrons from reaching the sensitive detector volume.
A totally absorbing Faraday cup, located at the very end of beam
line, was used to determine the integrated beam charge passing
through the target. The CLAS detector provides $\approx$ 80 \% of
$4\pi$ solid-angle coverage. The efficiency of detection and
reconstruction for single stable charged particles in the fiducial regions of
CLAS is greater than 95 \%. The combined information from tracking in the DC
and the SC systems allows us to reliably separate protons from
positive pions. Additional constraints for
event selection come from the overdetermined kinematics, which allows
use of the missing mass technique.

Due to possible slight misalignments in the DC positions and small inaccuracies in
the description of the torus magnetic field, the reconstructed momentum and angle of
particles may have small systematic deviations from the physical value.
To correct these deviations, elastic electron-proton scattering was
checked and the electron 3-momenta were corrected to ensure the proper mass peak
position for the recoil proton ~\cite{Golova}.
The proton energy losses in CLAS were estimated from a simulation of proton
propagation through the detector materials in kinematics corresponding to charged double pion electroproduction.

\subsection{Data taking and data reduction}

This analysis is based on data taken during the 1999
e1c run period. The 1.515~GeV electron beam at a current of 3~nA was incident
 on a
5-cm-long liquid-hydrogen target corresponding to an instantaneous
luminosity of $\sim 4 \times
10^{33}$~cm$^{-2}$s$^{-1}$. The size of the beam spot at the target was
$\sim 0.2$~mm, with position fluctuations $<$ $\pm0.04$~mm. The main
torus current was set at 1500~A, which created a magnetic field of
about 0.8 T at polar angles of 20$^{\circ}$ that decreased with
increasing polar angle. The CLAS event
readout was triggered by a coincidence of signals from an
electromagnetic calorimeter module and a threshold gas Cherenkov
counter in one of the six sectors, generating a total event rate of $\sim 2$~kHz. The
number of accumulated triggers at these detector settings was about
$4.2 \times 10^{8}$. These data were further analyzed to extract the
differential cross sections for the $e p \rightarrow e' p' \pi^{+}
\pi^{-}$ reaction.

\section{Event selection \label{ident}}

The $e p \rightarrow e' p' \pi^{+}
\pi^{-}$ reaction is selected by measuring the scattered electron, as well as the proton and
 $\pi^{+}$ in the hadronic final state. In the magnetic field configuration used in 
 this
measurement, the negatively charged pions have a smaller probability for detection
than the positively
charged particles. However, the process is kinematically overconstrained and the 
detection of all
particles in the final state is not required for an unambiguous identification of
the exclusive reaction. To retain maximum acceptance, we chose to not require 
detection
of the $\pi^-$, but infer the presence and kinematics of the undetected $\pi^-$
by computing the mass of the undetected particle from 4-momentum conservation and its
charge from charge conservation. The two other topologies, with either undetected proton
or undetected $\pi^+$, have considerably lower rates and were used for systematics studies
and for cross checks, but are not included in the determination of the final cross sections.

\begin{figure}[htp]
\begin{center}
\includegraphics[width=8.cm]{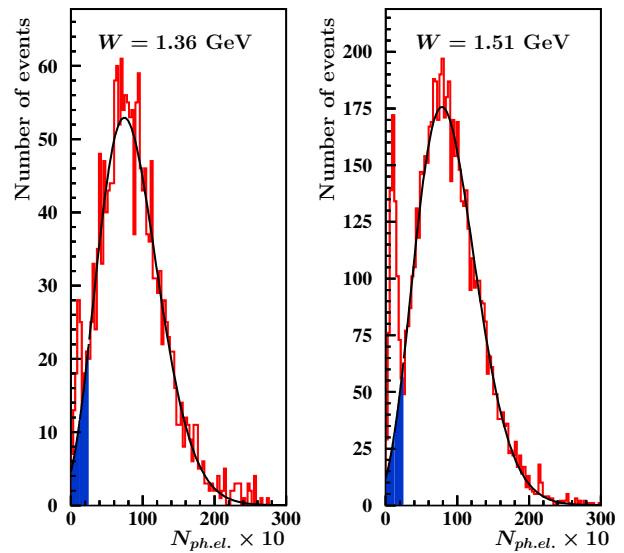}
\vspace{-0.1cm}
\caption{\small(color online) Distribution of photoelectrons in the 
Cherenkov counter for
0.3 $<$ $Q^{2}$ $<$ 0.4~GeV$^{2}$  in two $W$ bins centered at the values shown
in the plot. The curves represent a Poisson fit. \label{nphe}}
\end{center}
\end{figure}

\begin{figure}[h!]
\begin{center}
\epsfig{file=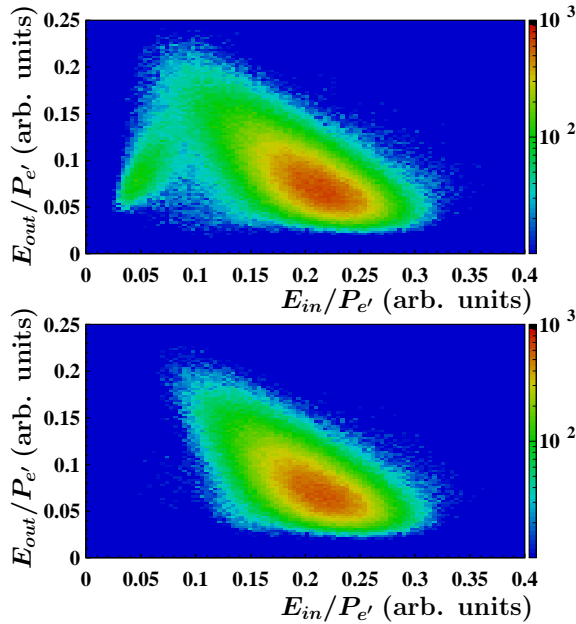,width=7.5cm} 
\caption{\small(color online) Distributions for
the energies deposited in the outer part ($E_{out}$) versus the energies deposited 
in the inner part ($E_{in}$) of the EC normalized to the momenta of the outgoing particles
($P_{e'}$). The distribution for the accumulated triggers is shown in the top plot,
while the distribution for the events selected, applying the photoelectron cut,
is shown in the bottom plot. \label{eoutvsein}}
\end{center}
\end{figure}

For each event we identify as  the electron candidate the first coming 
negatively charged
particle detected in the electromagnetic calorimeter and the Cherenkov counter. 
To select true electrons, we apply a cut
in the number of photoelectrons ($N_{phe} \geq 2.5$) produced by the Cherenkov
light signal in the photomultipliers. This cut also eliminates a small fraction of 
electrons ($<$ 6 \%), as shown in
Fig.~\ref{nphe}. The shadowed areas correspond to
the cut-out electrons. A special procedure was developed to account for these
electrons in the evaluation of the reconstruction efficiency, based on
the extrapolation of the photoelectron spectra into the cut-out areas
using a fit based on a Poisson distribution. The quality of electron
identification may be seen in Fig.~\ref{eoutvsein}, where we display the
energies deposited in the outer part of the calorimeter
versus the energies deposited in the inner part of the EC, normalized to the momenta of the
incoming particles.
A spot from minimum-ionizing pions, clearly seen in inclusive electron events
(top part of Fig.~\ref{eoutvsein})
disappears after applying the described cuts for electron selection
(bottom part of Fig.~\ref{eoutvsein}).

\begin{figure}[h!]
\begin{center}
\includegraphics[width=7.5cm]{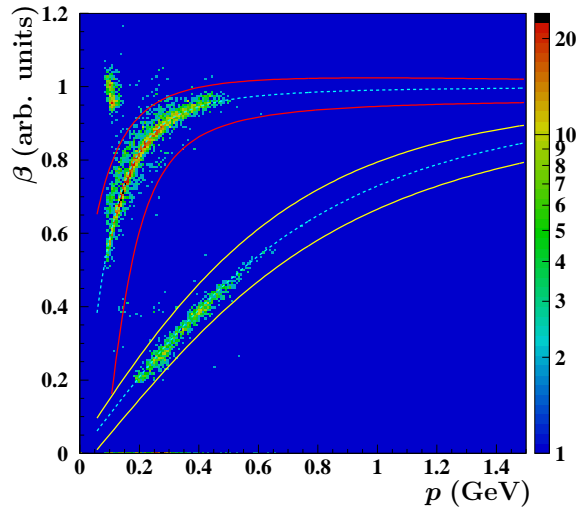}
\caption{\small (color online) $\beta$ versus momentum for positively charged hadrons;
solid
curves show the cuts used to identify pions and protons;
dashed curves show the explicit $\beta$ vs p relationships, using the pion and
proton mass, respectively. The events near $\beta$=1 and p=0.1 GeV are
positrons, e.g. from $\pi^{0}$ Dalitz decay. The data were collected in a single
scintillator bar. \label{bvspnorm}}
\end{center}
\end{figure}

Using information from the time-of-flight scintillators and the path length
determined by tracking, the particle's
velocity $(\beta)$ was determined. The information from the drift
chambers, combined with the known magnetic field, provides a measurement of the
particle momentum $(p)$. Relativistic relations between particle
mass, momentum and velocity were used in order to determine the particle's 
mass, allowing us
to separate
pions, kaons, and protons in the kinematic range covered by the measurement.
Fig.~\ref{bvspnorm} shows the velocity versus the momentum for
positively charged particles.

\begin{figure}[h!]
\begin{center}
\epsfig{file=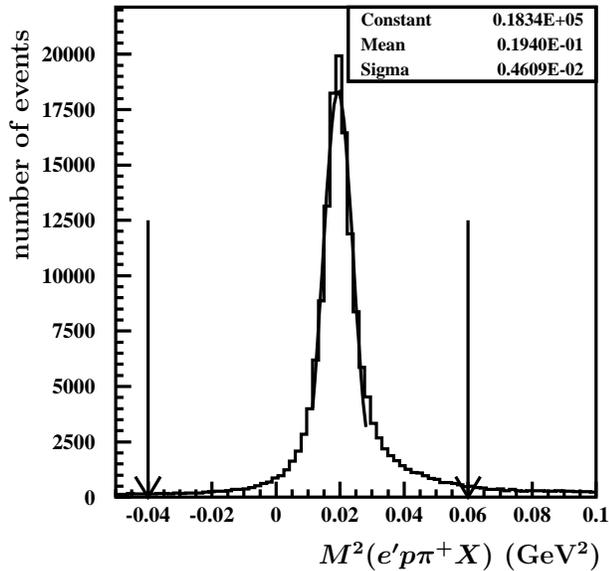,width=8cm} \caption{\small
The distribution of squared missing mass $M^{2}(e'p\pi^{+}X)$ (GeV$^{2}$). 
The arrows
show the exclusivity cut. \label{missmass}}
\end{center}
\end{figure}

After identification of the three particles,
events were selected with one electron,
one proton and one $\pi^{+}$. The squared missing mass distribution is
shown in Fig.~\ref{missmass}, which clearly shows the pion mass peak and also
indicates that multi-pion background ($>$ 2$\pi$) contributes
less than 1 \% to the total number of charged double pion events selected by the cuts.
The small background is related to the kinematic coverage of our experiment $W$ $<$
1.57 GeV, where the exclusive channels with more than two pions in the final state are
suppressed due to their thresholds.
The almost negligible contribution from the multi-pion background and the rather small 
radiative
effects (see Sect.~\ref{radiative}) allowed us to apply a wide
exclusivity cut over the squared missing mass distribution, shown in Fig.~\ref{missmass}
by the two arrows, in order to collect the majority of the 2$\pi$ events. 

The CLAS detector has an
active detection solid angle smaller than $4\pi$ due
to the space filled with the torus magnet coils. The angles
covered by the torus magnet coils are not equipped with any detection system and
therefore give rise to inactive areas. The boundaries
of the active areas are not well defined and do not provide regions for particle
reconstruction with full reconstruction efficiency. Therefore, for
the analysis we accept only events inside specific fiducial areas whose contours are
defined by parameterizations of the kinematic variables of each particle. 
Within these
well-defined regions, acceptances and track reconstruction efficiencies are well understood 
using Monte Carlo simulations.



\begin{figure}[h!]
\begin{center}
\epsfig{file=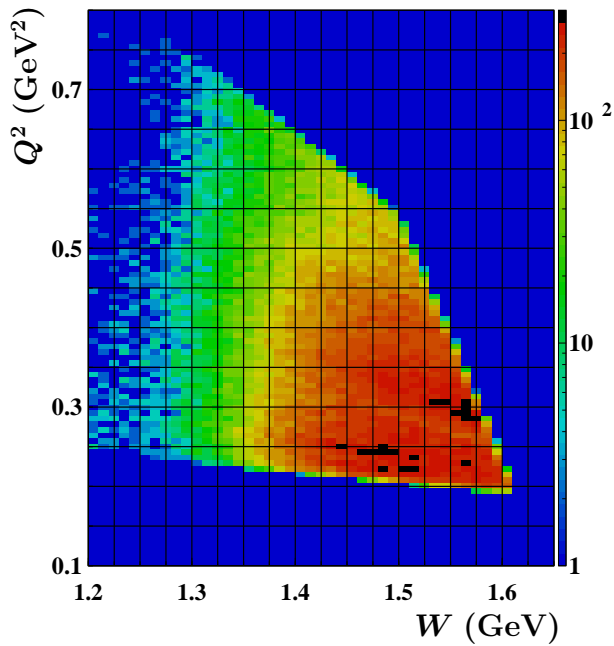,width=8cm} \caption{\small 
(color online)  $Q^{2}$
(GeV$^{2}$) versus $W$  (GeV) distribution for the selected $2\pi$ events. The grid
shows the binning used for the evaluation of the cross section. Only cells inside the allowed phase space were used for that purpose.
\label{q2vsw}}
\end{center}
\end{figure}

After all selections have been applied, there remain about
130,000 exclusive $p \pi^{+} \pi^{-}$ events. Fig.~\ref{q2vsw} shows
the $Q^{2}$ versus $W$ distribution for the selected  $2\pi$
events.

\section{Charged double pion electroproduction cross sections \label{xsect}}
The kinematics of the 3-body $p \pi^{+} \pi^{-}$ final state are unambiguously
determined by five independent variables~\cite{Byc}. However, the choice of
these variables is not unique. In this section we specify the kinematic variables
we employ to describe the $p \pi^{+} \pi^{-}$ final state and related phase space element for the
5-fold differential
cross section. Note, that the double pion cross sections for virtual photon 
absorption are
5-fold differential, while the double pion electroproduction cross sections are
7-fold differential,
since they contain the additional variables $W$ and $Q^{2}$. Then we
describe the procedure to
evaluate the 5-fold differential charged double pion cross section from
experimental data from the 7-dimensional event distributions.
The 5-fold differential cross section contains complete information on
double pion production at fixed $Q^2$ and $W$. However, the limited statistics
do not
allow direct study of the 5-fold differential cross sections. Therefore, for 
the physics
analysis we use various 1-fold differential cross sections, obtained by 
integrating the 5-fold differential cross section over the four other variables.

\begin{figure}[htp]
\begin{center}
\includegraphics[width=8cm]{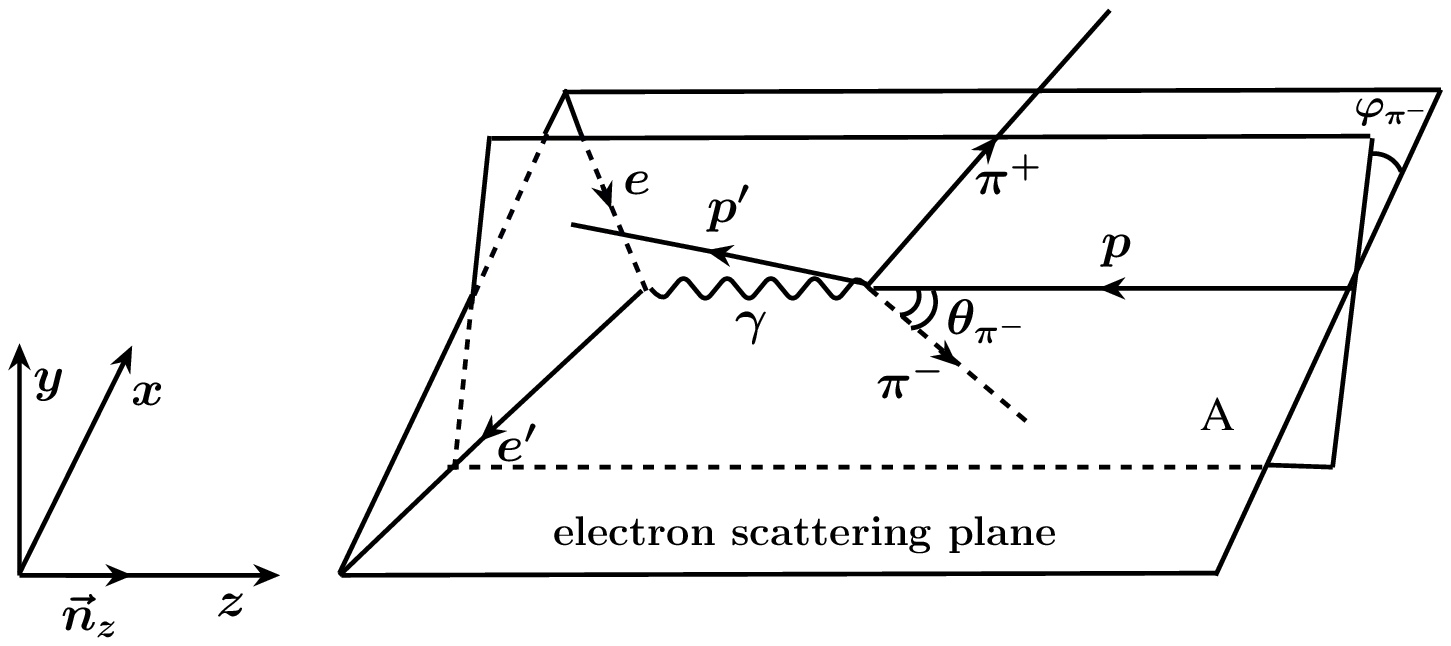}
\includegraphics[width=8cm]{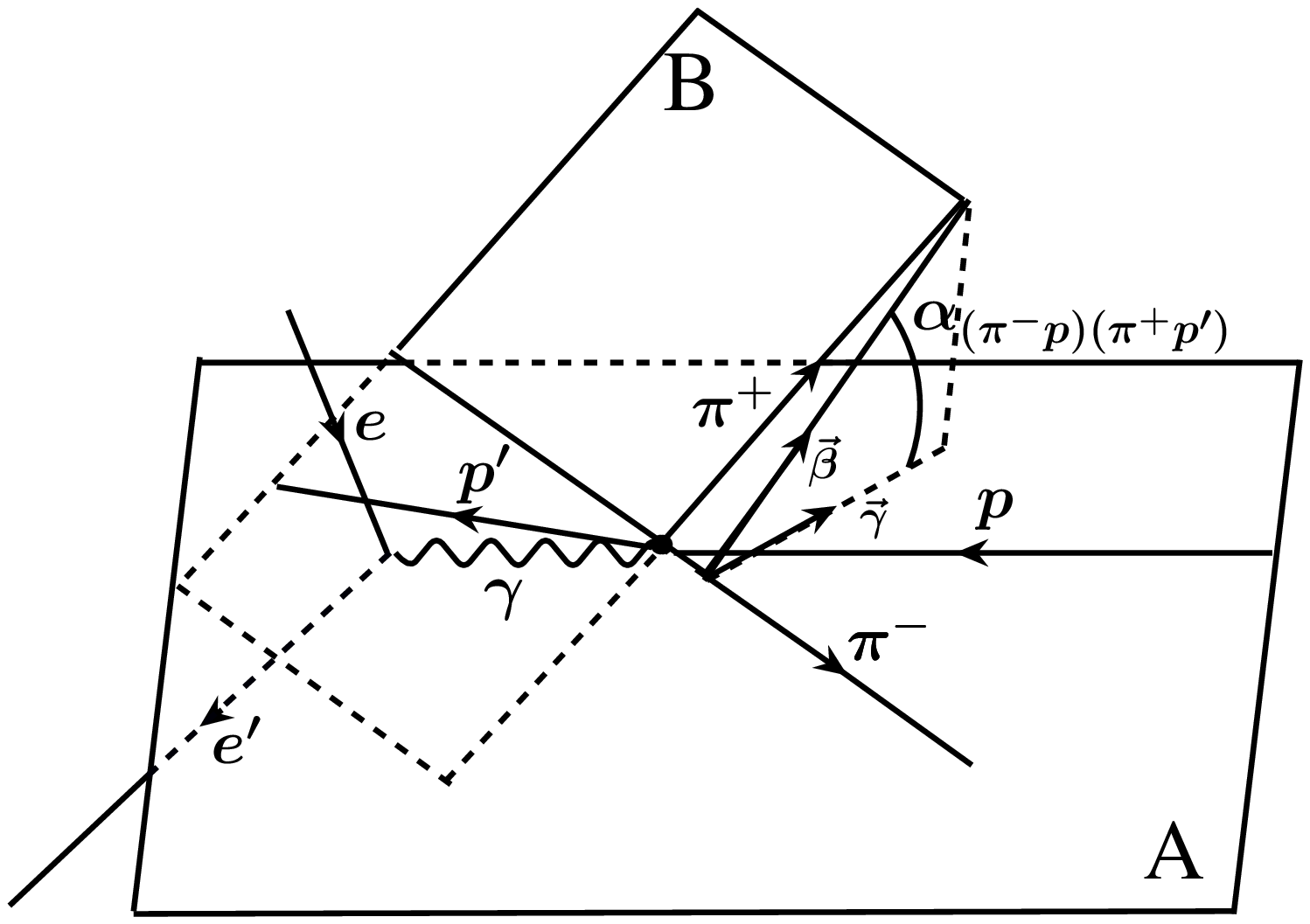}
\caption{\small Kinematic variables for the 
$e p \rightarrow e' p' \pi^{+} \pi^{-}$ reaction 
(choice 2 in Sect.~\ref{xsect_var}). The top plot shows the 
$\pi^{-}$ spherical angles $\theta_{\pi^{-}}$ and $\varphi_{\pi^{-}}$,
while the bottom plot
shows the angle $\alpha_{(\pi^{-} p)(\pi^{+} p')}$ of the plane defined by the 
momenta of 
the final $p$ $\pi^{+}$ pair
with respect to the plane comprised by the momenta of
the initial proton $p$ and $\pi^{-}$. \label{kinematic}}
\end{center}
\end{figure}

\subsection{Kinematic variables \label{xsect_var}}

We adopt the following set of variables to describe the 3-body final state:
\begin{itemize}
\item invariant mass of the first pair of
particles $M_{12}$;
\item invariant mass of the second pair of particles $M_{23}$;
\item the first particle solid angle $\Omega$;
\item the angle between two planes: one of
them (plane A) is defined by the 3-momenta of 
the virtual photon and the first hadron, and the second
plane (plane B) is defined by the 3-momenta of the 
two other hadrons (see Fig.~\ref{kinematic}).
\end{itemize}

\noindent
We use three different assignments for the first, second and
third final state hadrons:

\begin{itemize}
\item invariant mass of the $p\pi^{+}$ pair, invariant mass of the $\pi^{+}\pi^{-}$
pair, the final proton spherical angles $\theta_{p}$ and $\varphi_{p}$ and
the angle $\alpha_{(p p')(\pi^{+}\pi^{-})}$
between the two planes: B, composed by the momenta of 
the $\pi^{+}\pi^{-}$ pair, and A, composed by the momenta of the
initial and final protons (choice 1) ;
\item invariant mass of the $\pi^{+}\pi^{-}$
pair, invariant mass of the $p\pi^{+}$
pair, $\pi^{-}$ spherical angles
$\theta_{\pi^{-}}$ and $\varphi_{\pi^{-}}$
and the angle
$\alpha_{(p \pi^{-})(p' \pi^{+})}$
between the two planes:  B, composed by the momenta of
the final state proton $p'$ and  $\pi^{+}$, and A, composed by
the initial state proton $p$ and $\pi^{-}$ (choice 2);
\item invariant mass of the $p\pi^{+}$
pair, invariant mass of the $p\pi^{-}$
pair, $\pi^{+}$ spherical angles
$\theta_{\pi^{+}}$ and $\varphi_{\pi^{+}}$
and the angle
$\alpha_{(p \pi^{+})(p' \pi^{-})}$
between the two planes: B, composed of the momenta of
the final state proton $p'$ and  $\pi^{-}$ and A, composed by
the initial state proton $p$ and $\pi^{+}$ (choice 3).
\end{itemize}

The 5-fold differential cross sections were obtained for all three
sets of variables.
The emission angles for the final particles in the second set of
variables are shown in  Fig.~\ref{kinematic}. For the other sets the emission
angles are defined in a similar way.
In the physics analysis, described in Sect.~\ref{anal}, the second set of variables
is used. These variables are
suitable for the description of charged double pion electroproduction through a $\pi^{-}
\Delta^{++}$ intermediate state, which represents the main contributor
of all isobar channels in  the kinematic area covered by our data.
The relations between the four momenta of the final state hadrons and the kinematic
variables may be found in Appendix A.

\subsection{Evaluation of charged double pion cross sections \label{evsect}}
The selected double $\pi$ events were collected in
7-dimensional cells, composed of $W$, $Q^{2}$, invariant masses of the
first pair $M_{12}$ and the second
pair $M_{23}$ of the final state particles, solid angle for
 the first final state particle, and the angle
$\alpha_{\it{i}}$ between the two planes A and B. The cross sections were determined
only for those $W$ cells, that are fully inside the kinematically allowed area. Special
procedures were developed in order to evaluate the 1-fold differential cross sections for
the final state particle invariant mass values near edges of the reaction phase
space (see Sect.~\ref{genev},\ref{improve}).  
All frame-dependent variables
and the cross sections were evaluated in the center-of-mass frame.

For the second choice of kinematic
variables, the 7-fold differential cross sections
$\frac{d\sigma}{dWdQ^{2}d^{5}\tau_{2}}$,
$d^{5}\tau_{2}=dM_{p\pi^{+}}dM_{\pi^{+}\pi^{-}}d\Omega_{\pi^{-}}
d\alpha_{[p \pi^{-}] [p' \pi^{+}]}$ are given by the number of $p\pi^{+}\pi^{-}$
events $\Delta N$  and efficiencies  $\epsilon$ in the 7-dimensional cells as:

\begin{eqnarray}
\frac{d\sigma}{dWdQ^{2}d^{5}\tau_{2}} = \nonumber \\
\frac{1}{\epsilon \cdot \epsilon_{ch} \cdot R}
\frac{\Delta N }{\Delta W \Delta Q^{2} \Delta^{5} \tau_{2} L } \textrm{ .}
\label{expcrossect}
\end{eqnarray}
The number of events
inside the
7-dimensional cells were corrected for contamination from the target walls,
which was measured in separate runs with an empty target cell. The efficiency $\epsilon$ in
any 7-dimensional cell was determined in detailed Monte Carlo
simulations. The inactive zones of CLAS and all
 cuts on phase space used in the event selection were included in the
 efficiency evaluation. The factor $\epsilon_{ch}$ accounts for
the Cherenkov counter efficiency, which was determined separately.
 $\it{R}$  accounts for
radiative corrections. The integrated luminosity $L$ was determined from the total beam 
charge $Q$ 
measured in the Faraday cup, combined with the information on target length and target density:
\begin{equation}
L=Q\frac{l_{t}D_{t}N_{A}}{q_{e}M_{H}},
\label{lum}
\end{equation}
 where $q_{e}$ is the elementary charge,  $D_{t}$ is the
density of hydrogen ($D_{t} = 0.073 $ g/cm$^{3}$),
$l_{t}$ is the length of the target ($l_{t} = 5$ cm),
$M_{H}$ is the molar density of
hydrogen ($M_{H} = 1$ g/mol), and  $N_{A}$ is Avogadro's
number. The luminosity value was verified by reproducing elastic $\it{ep}$ 
cross sections with the same data set.
A comparison of the elastic ep scattering cross sections determined from our data with
a parameterization of world data given in Ref.~\cite{Bos} showed agreement within
better than 5 \%.
$\Delta
W$ and $ \Delta Q^{2}$ are bins over $\it{W}$ and $Q^{2}$, and
 $\Delta^{5} \tau_{2}$ represents the element of
 hadronic 5-dimensional phase space for the second choice of kinematic variables:

\begin{equation}
\Delta ^{5}\tau_{2} = \Delta M_{p\pi^{+}} \Delta
M_{\pi^{+}\pi^{-}} \Delta
\rm{cos}(\theta_{\pi^{-}}) \Delta
\varphi_{\pi^{-}} \Delta \alpha_{(p \pi^{-})(p' \pi^{+})}.
\label{multibin} 
\end{equation}

The 7-fold differential cross sections for the other two choices of 
kinematic variables
listed in
Sect.~\ref{xsect_var} may be obtained from eq. (\ref{multibin}), by 
substituting the phase space element $\Delta^{5} \tau_{2}$ with:
\begin{eqnarray}
\Delta ^{5}\tau_{1} = \Delta M_{p\pi^{+}} \Delta
M_{\pi^{+}\pi^{-}} \Delta
\rm{cos}(\theta_{p}) \Delta
\varphi_{p} \Delta \alpha_{(p p')(\pi^{-} \pi^{+})}; \nonumber \\
\\
\Delta ^{5}\tau_{3} = \Delta M_{p\pi^{-}} \Delta
M_{p \pi^{+}} \Delta
\rm{cos}(\theta_{\pi^{+}}) \Delta
\varphi_{\pi^{+}} \Delta \alpha_{(p \pi^{+})(p' \pi^{-})}. \nonumber
\label{others5d}
\end{eqnarray}

In the single photon exchange approximation, the 7-fold differential
electron scattering cross section is related
to the hadronic 5-fold differential cross section as \cite{Am89}:

\begin{eqnarray}
\frac{d\sigma}{dM_{p\pi^{+}}dM_{\pi^{+}\pi^{-}}d\Omega_{\pi^{-}}
d\alpha_{(p \pi^{-})(p' \pi^{+})}} = \nonumber \\
\frac{1}{\Gamma_{v}}
\frac{d\sigma}{dWdQ^{2}dM_{p\pi^{+}}dM_{\pi^{+}\pi^{-}}d\Omega_{\pi^{-}}
d\alpha_{(p \pi^{-})(p' \pi^{+})}}  \textrm{ ,}
\label{fulldiff}
\end{eqnarray}
where $\Gamma_{v}$ is
virtual photon flux, given by

\begin{equation}
\Gamma_{v} =
\frac{\alpha}{4\pi}\frac{1}{E_{b}^{2}M_{p}^{2}}\frac{W(W^{2}-M_{p}^{2})}
{(1-\varepsilon)Q^{2}} \textrm{ ,}
\label{flux}
\end{equation}
and $\alpha$ is the fine structure constant, $E_{b}$ is the beam energy, $M_{p}$ is the proton
mass, and $\varepsilon$ is the virtual photon transverse polarization given by

\begin{equation}
\varepsilon = \left( 1 + 2\left( 1 +
\frac{\nu^{2}}{Q^{2}} \right)
\rm{tan}^{2}\left(\frac{\theta_{e}}{2}\right) \right)^{-1} \textrm{ ,}
\label{polarization}
\end{equation}
where $\nu$ is the virtual photon energy and $\theta_{e}$ is 
the electron scattering angle in the
laboratory frame. $W$, $Q^{2}$ and $\theta_{e}$ were
taken at their respective bin centers.




The limited
statistics do not allow use of correlated
multi-fold differential cross sections for physics analysis.
In the physics analysis we instead used 1-fold differential
cross sections,  obtained after integration of the 5-fold
differential cross sections over four kinematic variables. The number of
5-dimensional bins contributing to the individual bins of the 1-fold differential
cross sections, range from 375 at $W=1.31$~GeV to 1600 at $W > 1.38$~GeV.
 Summing up events in all 5-dimensional bins, reasonable
statistical accuracy is achieved for the 1-fold
differential cross sections 
(see Figs. ~\ref{empty_zone_effect},~\ref{comparinitfin},~\ref{2_3_grids_mass},~\ref{9sectok},~\ref{9secnstbck}).
 We obtained in each ($W$, $Q^{2}$) bin
 covered by measurements a set of nine 1-fold differential cross sections, consisting of
 $\pi^{+} \pi^{-}$, $p \pi^{+}$, and $p \pi^{-}$ mass distributions,
 $\theta_{\it{i}}$ angular distributions, as well
 as 3 distributions over angles $\alpha_{\it{i}}$ ($i$=1,2,3), where the index
 $\it{i}$
 stands for the  $\it{i}$-th set of kinematic variables, defined in 
 the Sect.~\ref{xsect_var}.
These 1-fold differential cross sections represent the integrals from the 
5-fold differential cross section over four variables as:

$$
\frac{d\sigma}{dM_{\pi^{+}\pi^{-}}} =
\int\frac{d^{5}\sigma}{d^{5}\tau_{2}}d^{4}\tau_{\pi^{+}\pi^{-}}
$$
$$
d^{4}\tau_{\pi^{+}\pi^{-}} =
dM_{\pi^{-}p}d\Omega_{\pi^{-}}d\alpha_{(p \pi^{-})(p' \pi^{+})};
$$
$$
\frac{d\sigma}{dM_{\pi^{+}p}} =
\int\frac{d^{5}\sigma}{d^{5}\tau_{2}}d^{4}\tau_{\pi^{+}p}
$$
\begin{equation}
\label{inegr5diff}
 d^{4}\tau_{\pi^{+}p} = \\
dM_{\pi^{+}\pi^{-}}d\Omega_{\pi^{-}}d\alpha_{(p \pi^{-})(p' \pi^{+})} ;
\end{equation}
$$
\frac{d\sigma}{dM_{\pi^{-}p}} =
\int\frac{d^{5}\sigma}{d^{5}\tau_{3}}d^{4}\tau_{\pi^{-}p}
$$
$$
 d^{4}\tau_{\pi^{-}p} =
dM_{\pi^{+}\pi^{-}}d\Omega_{\pi^{+}}d\alpha_{(p \pi^{+})(p' \pi^{-})};
$$
$$
\frac{d\sigma}{d(-\rm{cos}\theta_{i})} =
\int\frac{d^{5}\sigma}{d^{5}\tau_{i}}d^{4}\tau_{i};\;
 d^{4}\tau_{i} =
dM_{\pi^{+}\pi^{-}}dM_{\pi^{+}p}d\varphi_{i}d\alpha_{i}
$$
$$
\frac{d\sigma}{d(\alpha_{i})} =
\int\frac{d^{5}\sigma}{d^{5}\tau_{i}}d^{4}\tau_{i};\;
 d^{4}\tau_{i} =
dM_{\pi^{+}\pi^{-}}dM_{\pi^{+}p}d\Omega_{i}.
$$
In the actual cross section calculations the
integrals in eq.~(\ref{inegr5diff}) were
substituted by the respective sums over
the 5-dimensional kinematic bins for the hadronic
cross sections.

 All cross sections represent
independent 1-dimensional projections of the 5-fold differential cross sections.
Any of the nine 1-fold differential cross sections provides independent information
and cannot be computed from the 8 remaining projections.

\begin{figure*}
\begin{center}
\epsfig{file=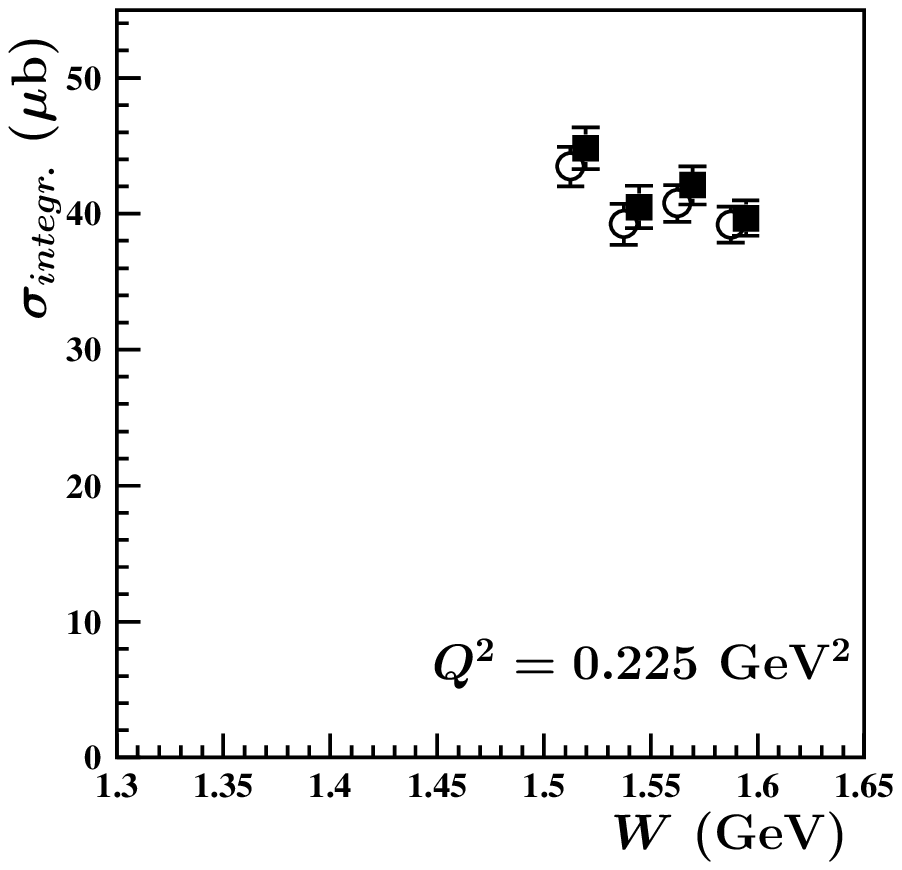,width=6.1cm}
\epsfig{file=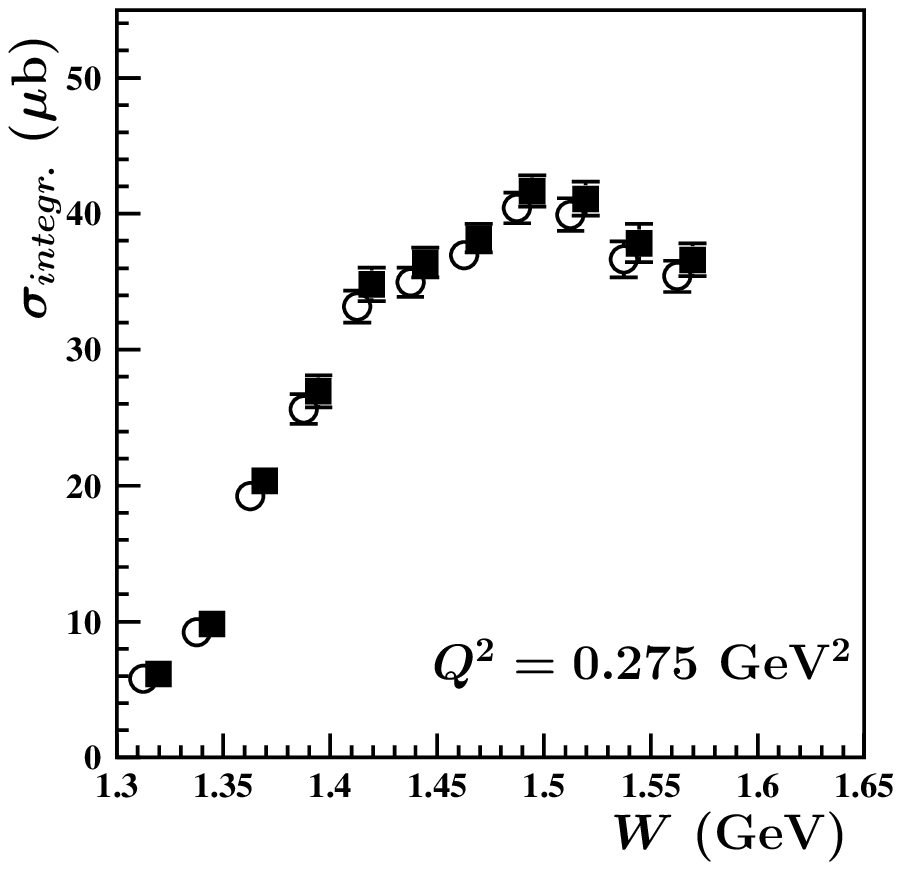,width=6.1cm}
\epsfig{file=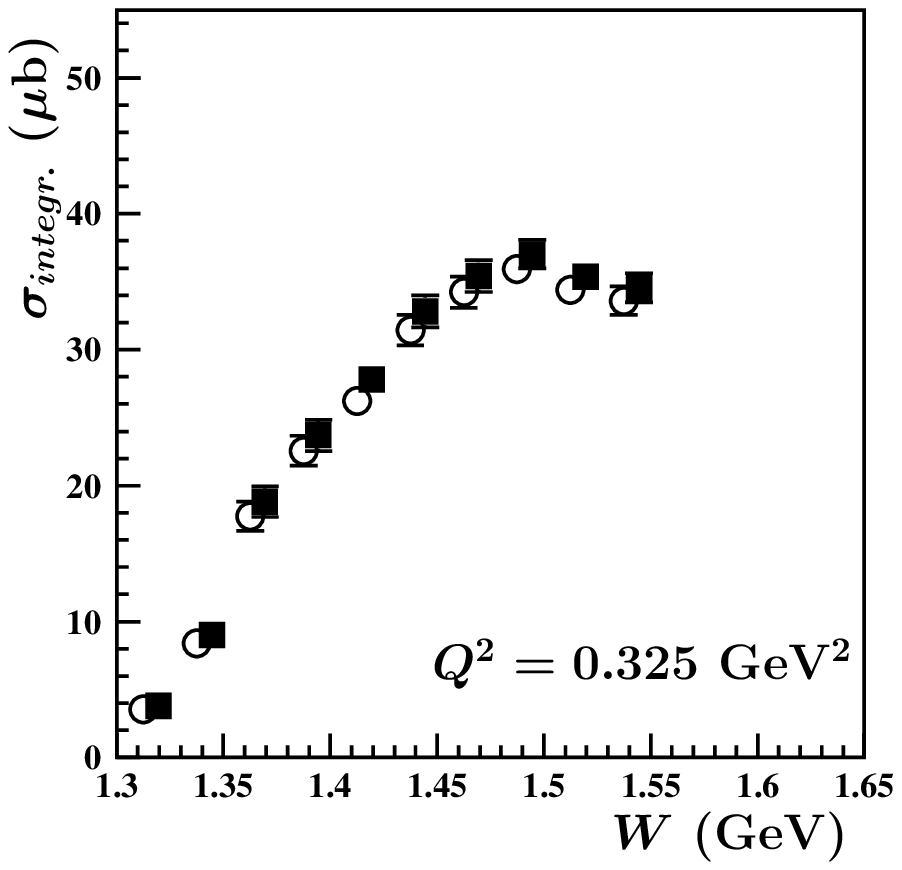,width=6.1cm}
\epsfig{file=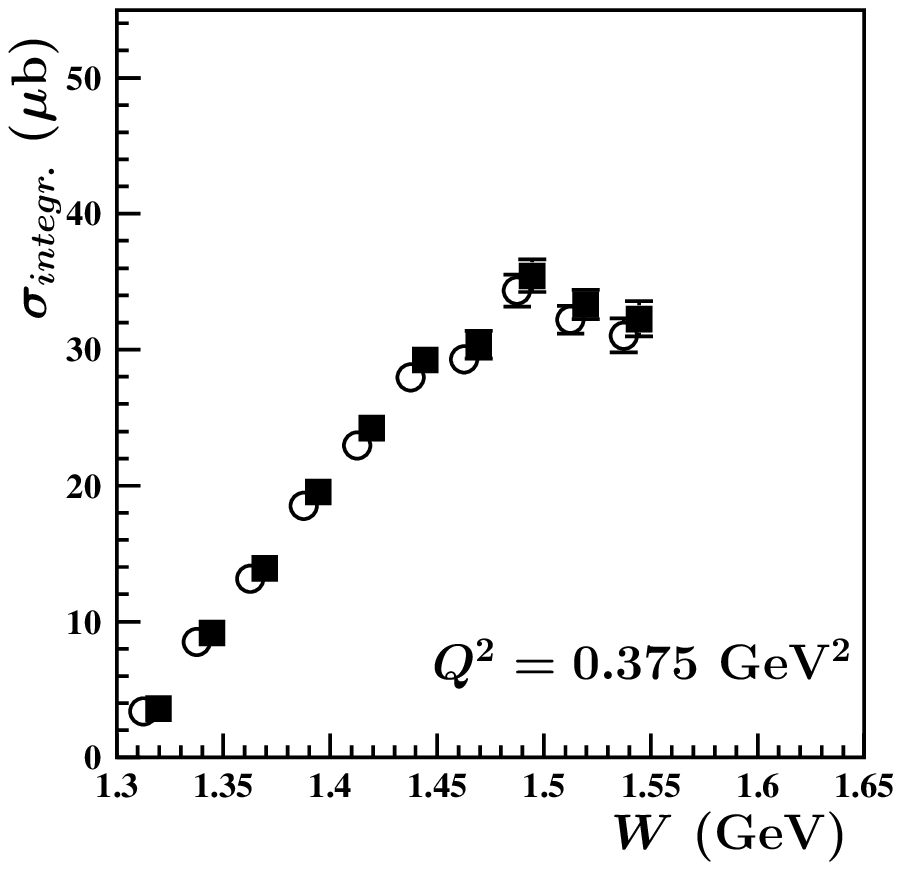,width=6.1cm}
\epsfig{file=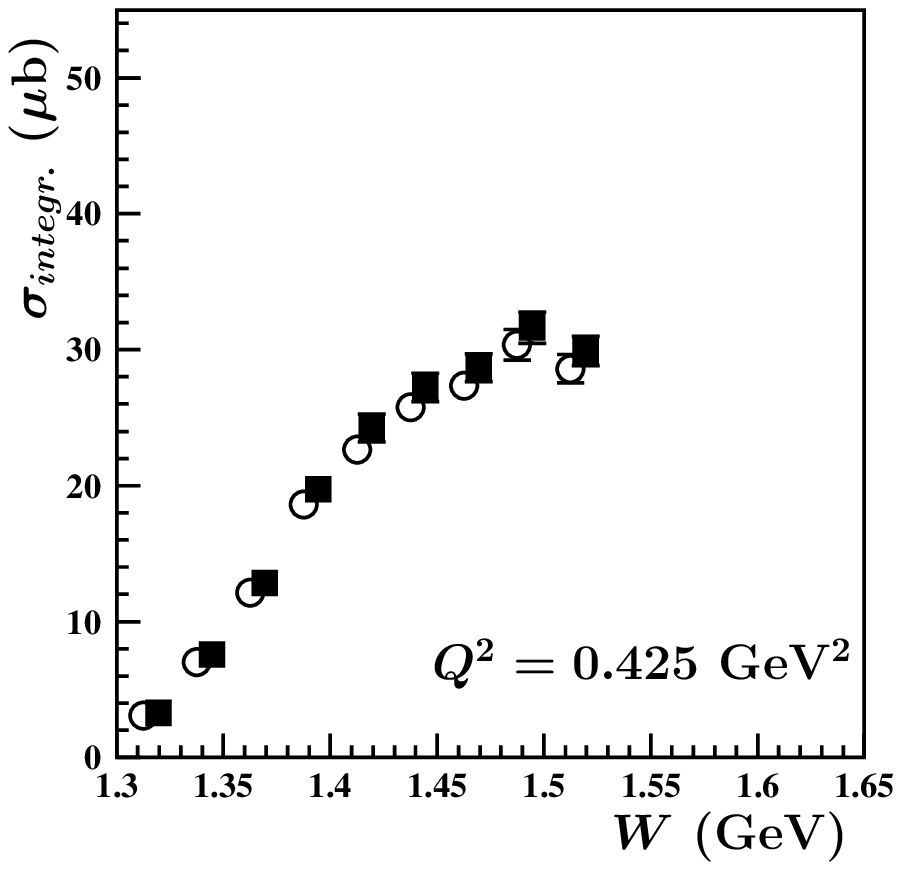,width=6.1cm}
\epsfig{file=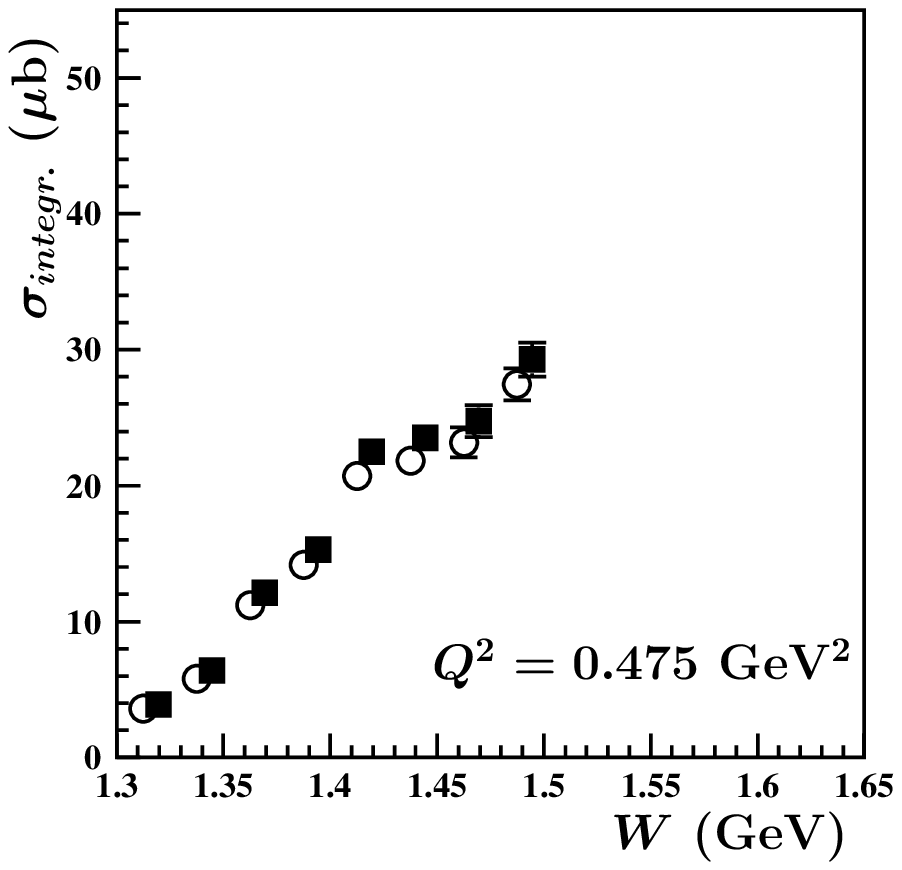,width=6.1cm}
\epsfig{file=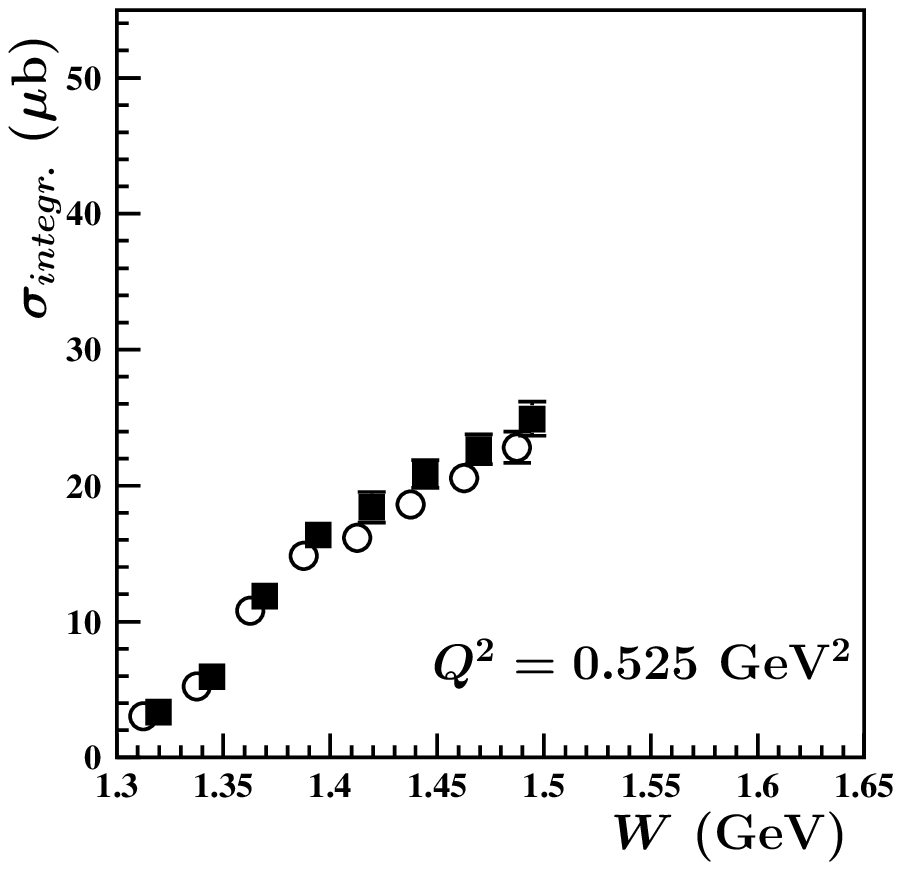,width=6.1cm}
\epsfig{file=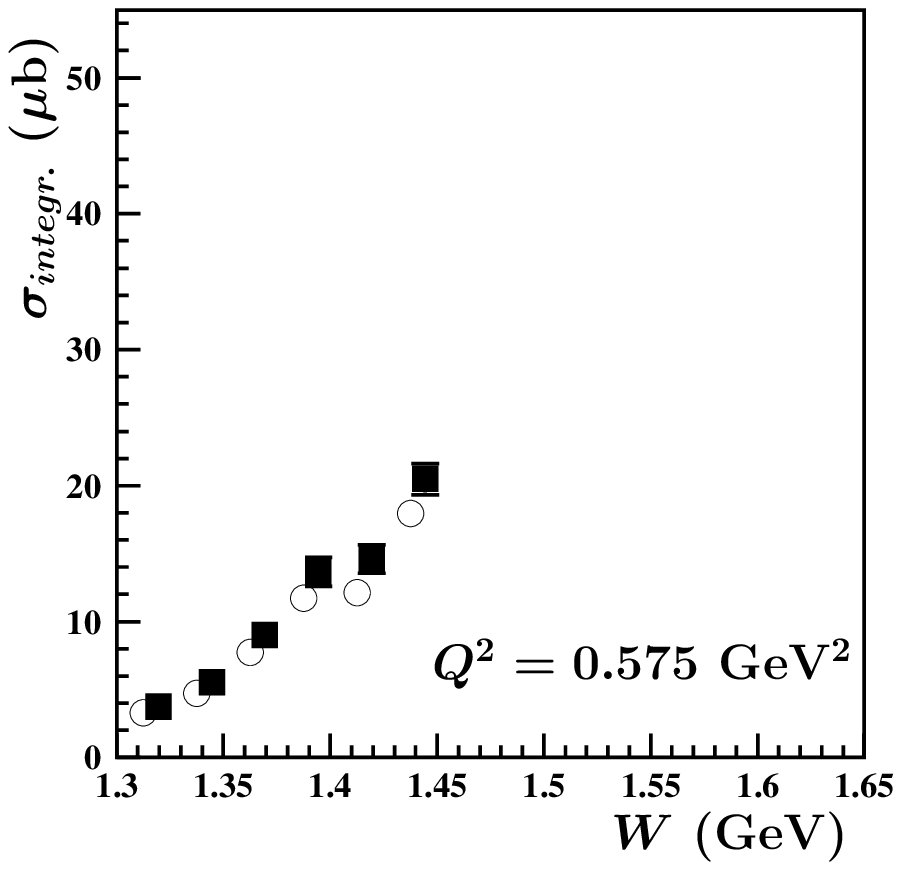,width=6.1cm}
\end{center}
\vspace{-0.6cm}
\caption{\small Comparison of fully integrated charged double pion cross sections, 
obtained
with (squares) and without (open circles) accounting for contributions
from the CLAS areas of zero acceptance, as described in the text of
Sect.~\ref{interpol_ineff}. \label{empty_zone_total}}
\end{figure*}

\subsection{Interpolation of 5-fold differential cross section into the CLAS detector 
areas of zero acceptance  \label{interpol_ineff}}

As discussed in Sect.~\ref{ident} the CLAS detector
has areas with zero acceptance. The
contributions from the 5-fold differential cross sections in such areas
must be taken into account in order to obtain the integrated 1-fold differential
cross sections. We developed a special procedure to extend the 5-fold differential
cross sections in the areas of CLAS with zero acceptance.

\begin{figure}[htp]
\begin{center}
\includegraphics[width=7.cm]{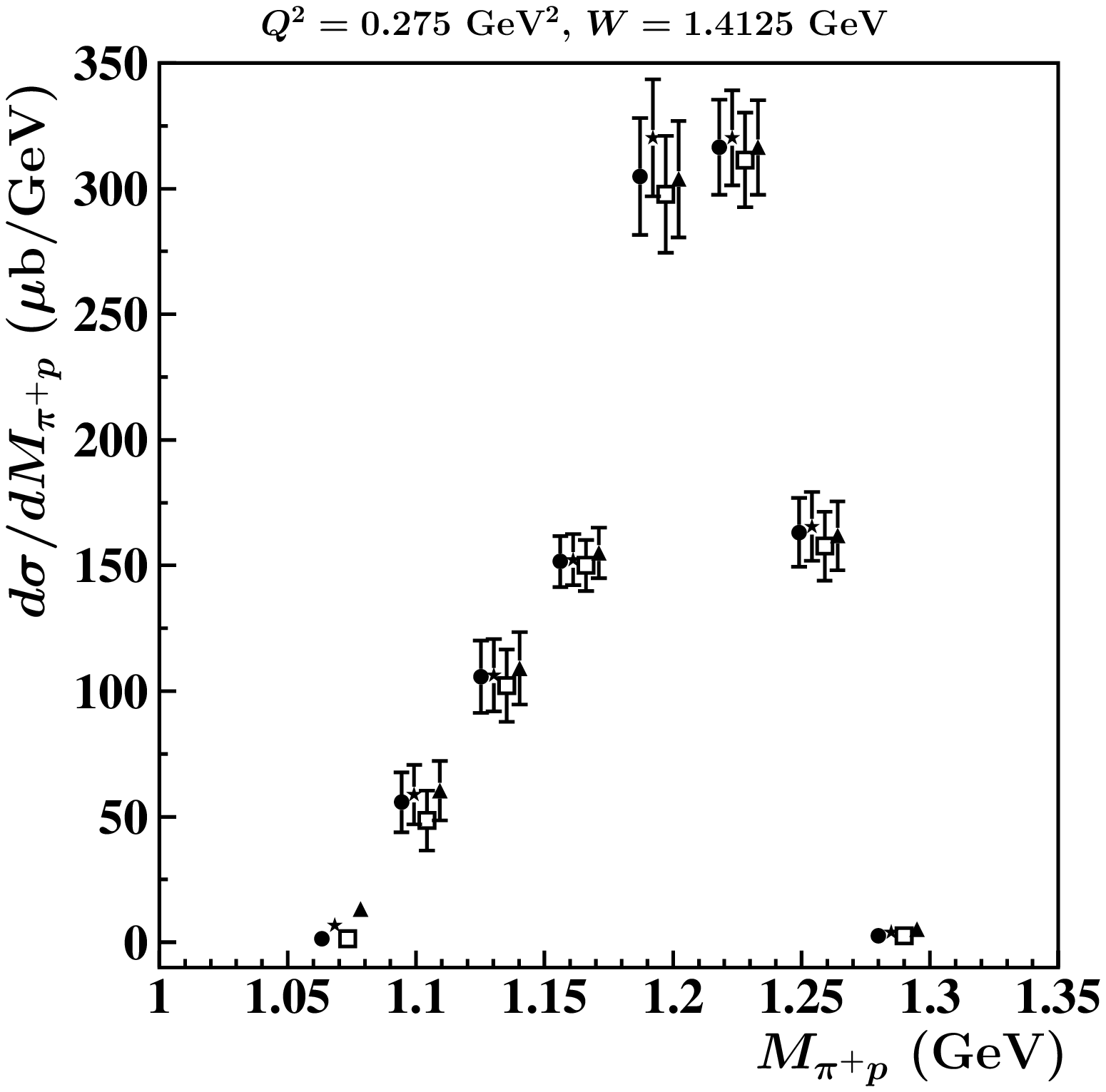}
\includegraphics[width=7.cm]{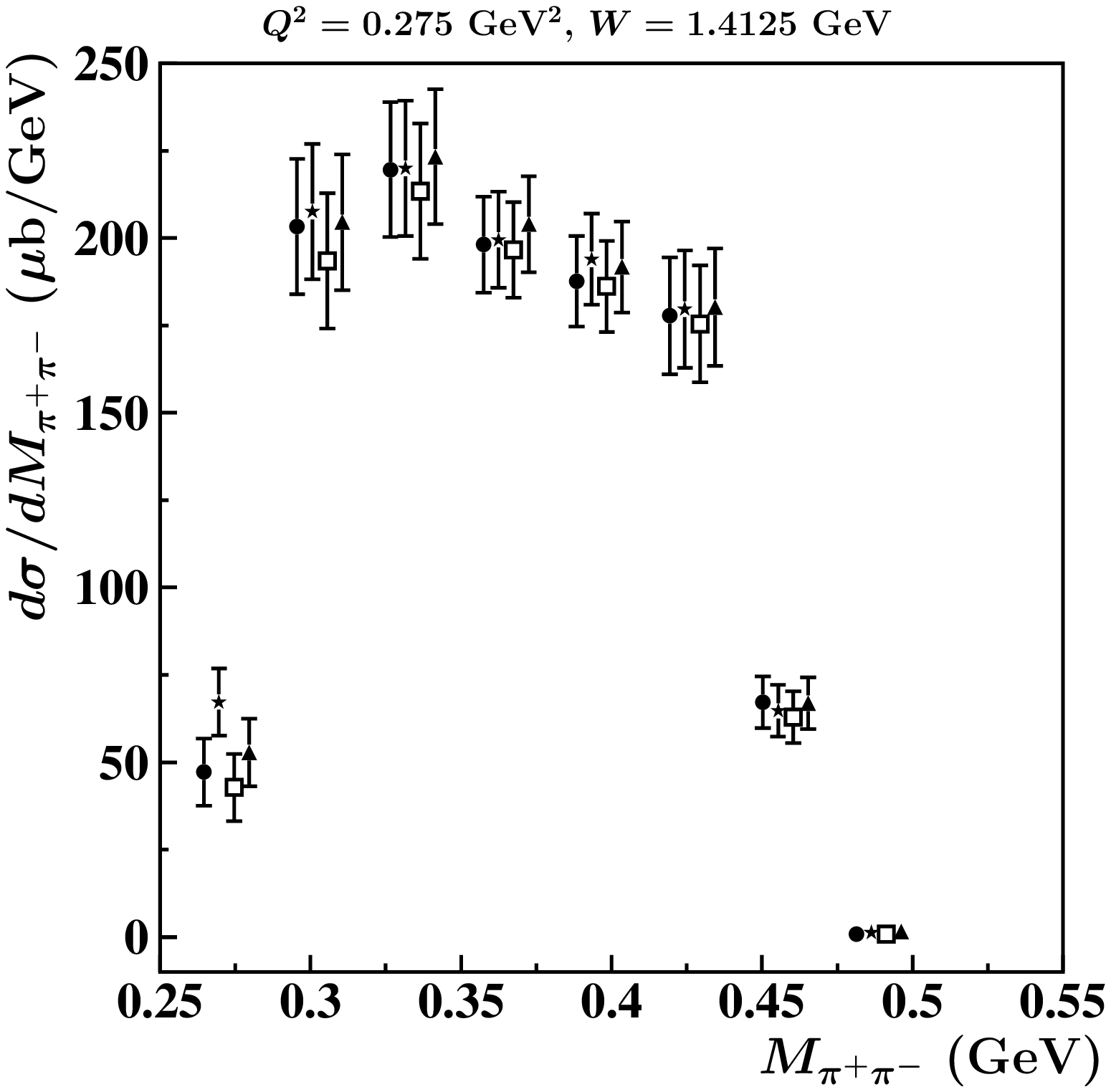}
\includegraphics[width=7.cm]{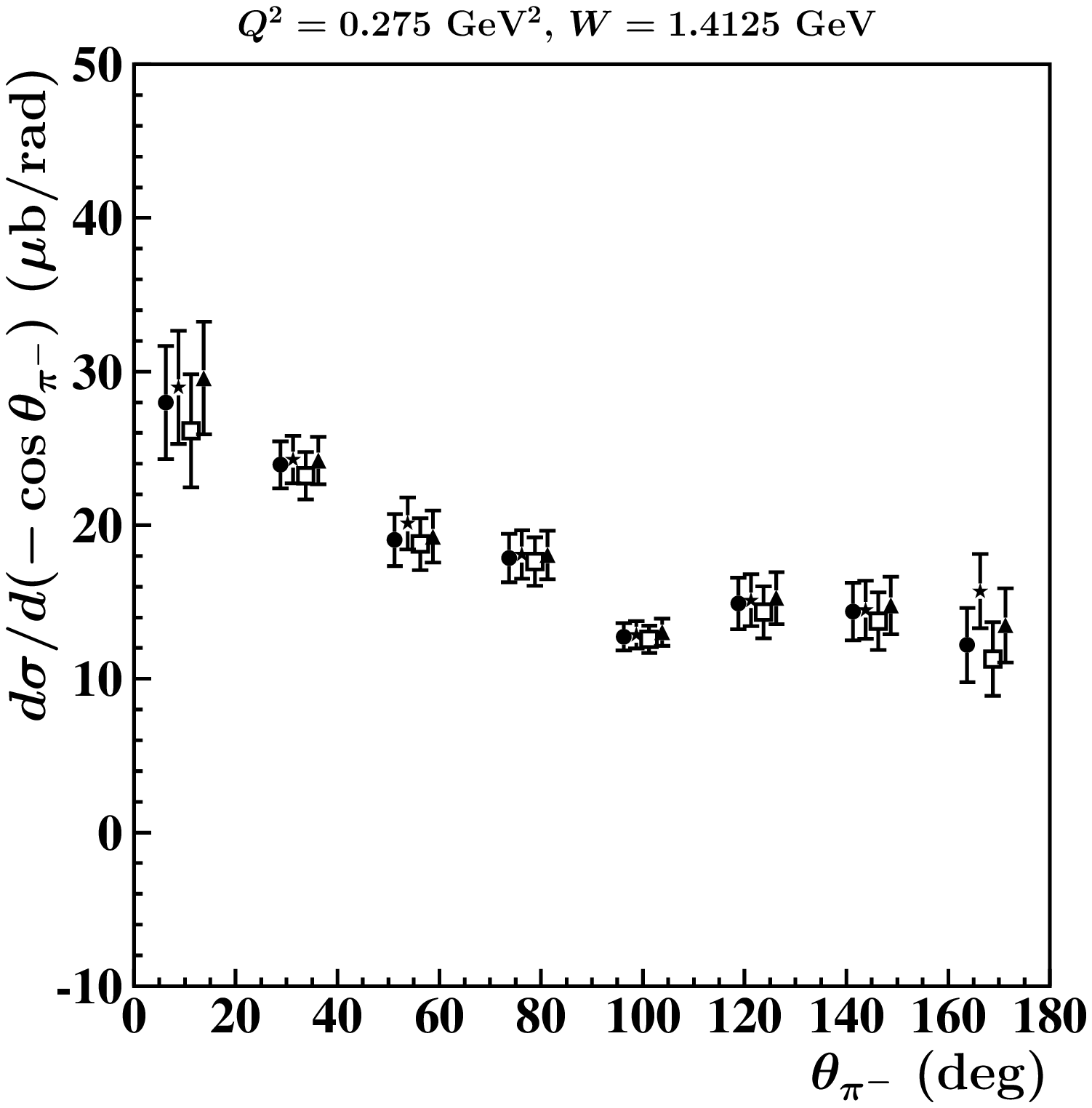}
\caption{\small Mass distributions of $p \pi^{+}$ (top) and $\pi^{+} \pi^{-}$ (middle), 
and $\pi^{-}$ angular (bottom) distributions, obtained from different ways of
interpolating the 5-fold differential cross sections into the CLAS areas of zero acceptance.
Open squares correspond to 1-fold differential cross sections estimated without
the contributions from the CLAS areas of zero acceptance. 1-fold differential
cross sections, obtained with the $R_{\it{j}}$ coefficients in eq. (\ref{determinea})
calculated within the framework of the JM03 and JM05 models, are shown by triangles
and
stars, respectively. 1-fold differential cross sections, estimated
by excluding contributions from $\phi$-dependent parts in eq. (\ref{phidep})
are shown by full circles. Note that points are spread out for clarity. \label{empty_zone_effect}}
\end{center}
\end{figure}

We used the general
$\varphi_{\it{i}}$ ($i$=1,2,3) dependence of the 5-fold differential 
cross sections fixing the other 4 kinematic variables:

\begin{eqnarray}
\label{phidep}
\frac{d^{5}\sigma}{d^{5}\tau_{i}} =
 A + B
\rm{cos}2\varphi_{i}+ C \rm{cos}\varphi_{i}+\nonumber \\
B^{'}\rm{sin}2\varphi_{i} + C^{'} \rm{sin}\varphi_{i}.
\end{eqnarray}

The first three terms  are
valid for any exclusive channel and for any kind of particular reaction dynamics,
being a consequence of rotational invariance of the production amplitudes. The 
last two terms
appear in the 5-fold differential cross sections for 3-body final states. After integration over
the $\alpha_{\it{i}}$ angles, these two terms vanish as a consequence of parity conservation.
The statistics in the populated bins is too small to allow evaluation of the $A$, $B$, $C$, $B^{'}$, and $C^{'}$
coefficients from the data in the populated bins alone. Hence, we used both data and input from the models
fit to the data to evaluate these coefficients.
The coefficient ratios $R_{\it{j}}$ ($R_{1}=B/A$, $R_{2}=C/A$, $R_{3}=B^{'}/A$, $R_{4}=C^{'}/A$ ) were taken from
phenomenological models for charged double pion electroproduction fit to our data. The
coefficient $A$ was determined from the data on the 5-fold differential 
cross sections in the 
populated 5-dimensional bins $\frac{d \sigma_{meas.}}
{dM_{p\pi^{+}}dM_{\pi^{+}\pi^{-}}d\Omega d\alpha_{(p \pi^{-})(p' \pi^{+})}}$  as: \\

\begin{eqnarray}
\label{determinea}
\sum_{\Delta \varphi_{\pi^{-}}} \frac{d \sigma_{meas.}}
{dM_{p\pi^{+}}dM_{\pi^{+}\pi^{-}}d\Omega d\alpha_{(p \pi^{-})(p' \pi^{+})}}\Delta \varphi = \nonumber \\
A(2\pi-\Delta \tilde\varphi) -  R_{1}A\int\limits_{\Delta \tilde\varphi}{\rm{cos}(2\varphi_{\pi^{-}})}\,
d\varphi_{\pi^{-}} \nonumber \\
- R_{2}A\int\limits_{\Delta \tilde\varphi}{\rm{cos}(\varphi_{\pi^{-}})}\,
d\varphi_{\pi^{-}} -  R_{3}A\int\limits_{\Delta \tilde\varphi}{\rm{sin}(2\varphi_{\pi^{-}})}\,
d\varphi_{\pi^{-}} \\ - R_{4}A\int\limits_{\Delta \tilde\varphi}{\rm{sin}(\varphi_{\pi^{-}})}\,
d\varphi_{\pi^{-}} \textrm{ ,} \nonumber
\end{eqnarray}
where the sum is running over the populated 5-dimensional bins, while the integrals 
are taken over
the CLAS areas of zero acceptance $\Delta \tilde\varphi$. The 5-fold differential cross sections in
the CLAS areas of zero acceptance were estimated
from eq. (\ref{phidep}) with coefficients $A$,$B$,$C$,$B^{'}$,$C^{'}$
calculated from eq. (\ref{determinea}).

In order to determine the ratios $R_{\it{j}}$ within the framework of the phenomenological
models, we propagated the 5-fold differential cross sections
into the CLAS areas of zero acceptance, using the JM03 model predictions for the shape of the
5-fold differential
cross sections ~\cite{Ri00,Mo01,Mo03}. The parameters of the JM03 model were determined from
previous CLAS charged double pion data in the resonance region \cite{Ri03}. In this 
way 
preliminary estimates for the 1-fold differential cross sections were obtained.
Similar approaches to propagate the 5-fold differential cross sections into the 
CLAS
areas of zero acceptance were used in previous charged double pion data analyses, 
published in Refs.~\cite{Ri03, Hj05}.
In the next step, the parameters of JM03 were further adjusted to reproduce 
preliminary
estimates of the 1-fold differential cross sections. The $R_{\it{j}}$ coefficients 
were calculated within the framework of the JM03 approach after  
mentioned adjustment of the JM03 parameters. The 
coefficients $A$ for the $\varphi$ independent parts of the 5-fold differential
cross sections (see eq.~(\ref{phidep}))  were obtained from the data in the populated 
bins,
according to eq. (\ref{determinea}), using the improved estimates for $R_{\it{j}}$.
Finally, the 1-fold differential charged double pion cross sections were obtained 
as described
in Sect.~\ref{evsect}, using the 5-fold differential
cross sections in the CLAS areas of zero acceptance determined from eq. (\ref{phidep}) with
values of  $A$, $B$, $C$, $B^{'}$ and $C^{'}$ determined as described above.

\begin{figure}[htp]
\begin{center}
\includegraphics[width=5cm]{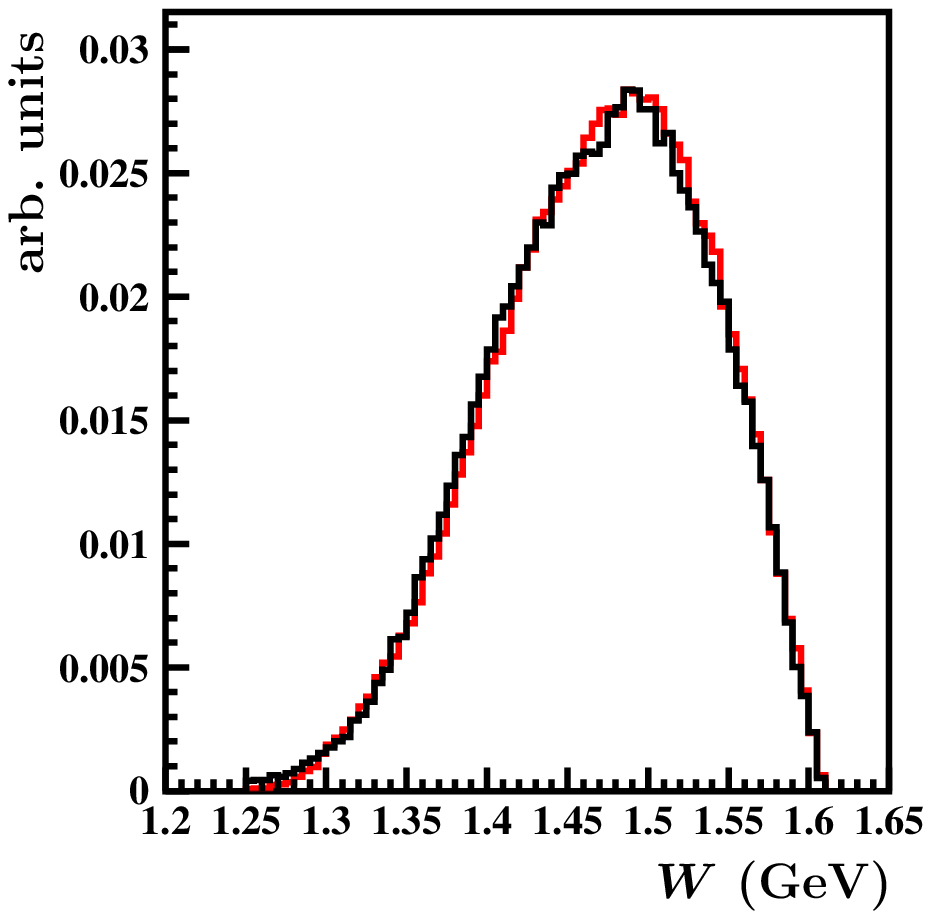}
\includegraphics[width=5cm]{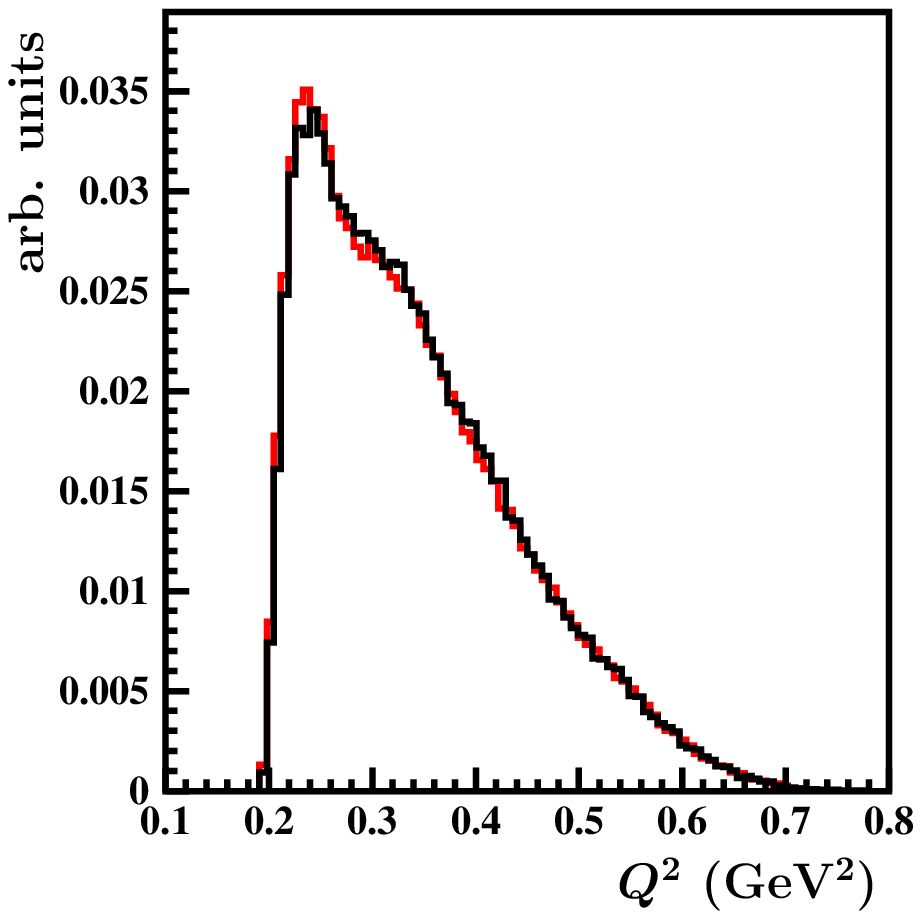}
\caption{\small(color on line) $W$ (top panel) and $Q^{2}$
(bottom panel) distributions. The black lines
represents data, and the red lines simulation. \label{jopa}}
\end{center}
\end{figure}

\begin{figure*}[htp]
\begin{center}
\includegraphics[width=15cm]{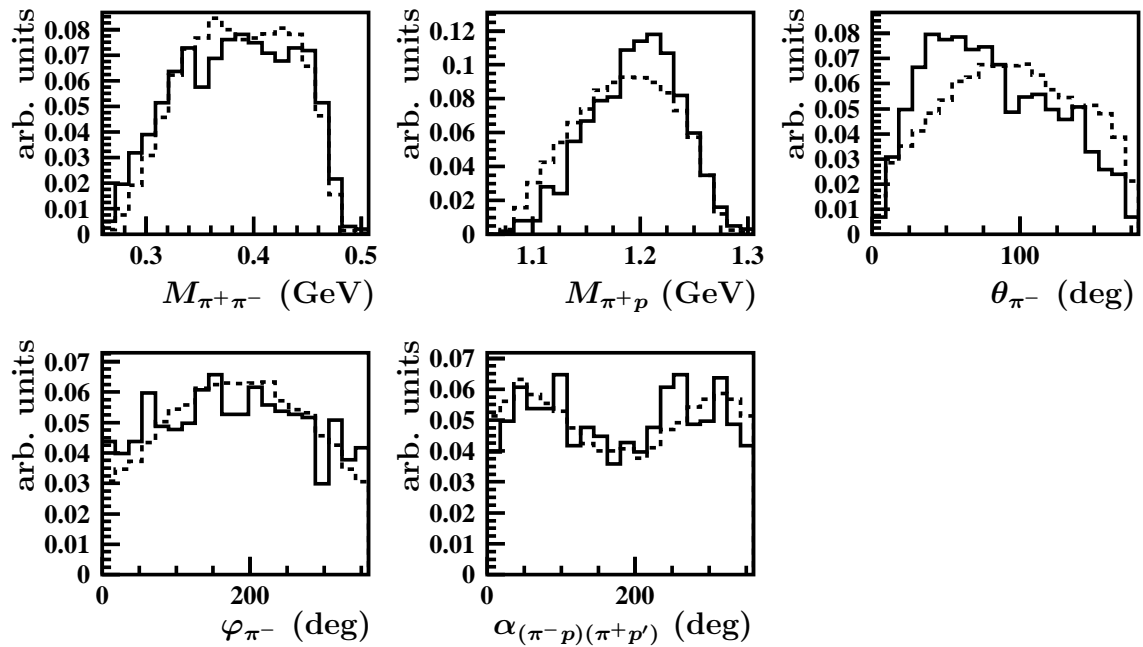}
\caption{\small Comparison between measured (solid lines) and
simulated (dashed lines) event distributions  for various final state 
variables. ($W$ =
1.4125~GeV, $Q^{2}$ =
0.525~GeV$^{2}$).\label{pizda}}
\end{center}
\end{figure*}

Since the model
was used to evaluate the ratios $R_{\it{j}}$, the model assumptions
affect mostly the $\varphi$-dependent parts in eq. (\ref{phidep}).
After integration over $\varphi_{\it{i}}$ angles, these parts disappear.
Nevertheless, as follows from eq. (\ref{determinea}), the model
assumptions used to interpolate the charged double pion cross sections
into the CLAS areas of zero acceptance will increase the uncertainties of
the 1-fold differential
cross sections obtained in our analysis.
Detailed studies to evaluate these uncertainties were carried out and
are described below.

First, we estimated the overall contribution from the inefficient
areas to the 1-fold differential cross sections, calculating them
 in two ways: (1) by including the contributions from
the areas of zero acceptance in CLAS, as described above, and (2) by excluding them.
In all cases, the two sets
of values were found to overlap well inside the statistical
 uncertainties in the
entire kinematic region covered in the experiment. One example is shown in
Fig~\ref{empty_zone_effect}.

In Fig~\ref{empty_zone_total} we show a comparison of the fully integrated 
charged double pion 
cross sections.
Again, within the entire kinematic area the differences between the two sets of
cross sections are well within the statistical uncertainties.

In the next step we investigate how the 1-fold differential cross sections may be affected by
the model assumptions used in the procedure described above.
Since only $\varphi$-independent
parts of eq. (\ref{phidep}) contribute to the 1-fold differential cross sections, we
need to know the model uncertainties just for the $A$ coefficients.  As follows 
from eq. (\ref{phidep}),
the influence of the $\varphi$-dependent parts on the $A$ coefficients
depend on: a) the relative
contributions of the $\varphi$-dependent parts to the 5-fold differential 
cross sections; b)
the ratio $\Delta
\tilde\varphi/2\pi$, where $\Delta \tilde\varphi$
is the overall $\varphi$ coverage of the CLAS areas of zero acceptance.
The relative contributions from the 
$\varphi$-dependent  parts is estimated by fitting  the
$\varphi_{\it{i}}$  angular distributions 
$^{[1]}$  \footnotetext[1]{${\it{i}}$=1,2,3 and stand for the set of kinematic variables, defined in 
 Sect.\ref{xsect_var}.},
using eq. (\ref{phidep}) with $A$, $B$ and $C$
coefficients as free parameters. The $\varphi_{\it{i}}$ angular distributions 
were obtained as integrals from the 5-fold differential cross sections over the 
other four variables. The  $B'$ and $C'$ terms should be equal
to zero since these terms are integrated over the $\alpha_{\it{i}}$ angles.
The contributions from $B$ and $C$ in eq.~(\ref{phidep}) range from 10 \% to 50 \%.
For the majority of bins in $Q^{2}$ and $\it{W}$, these contributions range from 15 \%
to 25 \%. The upper limit for the model dependence of the $A$ coefficients
has been estimated by
replacing in eq. (\ref{determinea}) all cosines by unity and
by assuming $\Delta \tilde\varphi$/$2\pi$  $\sim$0.2 for the
geometrical coverage of the zero acceptance areas. With these assumptions
we can calculate the model uncertainty of the $\it{A}$ coefficients, as 
the product
of the maximal contribution from the $\varphi$-dependent parts
to the 5-fold differential cross section (0.5) $^{[2]}$  
\footnotetext[2]
{ratio sum of $\varphi$-dependent parts over the full cross section. }
and the geometrical coverage
of the CLAS areas of zero acceptance (0.2), resulting in an upper limit of 10 \%.
However, this limit was obtained with extremely conservative estimates
for the integrands in eq. (\ref{determinea}), the CLAS areas of zero acceptance 
$\Delta \tilde\varphi$/$2\pi$, and the relative contributions of the
$\phi$-dependent parts. More realistic estimates, outlined below, result in uncertainties
of a few percent.

In Fig.~\ref{empty_zone_effect} we compare the results
obtained using various models to estimate $R_{\it{j}}$. The JM03 and JM05 models
are rather different in the description of the
5-fold differential cross sections. The JM05 approach provided a much 
improved treatment
for the direct charged double pion electroproduction mechanisms  ~\cite{Mo06}. It also
contains an additional contact term that was introduced to improve the description of the $\pi \Delta$ isobar
channels. The interpolations of the 
5-fold differential cross sections into the inefficient areas using these two
models for $R_{\it{j}}$, and in addition taking off
the contributions proportional to $\sin{2\varphi}$ and $\sin{\varphi}$,
result in minor modifications well inside the statistical uncertainties.

Finally we eliminate the contributions from the $\varphi$-dependent
parts in eq. (\ref{determinea}) and estimate
the $A$ coefficients from data in the populated bins.
 The results are shown in Fig.~\ref{empty_zone_effect} (diamonds).
Again, the estimated cross sections are well inside the statistical
uncertainties of the data.

\subsection{Event reconstruction efficiencies \label{genev}}

A Monte Carlo event generator \cite{Golova} was used to evaluate the event
reconstruction efficiencies. The event generator contains the
main meson production channels in the resonance region. The efficiency for
detection of the $p\pi^{+}\pi^{-}$ in the final state was studied in detailed
simulations that included the $2\pi$ as well as $3\pi$ final states.
The latter was needed to account for multi-pion background in the
selection of charged double pion events, when applying exclusivity cuts.
These events were processed using the same reconstruction program
$^{[1]}$ \footnotetext[1]{The correction factor that accounted for the events eliminated by 
the cut on the number of
photoelectrons in the Cherenkov counter was applied for the
measured events only.},
event selection procedures and fiducial cuts as for the events collected
in the experiment. Efficiencies in the 7-fold differential bins were determined
as the number of reconstructed events over the number of generated events,
and used in eq. (\ref{expcrossect}) to evaluate the 5-fold differential
cross sections.

As can be seen from Figs.~\ref{jopa} and ~\ref{pizda}, the $W$ and $Q^{2}$
dependencies are well reproduced, as are major features in the
event distributions over the final state hadronic variables. Differences 
between the measured and
simulated distributions seen for the $\pi^{+}p$ invariant mass and $\pi^{-}$ angular
distributions have little impact as  efficiencies inside these areas are
smooth. The event generator is therefore adequate to evaluate the event
reconstruction efficiency for major parts of the kinematic range covered in the experiment.

The limited phase space available for events in the low mass region
with $W$ $<$ 1.40~GeV
requires a different approach for determining the event reconstruction
efficiencies for the invariant mass distributions. In this region we found
rapid variations of 
efficiency from invariant masses inside the bins at the lowest edges
for  all mass distributions. The
use of an event generator that closely reflects the measured distributions
is crucial in this area, where no 2$\pi$ electroproduction 
data were previously available.
Therefore, we have used an iterative procedure starting with the model
event generator
described above, and extracted approximate cross sections for
the different mass distributions. These were
then used as a realistic input into the generated  event distributions over
invariant masses
$W_{gen}(M_{\it{k}})$ ($\it{k}$=$\pi^{+} p$, $\pi^{-} \pi^{+}$, $\pi^{-} p$) for
an accurate determination of
event reconstruction efficiencies and cross sections.

The improved estimates of event reconstruction efficiencies for the mass distributions
   $\epsilon_{imp}(M_{\it{k}})$ were
obtained as:

\begin{eqnarray}
\label{effimpr}
\epsilon_{imp}(M_{k})
=\frac{W_{meas}(M_{k})\epsilon(W,Q^{2})}{W_{gen}(M_{k})} ,
\end{eqnarray}
where $W_{meas}(M_{\it{k}})$ are event distributions
in the invariant mass $M_{k}$ and taken from the data. Both measured 
$W_{meas}(M_{\it{k}})$ and generated $W_{gen}(M_{\it{k}})$ ($\it{k}$=$\pi^{+} p$
event distributions were normalized to unity.
The quantity $\epsilon(W,Q^{2})$ is the event
reconstruction efficiency in a particular $(W,Q^{2})$-bin, and was
estimated using the event generator.

A comparison of the mass distribution obtained
using the event generator with those estimated from eq. (\ref{effimpr}) after further 
corrections, described
in the Sect.~\ref{improve}, is
shown in Fig.~\ref{comparinitfin}.

\subsection{Corrections for mass distributions \label{improve}}

After all of the previously discussed acceptance corrections, several mass distributions 
needed further corrections in order to account
for the rapid variation of the cross sections inside some of the mass bins.
For these bins, the cross sections were re-evaluated, using a binning size 
reduced by a factor of 4. Cross sections at the nominal grid 
were compared to those obtained 
at the grid of reduced bin size and interpolated into the nominal grid. In case of
discrepancies, interpolated values of
the cross sections were used, since they were
determined with better mass resolution.

\begin{figure}[htp]
\begin{center}
\includegraphics[width=8cm]{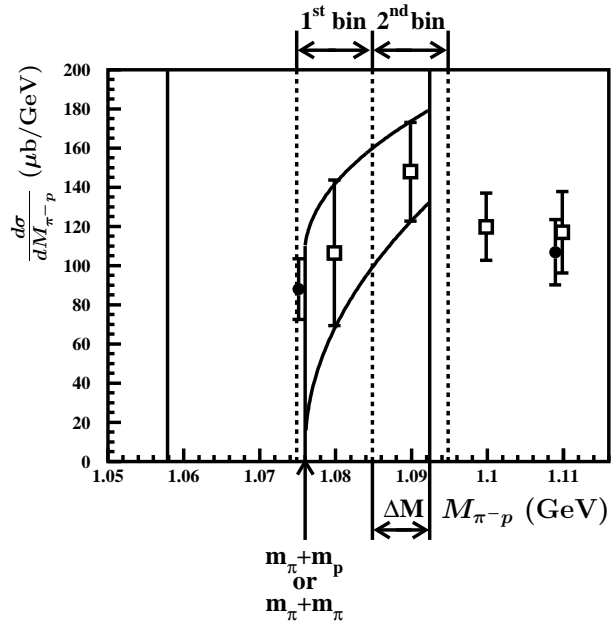}
\caption{\small Corrections for mass distribution in the cross sections 
 near threshold. The filled circles represent the 
preliminary cross sections obtained with the nominal bin size. The lowest mass 
bin 
of nominal size 
is shown by the 
solid vertical lines. Preliminary cross sections obtained using a bin
size reduced by a factor of 4  are shown 
by the open squares. The mass bins of reduced size are shown by the vertical dashed lines. 
The bin of lowest invariant masses $\Delta M$, where corrected cross sections 
were evaluated, covers the interval 
from the dashed 
to the solid lines connected by the double sided arrow. Corrected cross sections are given by 
the integrals from
interpolating curves inside the $\Delta M$  bin over the bin width. \label{massdistrth}}
\end{center}
\end{figure}

A special procedure was developed to evaluate the cross sections
at the smallest invariant masses using constraints on the amplitude
behavior near the phase space limits. All mass distributions at the smallest
invariant masses were re-evaluated using a binning size reduced by a factor of 4
and interpolated over invariant masses in a way compatible with general 
requirements on a power low  
amplitude behavior near threshold: 

\begin{eqnarray}
\hspace{-12.8cm}
& \frac{d \sigma}{dM_{k}}  \left[ \frac{\mu b}{GeV} \right]  =  \nonumber \\ 
& \hspace{-8.8cm} \Biggl\lbrace\!
\begin{tabular*}{0.03cm}{llll}
$C(M_{\pi^{-} p(\pi^{+} p)}-1.076)^{\alpha}$, &  $M_{\pi^{-}p(\pi^{+} p)} > 1.076$ GeV & \hspace{-0.3cm} \\
$0$,&  $M_{\pi^{-}p(\pi^{+} p)} < 1.076$  GeV & \\
$C(M_{\pi^{-} \pi^{+}}-0.276)^{\alpha}$, &  $M_{\pi^{+} \pi^{-}} > 0.276$ GeV & \hspace{-0.3cm} \\
$0$,   &  $M_{\pi^{+} \pi^{-}} < 0.276$ GeV ,&  \\
\end{tabular*} 
\label{xsect1bin}
\end{eqnarray}
where $C$ and $\alpha$ are free parameters fit to the cross sections obtained
with better mass resolution. The corrected cross sections were not evaluated in 
the mass areas closest to the threshold, which were affected considerably by the event migration.
 The bins of
smallest invariant masses cover the mass intervals 
from the left edge of the next-to-smallest
mass bin of reduced size to the right edge of the smallest mass bin of
regular size (Fig.~\ref{massdistrth}). Differential cross sections in these 
bins were computed as: integrals from interpolating curves inside the
bins divided by the bin size  $\Delta M$.
The two solid curves in Fig.~\ref{massdistrth} represent interpolating curves fit to the 
upper and lower
boundaries of preliminary cross sections obtained with these improved mass binnings. 
Differences in the corrected 
cross sections, calculated using these two interpolations, give us 
the systematic 
uncertainties. These corrections only affect the lowest mass bins near the phase space
limit. At larger
invariant masses the cross sections were determined with the nominal bin size as described in the 
Sect.~\ref{evsect}.

\begin{figure}[htp]
\begin{center}
\includegraphics[width=8cm]{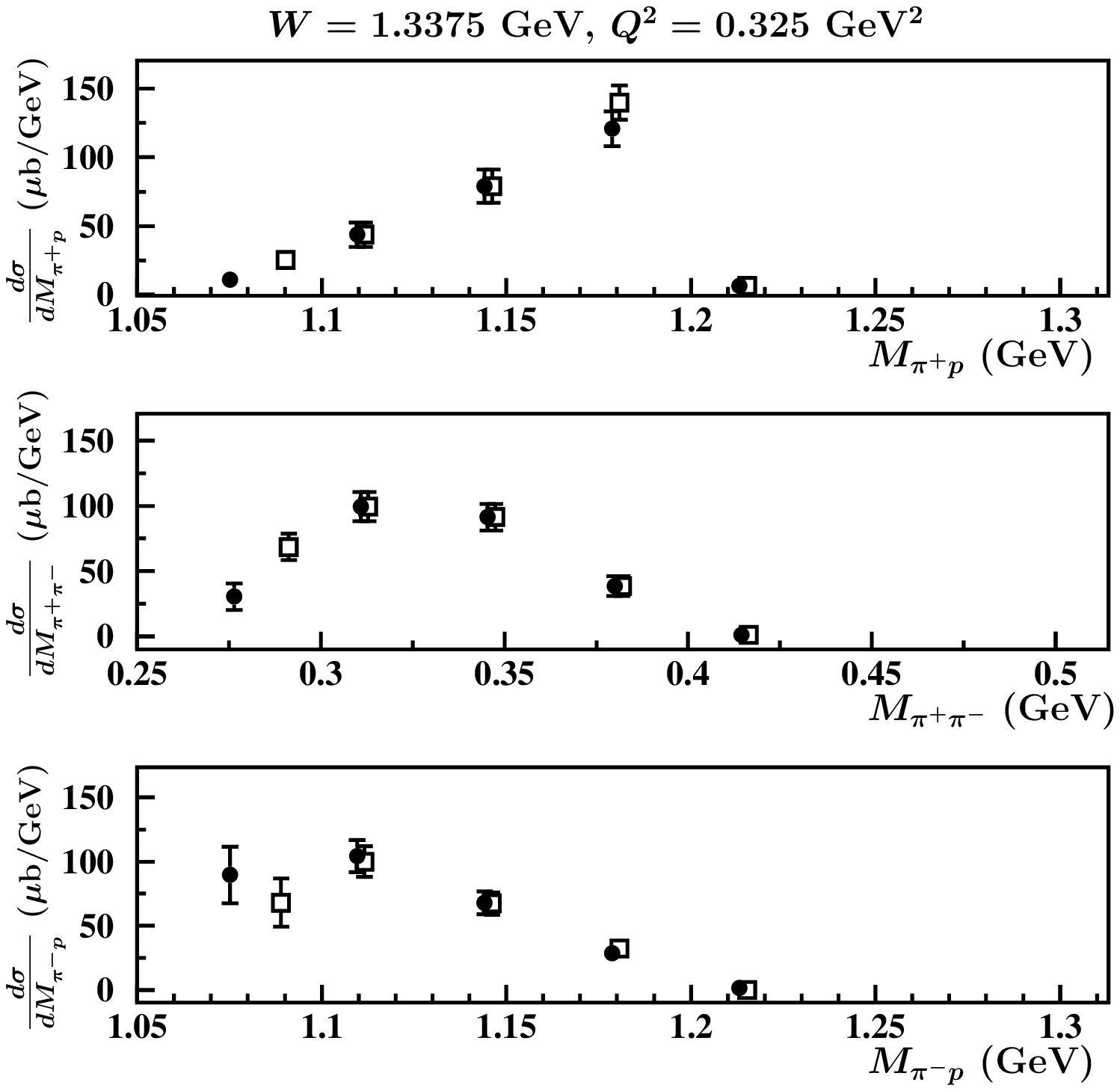}
\includegraphics[width=8cm]{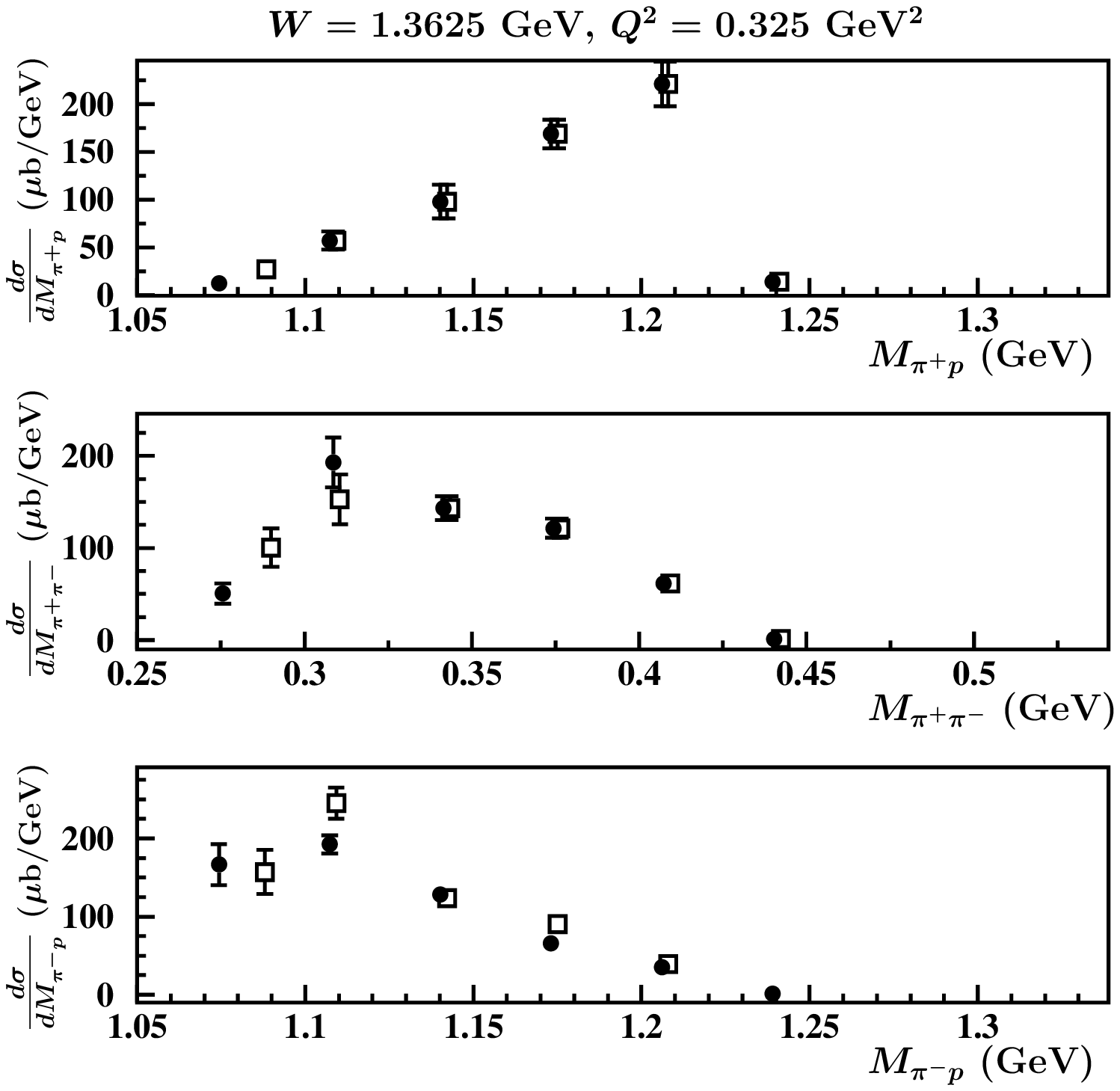}
\caption{\small Comparison of mass distribution cross sections before (filled circles) and 
after (open squares) improvements, described in
Sect.~\ref{genev},\ref{improve}. \label{comparinitfin}}
\end{center}
\end{figure}

 The comparison of mass
distributions before and after all corrections described 
in Sect.~\ref{genev},~\ref{improve} 
is shown in Fig~\ref{comparinitfin}.

\subsection{Radiative corrections \label{radiative}}

Radiative processes were evaluated using the procedure of Mo and
Tsai~\cite{MoTsai} developed for inclusive processes, and incorporated in the
event generator~\cite{Golova}. Approaches that are capable at describing radiative
processes in exclusive 2$\pi$ electroproduction, are not yet available. 

The radiative correction factor $\it{R}$ was determined as:
\begin{equation}
\label{radcorrformula}
R = \frac{N_{rad}}{N_{norad}} \textrm{ ,}
\end{equation}
where $N_{rad}$ and $N_{norad}$ are the numbers of generated
events in each $(W, Q^{2})$ bin with radiative effects switched on and off,
respectively. This factor $R$ was used in eq. (\ref{expcrossect}) for calculations of
the 5-fold differential
cross sections.

The particular hadronic tensor for exclusive 2$\pi$ electroproduction has impact
mostly on radiation of the hard photons by the ingoing and scattered electrons.
Moreover its influence on observables decreases after integration over the final
state kinematic variables \cite{Afanas}. 
The inclusive procedure for radiative corrections represents a
reasonable approximation for the case of our data, since we applied an  
exclusivity cut, which restricts the hardness for the emitted photons, and
all 1-fold differential cross sections represent integrals of the 5-fold 
differential
cross sections over four kinematic variables.

We found that
the relative contributions of hard photons to $R$ varied from 30 to 50 \%. It is only 
this contribution that could be affected by the hadronic tensor and may be 
different in
various exclusive channels.

In Ref.~\cite{Afanas} the effect of integration over kinematic variables 
for the case of the exclusive single pion electroproduction was studied, 
for which radiative processes have been
evaluated exactly with the hadronic tensor derived from the fit to data. It was 
found
that radiative corrections are reduced by factors of 2 to 4 after
integration over the $\varphi$ angle for the emitted pion. Therefore, integration
over four variables in the case of charged double pion electroproduction
 is expected to reduce the radiative correction factor considerably, at least 
 by a factor of 4 \cite{Afanas}.

Radiative corrections in the kinematics of this measurement
were found to be less than 20 \%. Therefore the contribution due to hard photon
emission to the
radiative corrections should be
less than 10 \%. They should be further reduced by more than a 
factor of 4 
after integration of the 
5-fold differential cross sections over four variables. Even a large uncertainty 
of 100 \% in the contributions
of hard photons would result in uncertainties of the overall radiative corrections to 
charged double pion
cross section of only a few percent.

The uncertainty in determining the cross sections caused by using the inclusive approximation for
radiative corrections is well below the statistical uncertainties of the data, and is
included in the systematic uncertainties $^{1}$ \footnotetext[1]{The information obtained 
on the hadronic tensor from our JM model \cite{Mo06-1} based on a fit
to the charged double pion data represents valuable input for the future development of a
fully exclusive radiative correction procedure for double pion electroproduction.}.

\begin{figure}[htp]
\begin{center}
\includegraphics[width=9.cm]{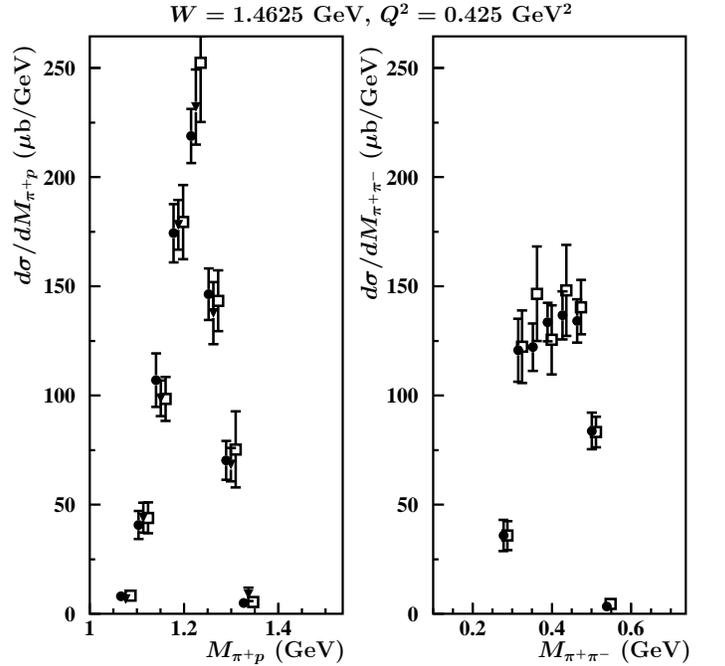}
\caption{\small Comparison between $\pi^{+} p$ and $\pi^{-} \pi^{+}$ mass
distribution cross sections, obtained by integration
of the 5-fold differential cross sections for various choices of the final
state kinematic
variables (choices from 1 to 3 are shown by file circles, open squares and triangle
respectively), defined in the Sect.~\ref{xsect_var}. }
\end{center}
\end{figure}

\begin{figure*}
\begin{center}
\epsfig{file=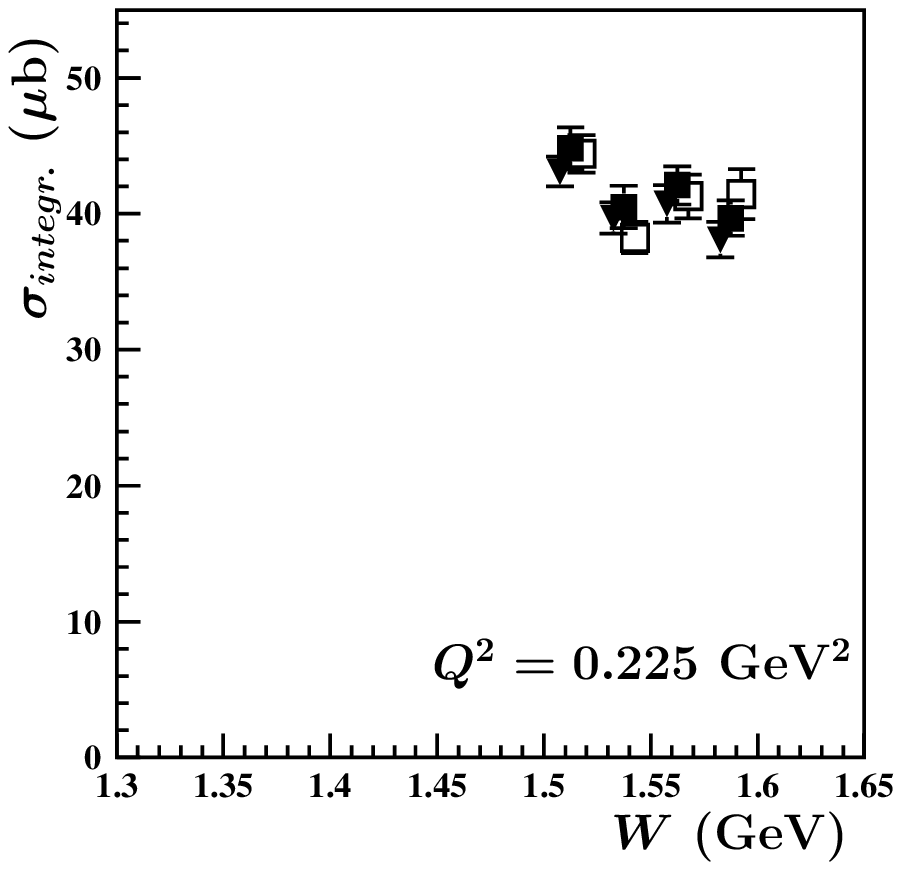,width=6.1cm}
\epsfig{file=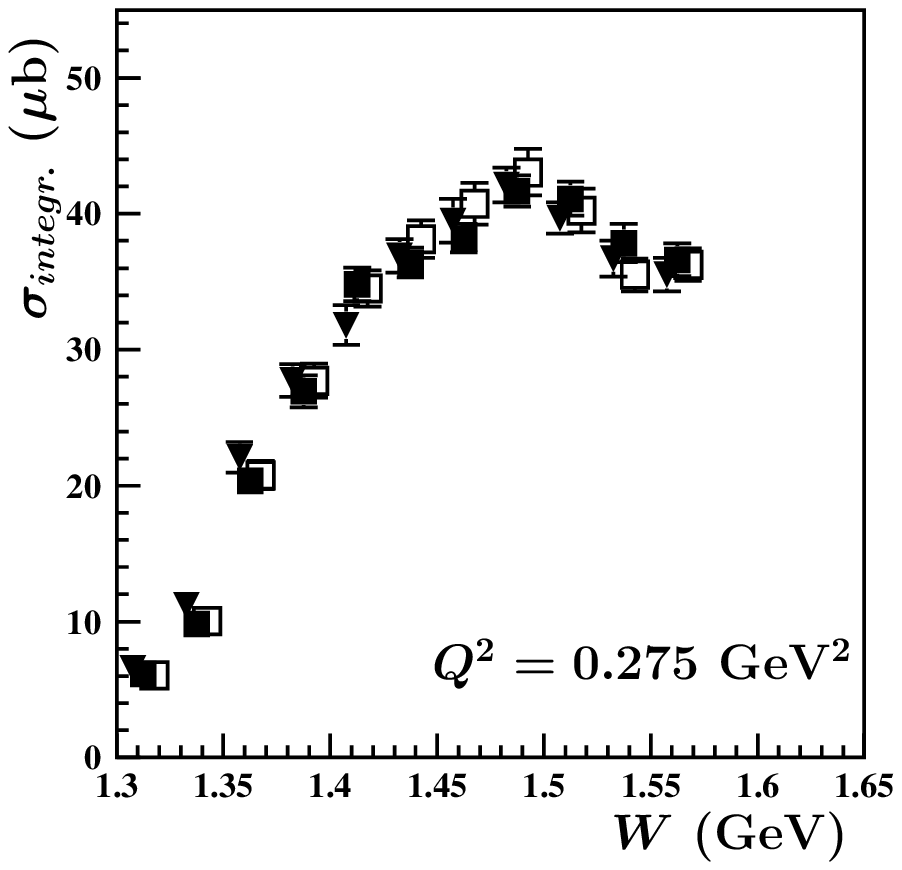,width=6.1cm}
\epsfig{file=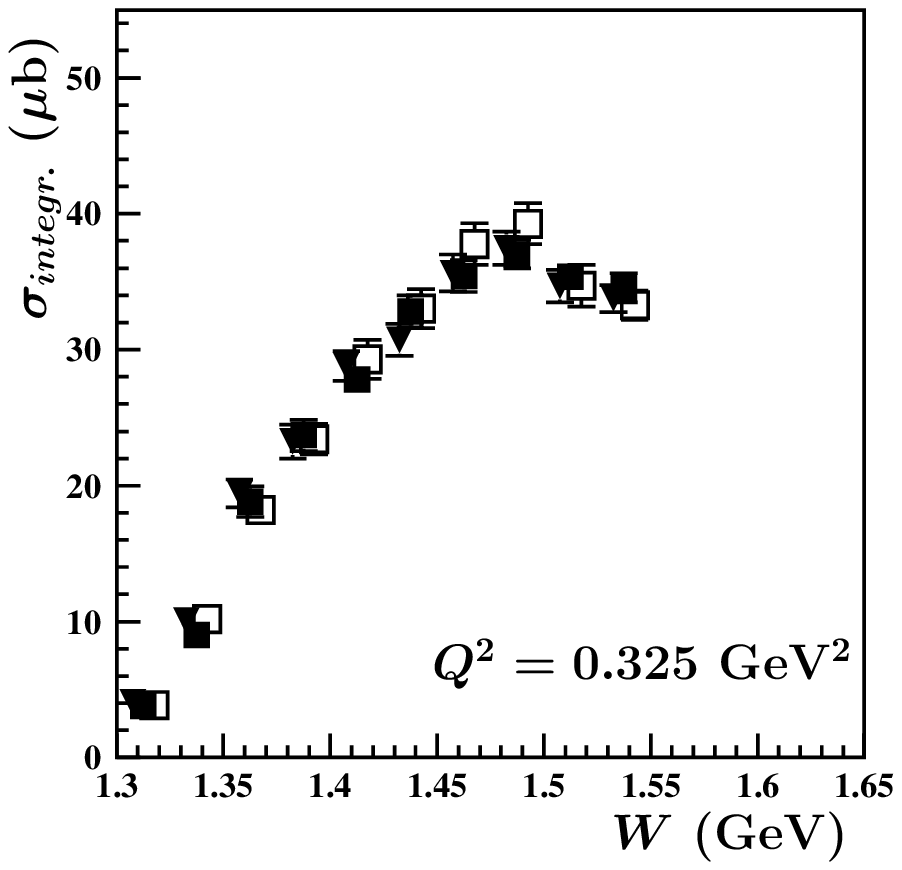,width=6.1cm}
\epsfig{file=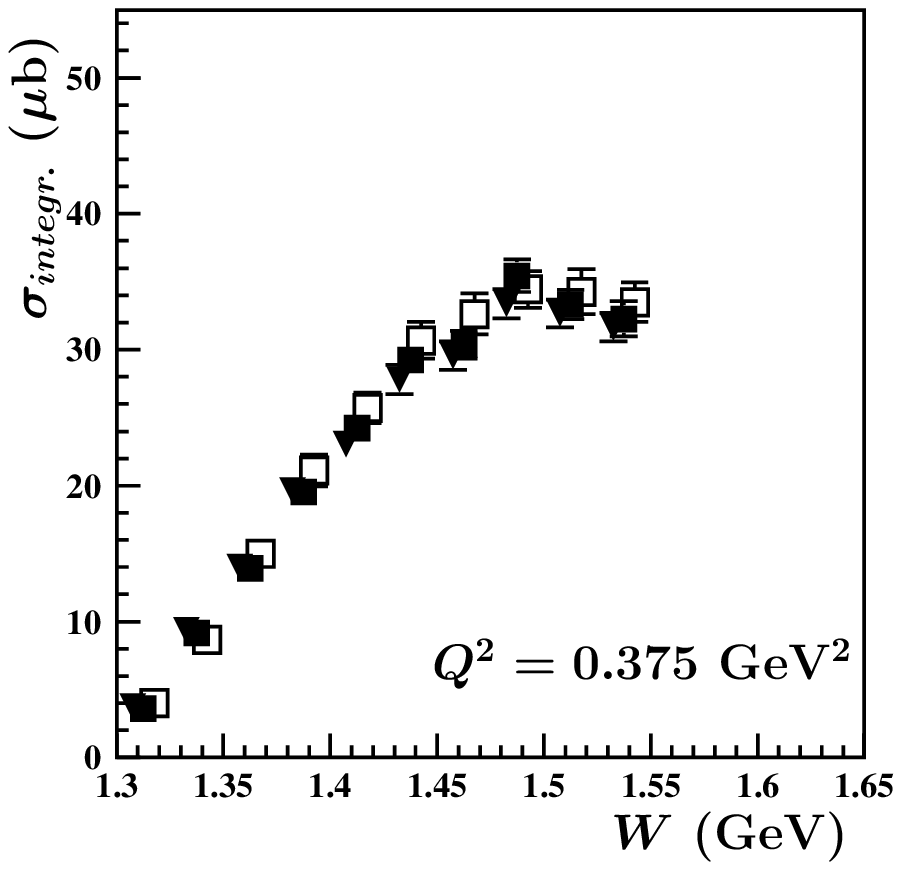,width=6.1cm}
\epsfig{file=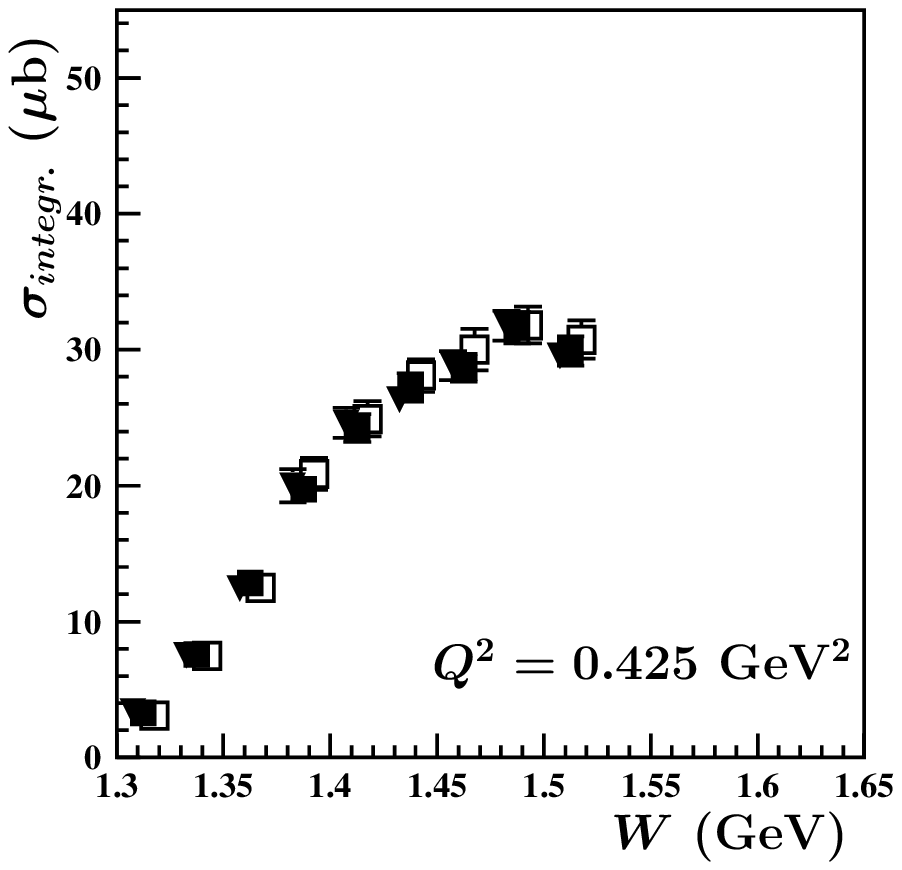,width=6.1cm}
\epsfig{file=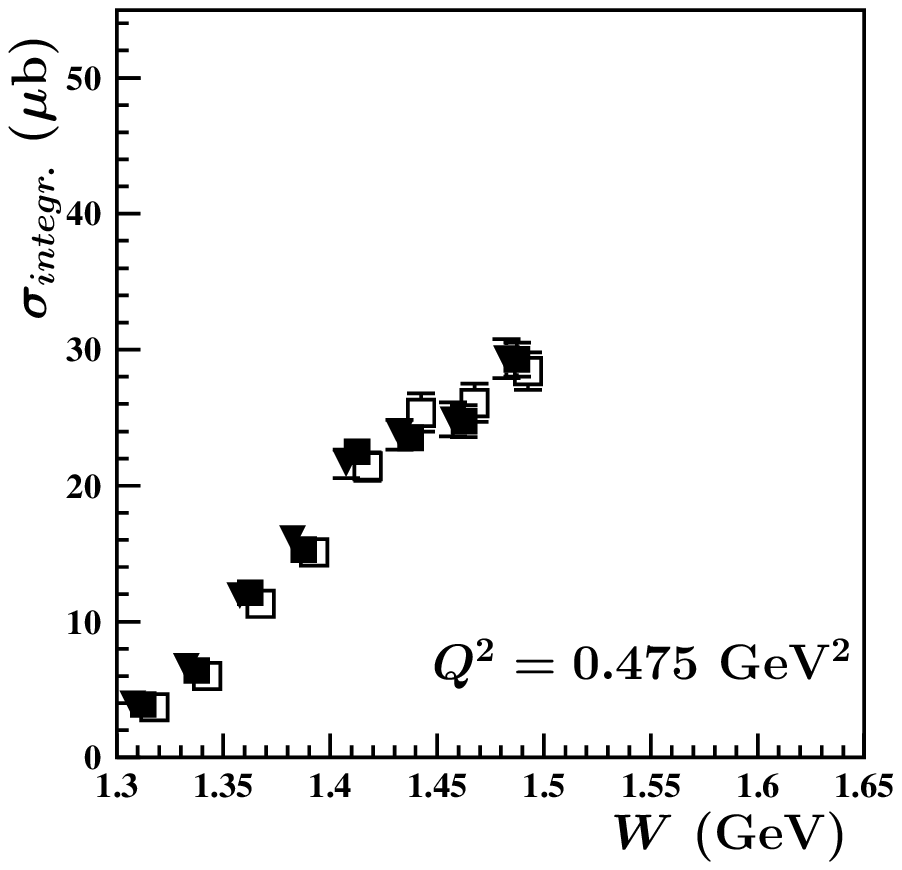,width=6.1cm}
\epsfig{file=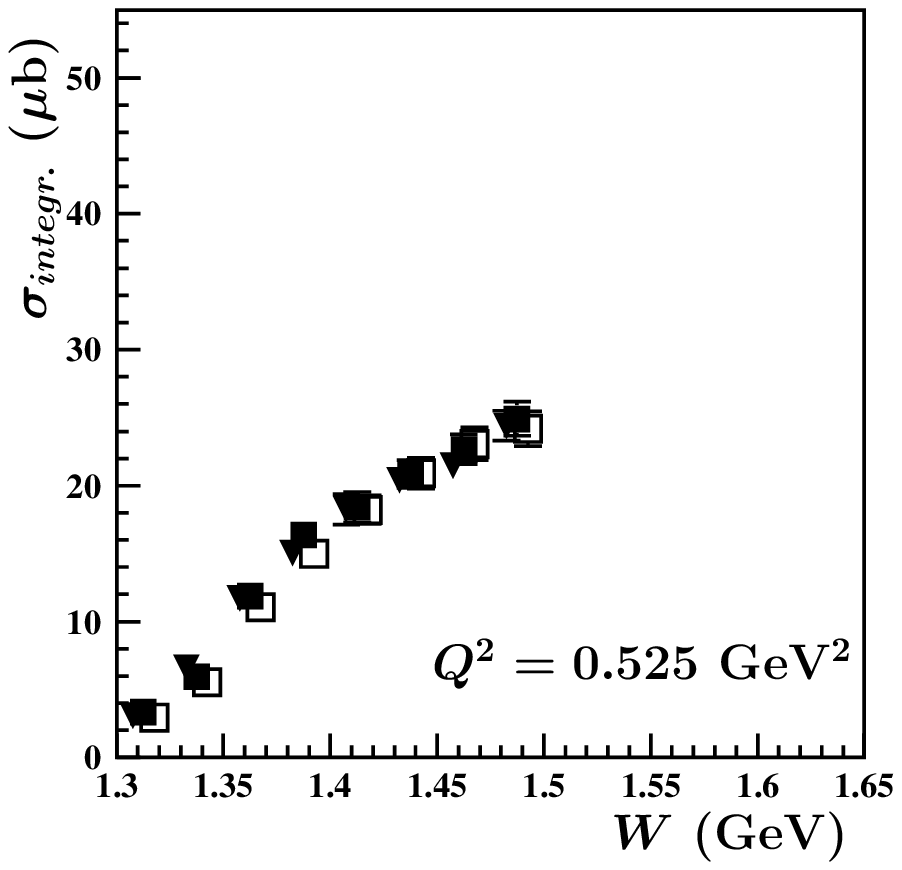,width=6.1cm}
\epsfig{file=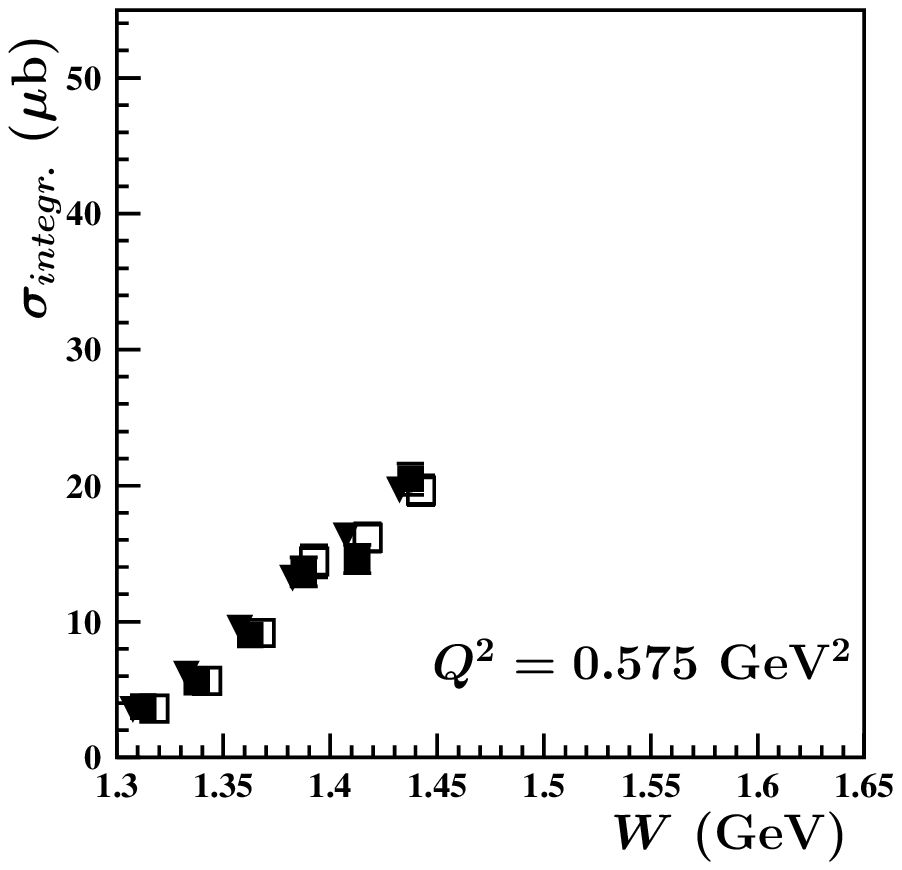,width=6.1cm}
\end{center}
\vspace{-0.6cm}
\caption{\small Comparison between fully integrated charged double pion cross sections,
obtained in the integration of the 5-fold differential cross sections over three various
choices of the final state variables, described in the Sect.~\ref{xsect_var}. The choices for
kinematical variables from 1 to three are shown by triangles, the open and full squares
respectively  \label{2_3_grids_total}}
\end{figure*}

\subsection{Results and systematic uncertainties} \label{experiment}
\begin{figure*}
\begin{center}
\epsfig{file=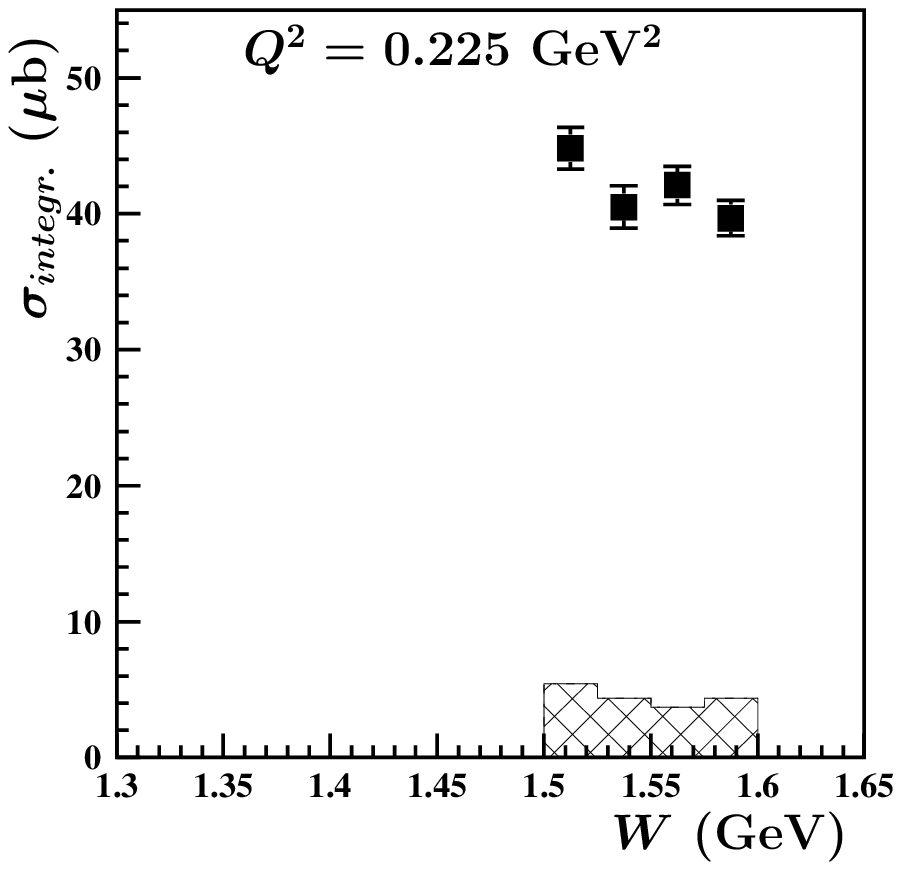,width=6.1cm}
\epsfig{file=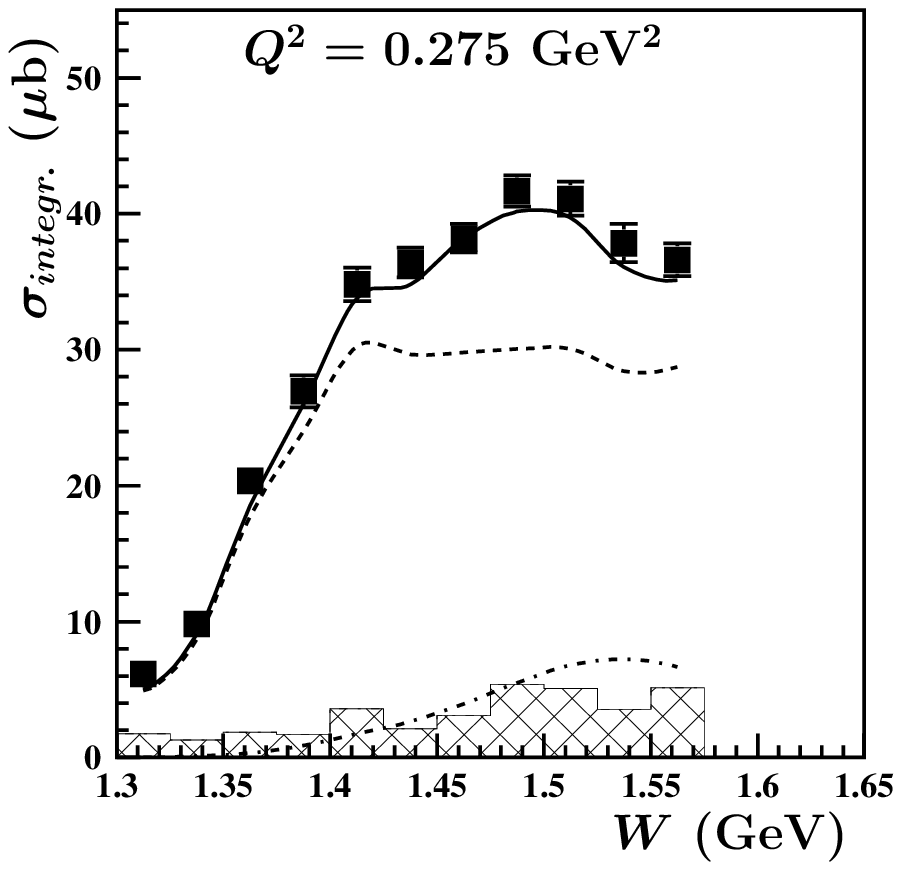,width=6.1cm}
\epsfig{file=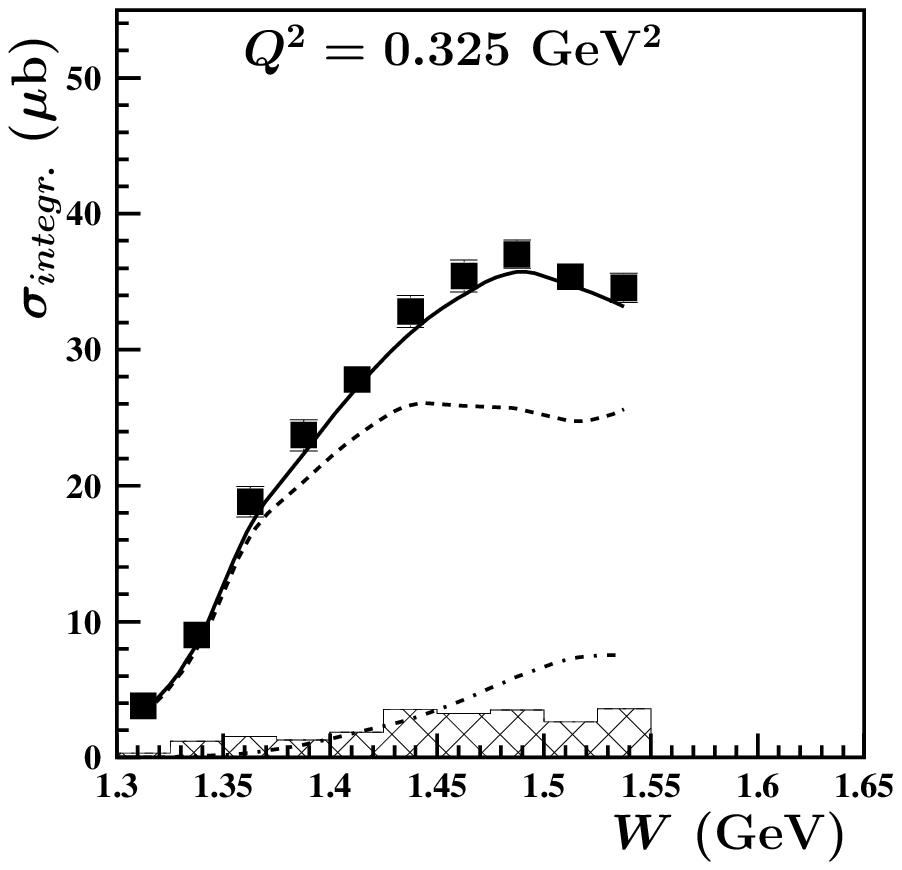,width=6.1cm}
\epsfig{file=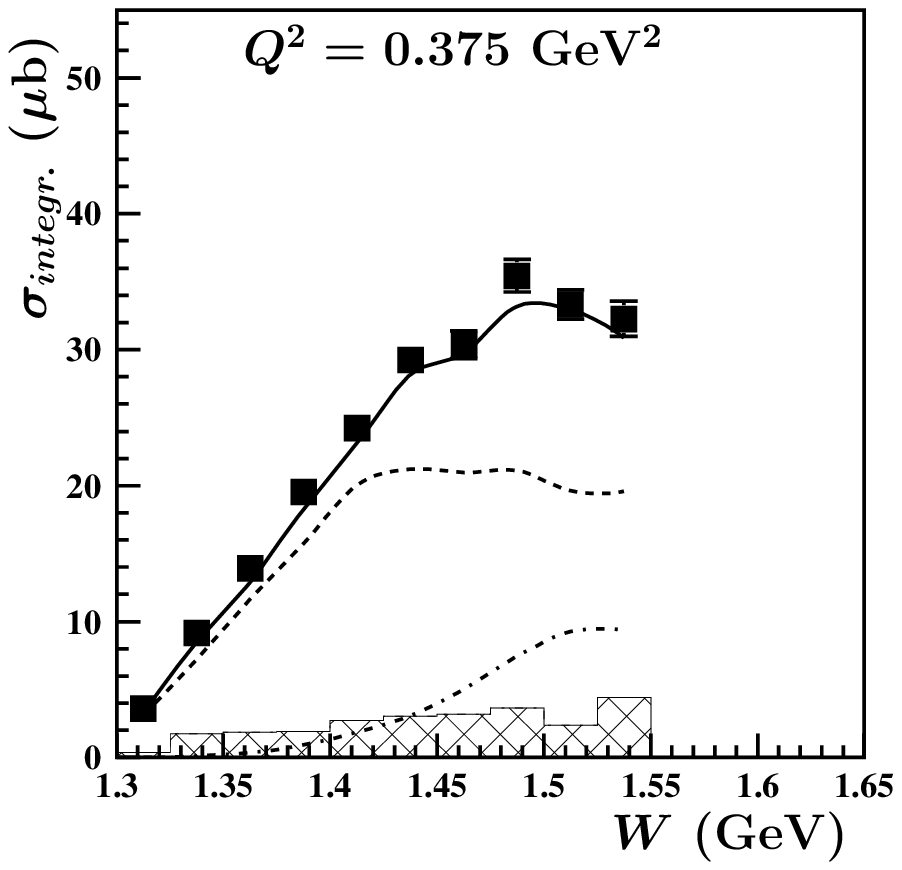,width=6.1cm}
\epsfig{file=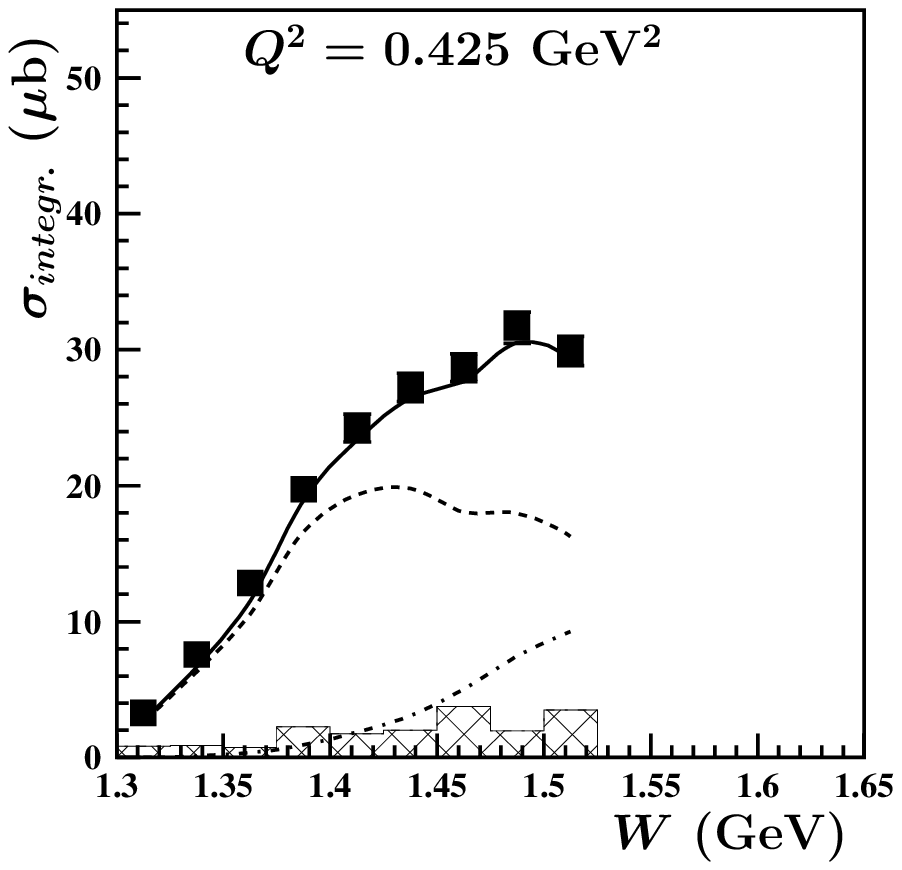,width=6.1cm}
\epsfig{file=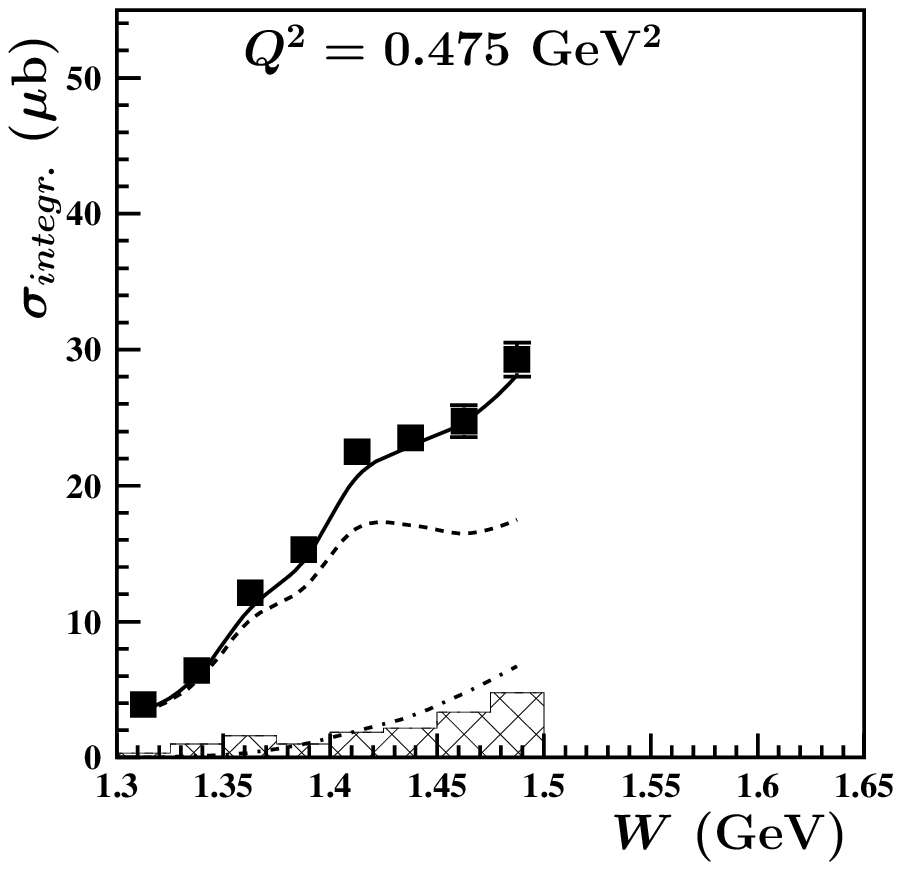,width=6.1cm}
\epsfig{file=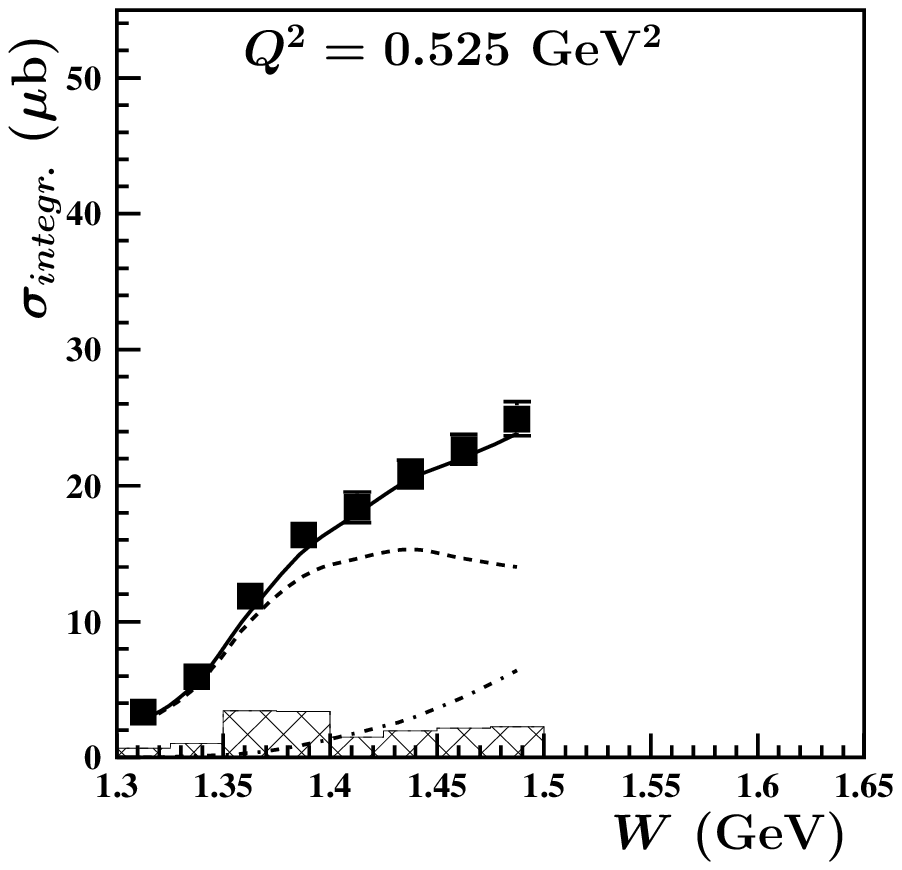,width=6.1cm}
\epsfig{file=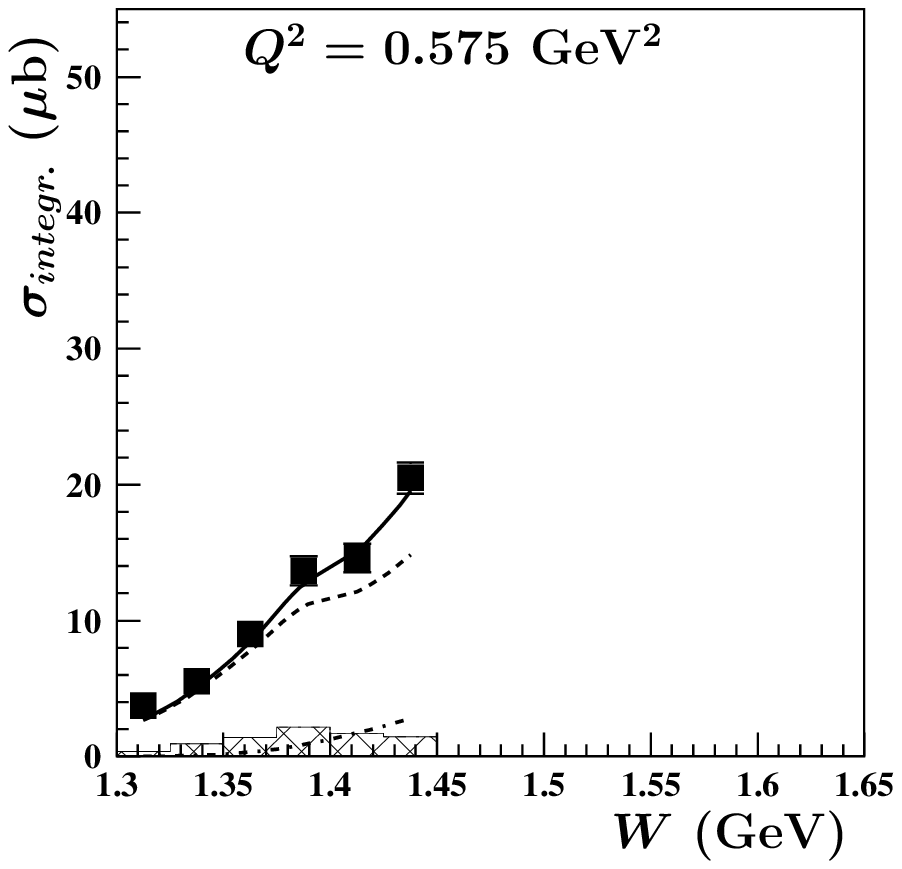,width=6.1cm}
\end{center}
\vspace{-0.6cm}
\caption{\small Fully integrated $2\pi$
cross section at various $Q^{2}$. Crossed
areas represent systematic uncertainties. Full calculations within the
framework of the JM06 model \cite{Mo06-1} are shown by solid curves. The
contributions from $s$-channel resonances and from non-resonant mechanisms are 
shown by the dot-dashed
and dashed curves respectively. \label{systemerr_nst}}
\end{figure*}

In our analysis we determined nine 1-fold differential cross sections in each ($W$, $Q^2$) bin at
invariant masses of the hadronic system from 1.30 to 1.57~GeV and
at photon virtualities from 0.2 to 0.6~GeV$^{2}$ with bins in $W$ of 25~MeV and in
 $Q^2$ of 0.05~GeV$^2$. The data consist of $\pi^{+} p$, $\pi^{-} \pi^{+}$, 
 and  $\pi^{-} p$ invariant mass
distributions, as well as $\pi^{-}$, $\pi^{+}$, and proton angular distributions,
and 3 distributions over angles $\alpha_{\it{i}}$ ($i$=1,2,3) defined in 
Sect.~\ref{xsect_var}.
The full data set, consisting of 4695 cross sections, may be found in Ref. \cite{db07}.
Fully integrated cross sections are shown in Fig.~\ref{2_3_grids_total} and Fig.~\ref{systemerr_nst}.
These results represent the first comprehensive data set for charged double 
pion electroproduction at
$Q^2$ $<$ 0.6~GeV$^2$. With respect to the previous data
\cite{Wa78}, the bin size in $W$ is reduced by almost a factor of 10, while the 
binning in $Q^{2}$ is reduced by a factor of 5.

\begin{table}
\caption{Summary of systematic uncertainties for the fully integrated cross sections.
The values represent averages over the kinematic areas covered by the data. \label{syserrtabl}}

\begin{tabular}{|l|c|}
\hline
Source & Systematic\\     
 & uncertainty   \\
\hline
Integration & 3 \% \\
\hline
Fiducial cuts & 3 \% \\
\hline
Missing mass cut & 2.5 \% \\
\hline
Normalization & $<$ 5 \% \\
\hline
Binning effects & 3 \% \\
\hline
Radiative correction & $<$ 3.5 \% \\
\hline
Acceptances & 3 \% \\
\hline
Total & $<$ 9 \% \\
\hline

\end{tabular}
\end{table}

Systematic uncertainties averaged over the kinematic range covered by the data
are presented in Table~\ref{syserrtabl}. In the following we discuss the various contributions to the systematic uncertainties in more detail.

To estimate uncertainties in the integration procedure, we
compared the values of the fully integrated cross sections obtained by integration
over three sets of kinematic variables, defined in Sect.~\ref{xsect_var}.
The ratios of r.m.s values for the
integrated cross sections over their mean values were treated as
systematics uncertainties
related to the integration procedure. They are maximal at the lowest
$W$ value for all photon virtualities, and range from 5 to 7 \%. As $W$ increases,
they drop to $\sim$ 1 \% at the highest $W$ value.

In order to estimate the uncertainties in the
integration procedure  for the 1-fold differential cross section, we compared 
their values  
determined in integration of the 5-fold
differential cross sections, obtained with three different choices of the final 
state variables, described in Sect.~\ref{xsect_var}. 
 When calculating these integrals
for various kinematic variables,
different sets of 5-dimensional bins contribute to
the respective integrals. The efficiency was
estimated for each set independently. Moreover, different inefficient areas 
contribute to
the same cross sections estimated from integration over various kinematic
grids.
Therefore, a comparison of fully integrated and 1-fold differential cross sections,
obtained by integration over various kinematic variables, allows us to check 
the accuracy of the
detector efficiency evaluations and propagation of the 5-fold differential
cross sections in the CLAS areas of zero acceptance.

Each of the three kinematic grids, discussed in Sect.~\ref{xsect_var},
contains the $\pi^{+} p$ invariant mass distribution, while  just two grids contain
$\pi^- \pi^+$ invariant masses. The angles describing the final state particles have unique
assignments for each of three kinematic grids. We can therefore compare the results of integrations
over 3 different kinematic grids for the $\pi^{+} p$ mass distributions.
For the $\pi^{-} \pi^{+}$ mass distributions the integration over two grids
can be compared. We found that these 1-fold differential
cross sections coincide well within their
statistical uncertainties and in the entire range of kinematics covered by
measurements. As an example, in Fig.~\ref{2_3_grids_mass}  we show the comparison of
mass distributions at $W = 1.41$~GeV and $Q^{2}=$ 0.425~GeV$^{2}$, with
the ones obtained from integrating the  5-fold differential cross sections 
over different sets of kinematic
variables. The comparison of the fully integrated cross sections is shown in
Fig.~\ref{2_3_grids_total}. The results differ by only a fraction of the statistical uncertainties.

The main contributions to the uncertainty in the overall cross section
normalization are given by uncertainties in the integrated luminosity and the electron
detection and reconstruction efficiencies. These contributions have been estimated by
measuring the well known elastic $ep$ scattering cross sections. The comparison with a
parameterization~\cite{Bos} of the world data shows that the overall normalization
is within a $\sim$5 \% uncertainty.

In order to evaluate the systematics involved in defining the final state exclusive process,
we varied the missing mass cut used to identify the unmeasured $\pi^-$,
and modified the fiducial regions where final state particles are selected.
The average uncertainties are shown in Table~\ref{syserrtabl}.

In Sect.~\ref{interpol_ineff} we concluded that the contributions of the zero acceptance
regions in CLAS affect the extracted differential and integrated cross sections well within
the statistical uncertainties. Systematics uncertainties related to propagating the data
into the inactive areas of CLAS were estimated, assuming a 50 \% uncertainty in the
extrapolation of the 5-fold differential cross sections, resulting in 2 \% to 5 \% uncertainties, and
increasing in $Q^2$.

The global systematics for radiative corrections, listed in the Table~\ref{syserrtabl}, were
calculated assuming the individual contributions are uncorrelated. The factor
$\it{R}$, obtained from our Monte Carlo simulation (Sect.~\ref{radiative}), revealed no $Q^{2}$ dependence in the
entire kinematic range of our measurements. The root-mean-square values for the $R$
 factors
calculated at various $Q^{2}$ were assigned to the uncertainties for the radiative correction factor.
Based on the estimates described in Sect.~\ref{radiative}, we assigned
an upper limit of 2.5 \% to the uncertainties related to the
hard photon emission.

The overall systematic uncertainty shown in Table~\ref{syserrtabl}
was obtained as the square root of the quadratic sum over the individual
contributions. Applying the described
procedures in each bin of $\it{W}$ and $Q^2$ individually, we obtain the systematic
uncertainties for the integrated cross sections as shown by the shaded areas in
 Fig.~\ref{systemerr_nst}.



\section{\label{anal} Phenomenological analysis}

\begin{figure*}[ht]
\begin{center}
\framebox{\epsfig{file=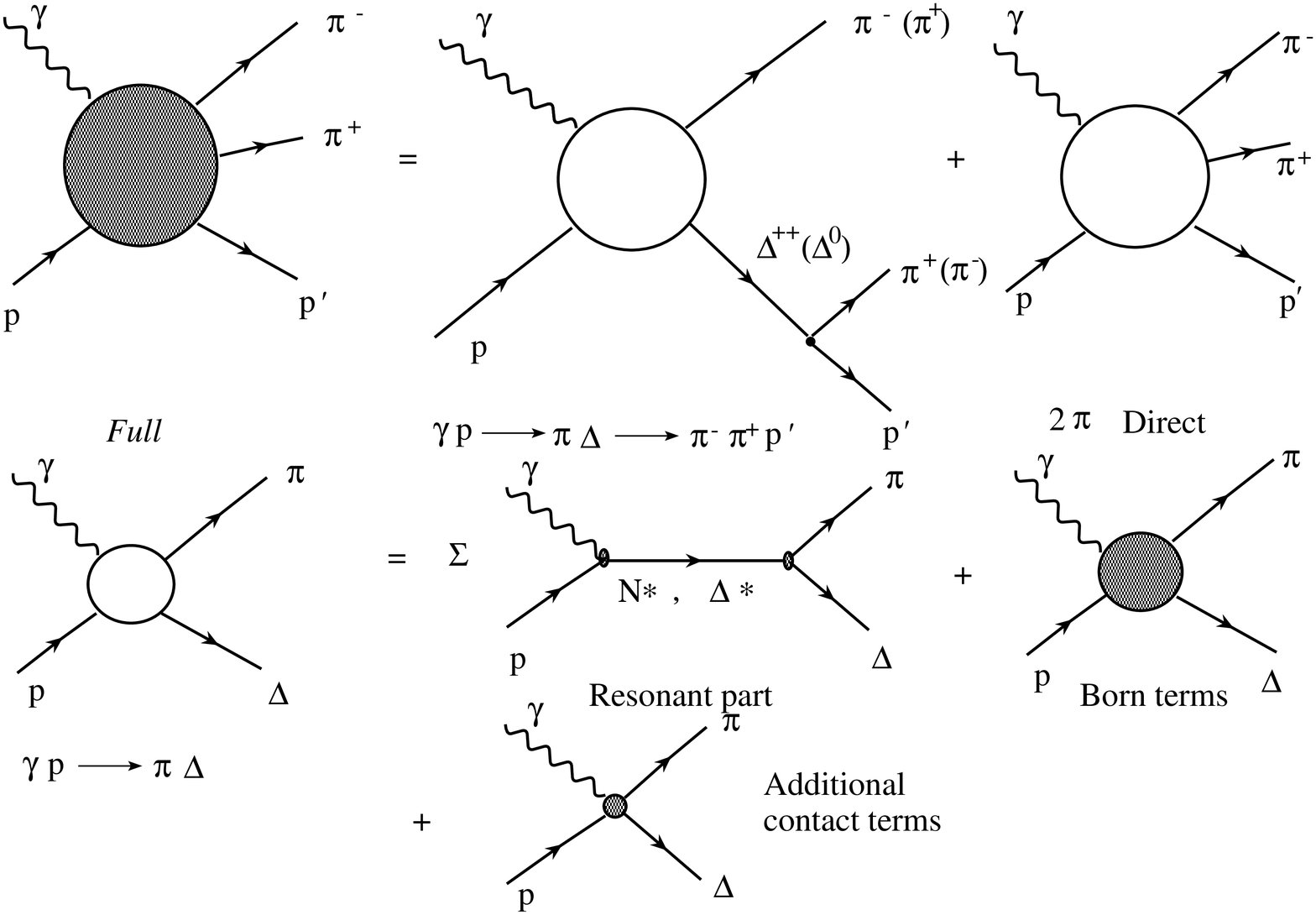,width=12cm}}
\end{center}
\vspace{-0.6cm}
\begin{center}
\framebox{\epsfig{file=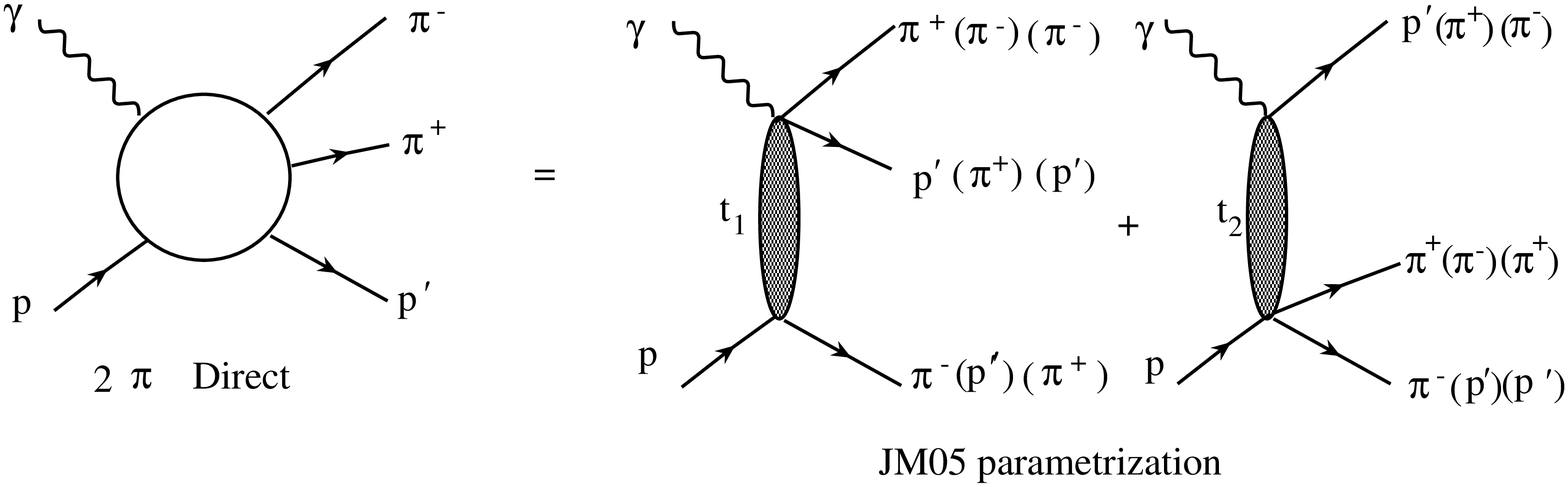,width=12cm}}
\end{center}
\vspace{0.6cm}
\caption{\small The mechanisms of JM05 model \cite{Mo06,Az05} contributing to
2-$\pi$ electroproduction at low W and $Q^{2}$.}
\label{mech_low_06}
\end{figure*}

\begin{figure}[htbp]
\begin{center}
\includegraphics[width=8.5cm]{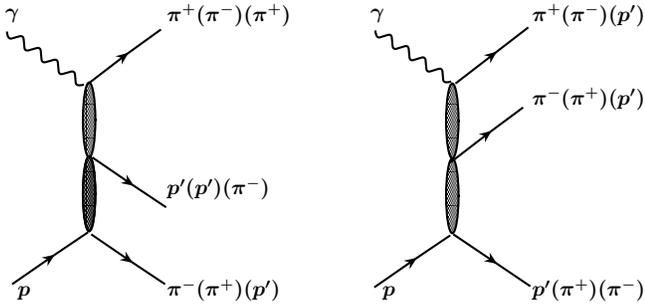}
\caption[]{\small{Direct charged double pion production mechanisms in JM06 model.}\label{diag}}
\end{center}
\end{figure}

The comprehensive information on the 1-fold
differential cross sections for charged double pion production provided by 
the data
enabled us to carry out a combined analysis of the nine 1-fold
differential cross sections in each $W$ and $Q^2$ bin,
and to establish all significant mechanisms contributing
to charged double pion electroproduction in the range $W$ $<$ 1.6~GeV and $Q^2$ = 
0.2-0.6~GeV$^2$. Before our data were available, the information on 2$\pi$ electroproduction
mechanisms in this kinematic area was rather scarce and quite uncertain.
Reasonable knowledge on 2$\pi$ electroproduction mechanisms, achieved in our data
analysis, also allowed the separation of resonant and non-resonant contributions needed for
the evaluation of nucleon resonance electrocouplings.

\begin{figure*}[htp]
\begin{center}
\includegraphics[width=16cm]{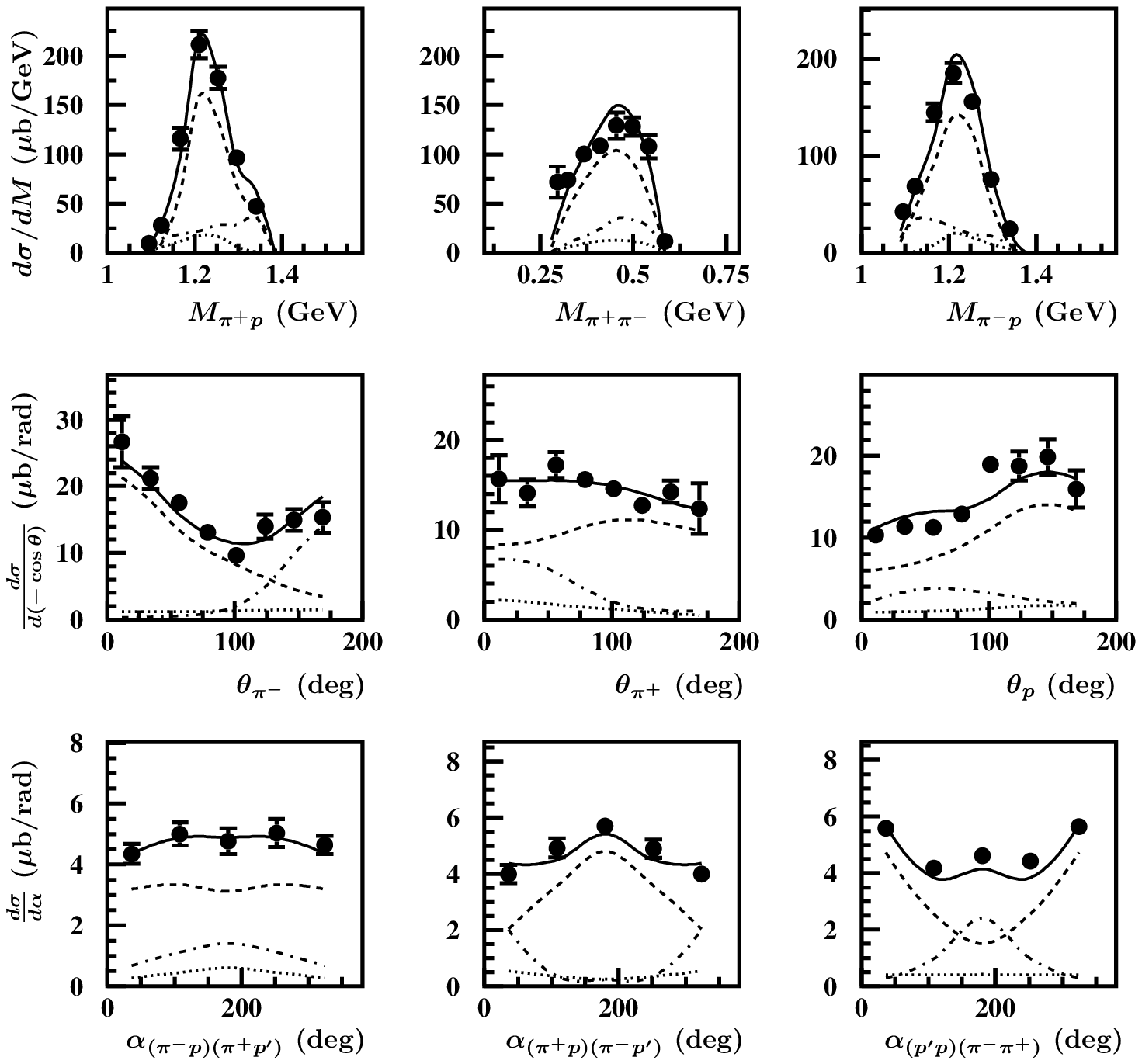}
\caption{\small Description of CLAS charged double pion differential cross sections at $W = 1.51$~GeV
 and $Q^{2}$ = 0.425~GeV$^{2}$ within the framework of the JM06 model.
The full calculations are shown by the solid lines. The contributions from
$\pi^{-} \Delta^{++}$ and $\pi^{+} \Delta^{0}$ isobar channels are shown by the dashed
and dotted lines, respectively. The contributions from direct
charged double pion production processes are shown by the dot-dashed lines.
$\alpha_{\it{i}}$
angular distributions were calculated with the JM06 parameters fit to the other 6
differential cross sections. \label{9sectok}}
\end{center}
\end{figure*}


\begin{figure*}[htp]
\begin{center}
\includegraphics[width=17cm]{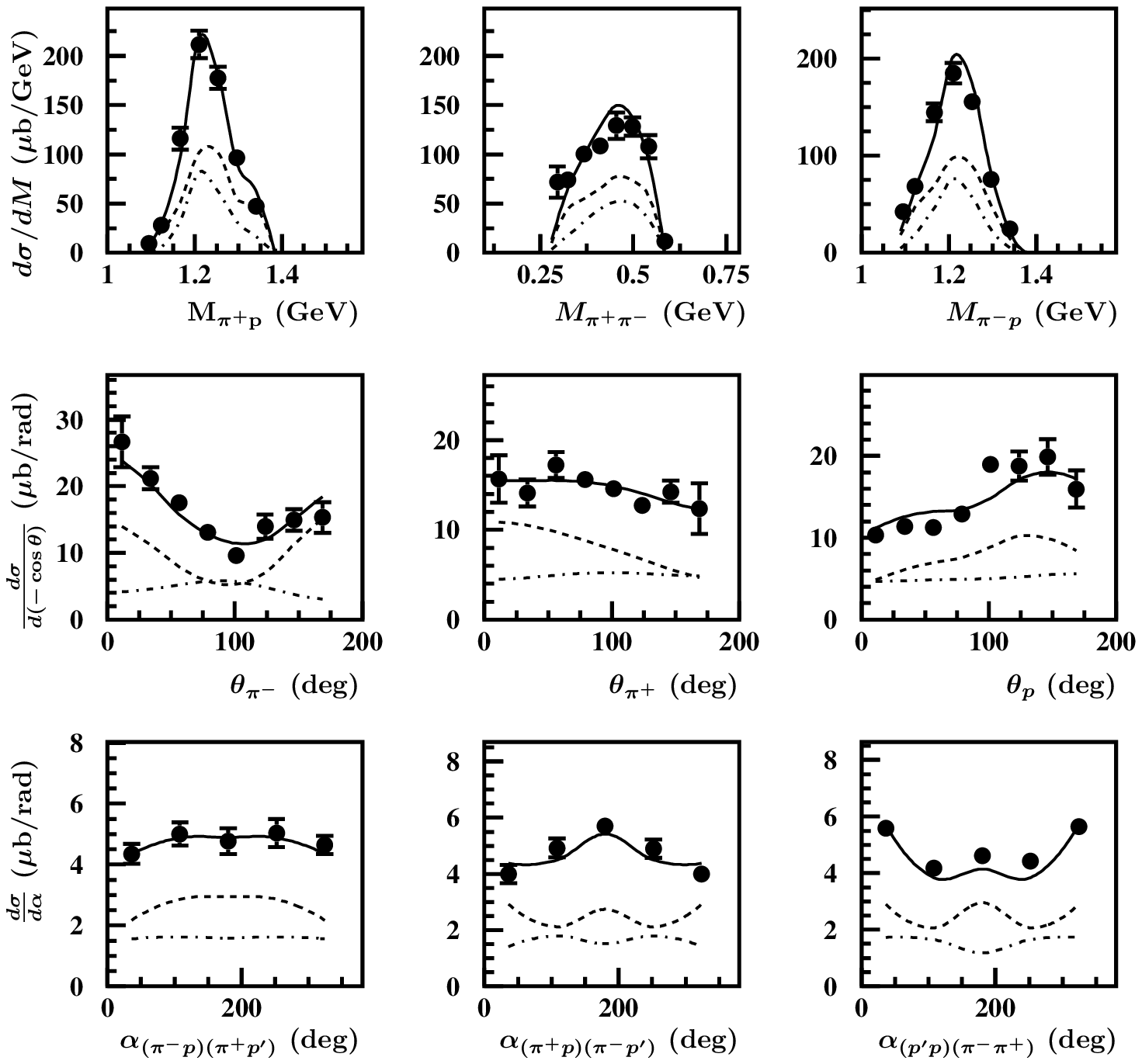}
\caption{\small Resonant (dot-dashed lines) and non-resonant (dashed lines)
contributions to the charged double pion differential cross sections at $W$=1.51 GeV and
$Q^{2}$=0.425 GeV$^{2}$. The full JM06 calculations are shown by black solid lines. 
\label{9secnstbck}}
\end{center}
\end{figure*}

The presence and strengths of the contributing
2$\pi$ electroproduction mechanisms were established
by studying the kinematical dependencies in differential cross sections 
and their correlations in a variety of available observables.
The analysis was carried out using a phenomenological model developed in 
the past few years by 
the  Jefferson Laboratory - Moscow
State University (JM) collaboration
\cite{Mo06,Mo06-1,Mo07,Fe07,Mo08}. Within the JM model developed up to 2005, 
called JM05,
the major part of $\pi^+\pi^-$ production  at $W$ $<$ 1.6 GeV is due to 
the contribution from the  $\pi \Delta$ isobar channels.
This is supported by the data in that the peaks from $\Delta^{++}$ (1232) resonance
were clearly seen in all $\pi^{+} p$ mass distributions at $W$ $>$ 1.4
GeV, while other mass distributions did not show any visible structures. 
The contributions from all other isobar channels included in JM05 
become negligible in this deeply sub-threshold region of $W$ $<$ 1.6 GeV. 
Analysis of the recent CLAS 2$\pi$ data, presented in this paper, allowed us
to study manifestations  of the JM05 mechanisms and search for other contributing
processes in the still unexplored kinematics area of 
photon virtualities from 0.25 to 0.6 GeV$^{2}$. This analysis also enabled us 
to access the dynamics of the contributing mechanisms at a phenomenological level, 
fitting them to all measured  observables combined. Our knowledge on the
contributing mechanisms was extended considerably, resulting in the recent version
of the JLAB-MSU model, which we refer for as JM06. 
 A detailed description of the JM06 model 
version may be found in a separate paper~\cite{Mo07}. Here we
discuss the basic ingredients of JM06 and the major results.

The $\gamma^* p \rightarrow \pi^+\pi^- p$ 
production amplitude within the JM05/JM06 model versions can be written as:

\begin{eqnarray}
T_{\gamma^* N,\pi\pi N} =  T^{\pi\Delta}_{\gamma^*N, \pi\pi N} 
+T^{dir}_{\gamma^*N, \pi\pi N}
\label{eq:full-t}
\end{eqnarray}
with
\begin{eqnarray}
T^{\pi\Delta}_{\gamma^* N,\pi\pi N} = [t^R_{\gamma^* N,\pi\Delta}+
 t^{Born}_{\gamma^* N,\pi\Delta} 
+ t^{c}_{\gamma^*N,\pi\Delta}]\cdot\\ \nonumber
G_{\Delta}\Gamma_{\Delta,\pi N},\,
\label{eq:pid-t}
\end{eqnarray}
where $G_{\Delta}$ is a propagator of the $\Delta$ intermediate
state, and the vertex function
$\Gamma_{\Delta,\pi N}$ describes the
$\Delta (1232) \rightarrow \pi N$ decay.
The mechanisms in the above equations for the JM05 model version are illustrated
in Fig.\ref{mech_low_06}. The diagram 
$\gamma N \rightarrow N^*,\Delta^* \rightarrow \pi\Delta$ in the second row
of the figure is the
 resonant term $t^{R}_{\gamma^* N,\pi\Delta}$ 
in eq. (\ref{eq:pid-t}).
It is parameterized as a  Breit-Wigner
form \cite{Ri00}  and calculated from
all well established $N^*$ states with
masses less than 2.0 GeV that have hadronic decays to the $\pi$$\pi$$N$ final 
states. In the kinematic area covered in our measurements, only the 
$P_{11}(1440)$ and
$D_{13}(1520)$ nucleon resonances have the contributions in 1-fold differential
cross-sections, that are outside of the data uncertainties. 
The non-resonant term $t^{Born}_{\gamma^* N,\pi\Delta}$
is calculated from the
 well established Born terms 
 of $\gamma^* N \rightarrow \pi \Delta$~\cite{So71,Ri00}. Their amplitudes are
 presented in  Ref.~\cite{Ri00,Mo07}. The final state interactions were 
 treated effectively
within the framework of the absorptive approximation \cite{Ri00}. 
Additional contact terms
were implemented in order to account for possible contributions from other
mechanisms to $\pi \Delta$ production, as well as for hadronic
interactions of $\pi \Delta$ states
with other open channels \cite{Mo06,Mo08}. These extra contributions to the $\pi
\Delta$ isobar channels were previously established \cite{Mo06} in analysis of CLAS 
2$\pi$ data
\cite{Ri03} and confirmed by the data of this paper. A parametrization for these
amplitudes will be presented in Ref.~\cite{Mo07}. The contributions from all
isobar channels combined account for from 70 to 90 \% of the 2$\pi$ fully integrated
cross sections in the kinematic area covered in our measurements.
A remaining part of cross sections comes from 
direct charged double pion
production mechanisms, when the $p \pi^{+} \pi^{-}$ final state is created 
without the formation
of the intermediate quasi-two-body states with unstable hadrons. In the JM03 model version 
we started with a
simplest parametrization for direct 2$\pi$ mechanisms as the three body phase space with 
the amplitudes that were independent from the final state kinematic variables and  
fit to the data in each bin of
$W$ and $Q^{2}$ independently \cite{Mo01,Mo03}. However, this parametrization did not
allow us to reproduce steep dependencies in $\pi^{-}$ angular distributions at the
backward angles, clearly seen
both in the previous CLAS 2$\pi$ electroproduction data \cite{Ri03} and in the data of
this paper. The example is shown in Fig.~\ref{9sectok},\ref{9secnstbck}. In order to
reproduce such data behavior, the direct 2$\pi$ production mechanisms were  parameterized 
in JM05 \cite{Az05,Mo06-1}
in terms of a contact vertex and a unspecified particle-exchange amplitude,
described by the effective propagator, which depends exponentially from a 
running four-momenta squared.
The diagrams in the bottom of Fig.\ref{mech_low_06} represents
the direct term $T^{dir}_{\gamma^* N,\pi\pi N}$ in
eq.(\ref{eq:full-t}) that was introduced in Refs.\cite{Az05,Mo06-1} to
describe the direct $\gamma N \rightarrow
\pi\pi N$ mechanisms.   
The parameterization of this term in JM05 was given explicitly 
in Ref.~\cite{Az05}.

This parametrization allowed us to describe successfully the previous
CLAS 2$\pi$ data \cite{Ri03} and the data of our paper on three
invariant masses and $\pi^{-}$ C.M. angular distributions. Analysis of nine
1-fold differential cross-sections, which was carried out for the first time, 
demonstrated the shortcomings in description of $\pi^{+}$ and $p$ C.M. angular
distributions related to the parametrization of the direct 2$\pi$
electroproduction in the JM05 model \cite{Mo08}. Our data offer a compelling evidence for
necessity to modify a description of the direct 2$\pi$ production mechanisms. 
In the recent JM06 model version they were parameterized as two
subsequent unspecified particle-exchanges amplitudes, 
shown in Fig.~\ref{diag} for various assignments of the final state particles.
  Both propagators describing unspecified particle exchanges were parametrized by the
same exponential dependence from a running four-momenta squared. Explicit parametrization
for the amplitudes of these
mechanisms will be presented in a separate paper of Ref. \cite{Mo07}.

Within the framework of the JM06 approach we were able
to describe the 2$\pi$ data of our paper in the entire kinematic range covered by
the measurements.  As a typical example, the model description of the nine
1-fold differential charged double pion
cross sections at $W = 1.51$~GeV and $Q^{2}$ = 0.425~GeV$^{2}$ is shown in
Fig.~\ref{9sectok} together with the contributions of
all mechanisms incorporated in the JM06 description.

The shapes of the cross sections for the different mechanisms are
substantially different in the various observables, but highly correlated by the reaction
dynamics. Moreover, we found no need to implement
additional mechanisms beyond the ones already included in JM06.
Therefore, the successful description of all nine 1-fold differential charged
double pion
cross sections allowed us to identify all significant
contributing processes and access their dynamics at the phenomenological level.
To check the robustness of the results obtained within the framework of the JM06
model,
we fit the model parameters to a limited set of data that included only six
differential cross sections$:$ all three invariant
masses and three angular $\theta_{\it{i}}$ ($i$=$\pi^{-}$, $\pi^{+}$, $p$) distributions
 for the final state hadrons. The remaining
three distributions
over the $\alpha_{\it{i}}$ angles were calculated, keeping the JM06 parameters fixed.
A good description of all of the $\alpha_{\it{i}}$ distributions was achieved 
throughout the
kinematics covered by the measurements, giving us confidence that the main processes
are described within the JM06 model.

The separated resonant and non-resonant contributions to the fully integrated
cross sections are also presented in Fig.~\ref{systemerr_nst}.
The resonant and non-resonant parts of the differential
cross sections at $W = 1.51$~GeV and $Q^2$ = 0.425~GeV$^2$ are shown in
Fig.~\ref{9secnstbck}. Non-resonant mechanisms represent a major contributor in
entire kinematic area covered by our measurements. The 
relative resonant contributions increase with $Q^2$ but remains below than 30 \%. 
However, the resonant 
contributions increase rapidly at
$W > 1.4$~GeV, where they become larger
than the data uncertainties and may be determined from the data fit. 
The dominant part of the resonant amplitudes comes from the $P_{11}(1440)$ and
$D_{13}(1520)$ states combined. The phase space limitations 
prevent $P_{33}(1232)$
decays to the final states with two pions. There is also no evidence for substantial decays 
of the $S_{11}(1535)$ resonance with two pion emission, while the tail from 
the nucleon excitations  with masses
above 1.6 GeV is well inside the
data uncertainties. 

The shapes of the resonant and non-resonant contributions are quite different 
for most differential cross sections, especially the angular distributions.
Moreover, the correlations between kinematical dependencies of the
resonant/non-resonant 
parts in various 1-fold differential cross-sections
 are also quite different. Therefore, a reasonable
description of all differential cross sections combined, which is achieved 
within the framework
of the JM06 model, provides an evidence that a proper isolation of 
the resonance contributions 
can be obtained from global fits to all available cross section data, even in a
case of relatively small ($<$ 30 \%) resonant contributions to the fully
integrated cross-sections. It may be
seen in Fig.~\ref{9secnstbck}, where derived from the data fit the resonant
contributions are presented.
 
While the evaluation of $N^{*}$ electrocouplings 
is outside the scope for this
paper, it is nevertheless noteworthy that 
 our data open up the possibility to determine
electrocouplings for $P_{11}(1440)$ and
$D_{13}(1520)$ states at $Q^2$ $<$ 0.6~GeV$^2$  from double pion
electroproduction for the first time. The preliminary results may be found in the
Ref.~\cite{Mo08}. The kinematical region of low photon
virtualities is expected to be particularly sensitive to meson-baryon dressing
effects in the resonance structure \cite{Sa01,BLee08}.

Moreover, in this kinematic region there are also CLAS data on single pion electroproduction
\cite{db07,Az05-1,Joo02,Joo03,Joo04,Joo05,egiyan06}.
At $W$ $<$ 1.5~GeV only single and double
pion exclusive channels contribute to the total meson production
cross sections off protons. The amplitudes for various  non-resonant mechanisms
contributing to single and double pion electroproduction, obtained in a 
phenomenological analysis of CLAS data with the framework of the JM06 model
\cite{Mo07},
represent valuable information
for the study of nucleon resonance transitions in combined analyses
of these major channels. Advanced coupled-channel approaches are under
development at the Excited Baryon Analysis Center (EBAC) at Jefferson Lab 
\cite{Lee07,Lee06}.
 Phenomenological analysis of the double pion data of this paper,
described in detail in Ref.~\cite{Mo07}, is of particular interest for
this activity.







\section{\label{concl} Conclusions}

\noindent In this paper we have presented a large body of electroproduction data for
the process $\gamma_{v} p \rightarrow p \pi^{+} \pi^{-}$.
The large acceptance of CLAS allowed extraction of the 1-fold differential and
fully integrated charged double pion cross sections. The results were tested
for robustness by using different integration grids, that showed that
consistent results are obtained independent of the specific integration procedure.

1-fold differential and fully integrated
cross sections were obtained for $W$ from 1.3~GeV to 1.6~GeV and $Q^2$
from 0.2~GeV$^2$ to 0.6~GeV$^2$. The high statistics and good momentum resolution of 
the measurements allowed us to use
bin sizes of $\Delta W = 25$~MeV and $\Delta Q^2$ = 0.05~GeV$^2$, which are at 
least a factor of 5
smaller than the ones used in previous measurements. For the first time, nine independent
differential cross sections in each bin of $W$ and $Q^2$  were measured.

The phenomenological analysis of the cross sections within the framework of the JM06
approach \cite{Mo06-1,Mo07} allowed us to establish significant mechanisms
contributing to charged double pion electroproduction for the kinematics 
covered by our
measurement. All differential and fully integrated cross sections obtained in
our measurement 
can be reasonably described
by the contributions from $\pi \Delta$ isobar channels and 
direct double pion production mechanisms, established from phenomenological 
analysis of our 
data. 
The analysis of resonance contributions to the differential
cross sections indicates that these data are sensitive to the 
resonant parts of cross-sections, allowing us to study the excitation of
$P_{11}(1440)$ and $D_{13}(1520)$ states by virtual photons off protons 
at $Q^{2}$ $<$ 0.6~GeV$^{2}$.

The data also allowed us to determine cross sections and amplitudes 
for isobar channels, offering valuable information for nucleon resonance studies in a global analysis of
the major meson electroproduction exclusive reactions within the framework of 
an advanced coupled-channel approach currently under development in EBAC at Jefferson Lab
\cite{Lee06,Lee07}.


\section{\label{cha:ack}Acknowledgments}
We would like to acknowledge the outstanding efforts of the staff of the
Accelerator and the Physics Divisions at JLab that made this experiment possible.
This work was supported in part by the Skobeltsyn Institute of Nuclear
Physics
 and Physics Department at Moscow State University, the U.S. Department of Energy and the National
Science Foundation, the U.K. Engineering and Physical Science Research Council,
the Istituto Nazionale di Fisica Nucleare, the
 French Centre National de la Recherche Scientifique,
the French Commissariat \`{a} l'Energie Atomique, and the Korean Science and
Engineering Foundation.
Jefferson Science Associates (JSA) operates the
Thomas Jefferson National Accelerator Facility for the United States
Department of Energy under contract DE-AC05-06OR23177.

\appendix

\section*{Appendix A: Kinematics variables for 5-differential 2-$\pi$ production
cross-sections}

In this appendix we present the final state kinematics 
for the second choice of variables defined in Sect.~\ref{xsect_var}.
Since all momenta are measured in the lab
frame, first we boost the 3-momenta of the
final state particles in the c.m. frame. All 3-momenta
used below, if not specified otherwise, are
defined in the c.m. frame. 
 
 $M_{\pi^{+}\pi^{-}}, M_{\pi^{+}p}$ and $M_{\pi^{-}p}$ invariant masses were 
 related to the four momenta of the final particles as:
\begin{flalign}
& M_{\pi^{+}\pi^{-}} = \sqrt{(P_{\pi^{+}} + P_{\pi^{-}})^{2}}  && \nonumber \\
& M_{\pi^{+}p'}  = \sqrt{(P_{\pi^{+}} + P_{p'})^{2}}  &&  \label{invmasses}\\
& M_{\pi^{-}p'}  = \sqrt{(P_{\pi^{-}} + P_{p'})^{2}} \textrm{ ,}  &&  \nonumber
\end{flalign}

where $P_{i}$  ($i$=$\pi^{-}$, $\pi^{+}$, $p$) stand for the final state 
particle four-momenta.

The angle $\theta_{\pi^{-}}$ between the 3-momentum of the initial photon and
the final state $\pi^{-}$ in the c.m. frame was calculated as:
\begin{flalign}
\theta_{\pi^{-}} = {\rm acos}\left( \frac{(\vec  P_{\pi^{-}} \vec P_{\gamma})}
{|\vec {P_{\pi^{-}}}| |\vec {P_{\gamma}}| }\right).  &&  
\label{flalign}
\end{flalign}

The angle $\varphi_{\pi^{-}}$ was determined as:
\begin{flalign}
& \varphi_{\pi^{-}} = {\rm arctg}\left( \frac{P_{y
\pi^{-}}}{P_{x\pi^{-}}} \right) ; \:
P_{x\pi^{-}} > 0 ; \:  P_{y\pi^{-}} > 0 \\
& \varphi_{\pi^{-}} = {\rm arctg}\left( \frac{P_{y
\pi^{-}}}{P_{x\pi^{-}}} \right) + 2\pi  ; \: 
P_{x\pi^{-}} > 0 ; \:  P_{y\pi^{-}} < 0 \\
& \varphi_{\pi^{-}} = {\rm arctg}\left( \frac{P_{y
\pi^{-}}}{P_{x\pi^{-}}} \right) + \pi ; \:  
P_{x\pi^{-}} < 0 ; \:    P_{y\pi^{-}} < 0 \\
& \varphi_{\pi^{-}} = {\rm arctg}\left( \frac{P_{y
\pi^{-}}}{P_{x\pi^{-}}} \right) + \pi ;  \: 
P_{x\pi^{-}} < 0 ; \:  P_{y\pi^{-}} > 0   \\
& \varphi_{\pi^{-}} = \pi/2 ;  \:
P_{x\pi^{-}} = 0 ;  \:  P_{y\pi^{-}} > 0 \\
& \varphi_{\pi^{-}} = 3\pi/2 ; \: 
P_{x\pi^{-}} = 0 ;  \:   P_{y\pi^{-}} < 0.  
\end{flalign}

The calculation of the angle $\alpha_{(\pi^{-}p)(\pi^{+}p')}$
between the two planes A and B (see Fig.~\ref{kinematic})
is more complicated. First we determine two
auxiliary vectors $\vec \gamma$  and
$\vec \beta$. The vector $\vec \gamma$ is the unit vector perpendicular to the 
$\vec P_{\pi^{-}}$ 3-momentum, directed toward the vector $-\vec n_{z}$ and situated in 
the plane composed by the
virtual photon 3-momentum and the $\pi^{-}$ 3-momentum $\vec
P_{\pi^{-}}$ (see Fig.~\ref{kinematic}). $\vec
n_{z}$ is the unit vector directed along the $z$-axis (see Fig.~\ref{kinematic}).
The vector $\vec \beta$ is the unit vector perpendicular to the 3-momentum 
of $\pi^{-}$,
directed toward the $\pi^{+}$ 3-momentum $\vec P_{\pi^{+}}$ and situated 
in the plane composed
by the $\pi^{+}$ and $p'$
3-momenta. Note that the 3-momenta of the $\pi^{+}$,
$\pi^{-}$ and  $p'$ are in the same plane, since in the center-of-mass 
their total 3-momentum should be equal to zero.
 Then the angle between the two planes  $\alpha_{(\pi^{-}p)(\pi^{+}p')}$ is:
\begin{flalign}
\alpha_{(\pi^{-}p)(\pi^{+}p')} = \rm{acos}(\vec \gamma \vec \beta), &&
\label{anglealpha}
\end{flalign}
with the acos function running between zero and
$\pi$, and the angle between the 
planes $A$ and $B$ running from zero to
$2\pi$. To determine $\alpha$ in a
range between $\pi$ and $2\pi$, we look at
the relative orientation of the vector $\vec
P_{\pi^{-}}$ and vector product $\vec \delta$ for
the auxiliary vectors $\vec
\gamma$ and $\vec \beta$:
\begin{flalign}
\vec \delta = \vec \gamma \times \vec \beta . &&
\label{vecprod}
\end{flalign}
If $\vec \delta$ is collinear to $\vec
P_{\pi^{-}}$, $\alpha_{(\pi^{-}p)(\pi^{+}p')}$ is determined
from eq.~(\ref{anglealpha}). In the case of anti-collinear 
vectors $\vec \delta$ and  $\vec
P_{\pi^{-}}$:
\begin{flalign}
\alpha_{(\pi^{-}p)(\pi^{+}p')} = 2\pi - \rm{acos}(\vec \gamma \vec \beta). &&
\label{anglealpha_var}
\end{flalign}
The vector $\vec \gamma$ may be expressed
through the particle 3-momenta as:
\begin{flalign}
& \vec \gamma = a_{\alpha}(-\vec n_{z}) + b_{\alpha}\vec n_{P_{\pi^{-}}} && \nonumber \\
& a_{\alpha} = \sqrt{\frac{1}{1 - (\vec n_{P_{\pi^{-}}} (-\vec n_{z} ) )^{2}}} && \label{alphavec}\\
& b_{\alpha} = - (\vec n_{P_{\pi^{-}}}(-\vec n_{z} ) ) a_{\alpha} \textrm{ ,} && \nonumber
\end{flalign}
where $\vec n_{P_{\pi^{-}}}$ is the unit vector directed along the 
$\pi^{-}$ 3-momentum (see Fig.~\ref{kinematic}).
Taking scalar products $(\vec \gamma \vec
n_{P_{\pi^{-}}})$ and $(\vec \gamma \vec  \gamma)$,
it is straightforward to verify that $\vec \gamma$ is the unit vector perpendicular to $\vec P_{\pi^{-}}$.

The vector $\vec \beta$ may be obtained as:
\begin{flalign}
& \vec \beta = a_{\beta}\vec n_{P_{\pi^{+}}} + b_{\beta}\vec n_{P_{\pi^{-}}}  && \nonumber \\
& a_{\beta} = \sqrt{\frac{1}{1 - (\vec n_{P_{\pi^{+}}}  \vec n_{P_{\pi^{-}}})^{2}}}  && \label{betavec}\\
& b_{\beta} = - (\vec n_{P_{\pi^{+}}} \vec n_{P_{\pi^{-}}}) a_{\beta} \textrm{ ,}  && \nonumber
\end{flalign}
where $\vec n_{P_{\pi^{+}}}$ is the unit vector directed along the
 $\pi^{+}$ 3-momentum.
Again, taking scalar products $(\vec \beta \vec
n_{P_{\pi^{-}}})$ and $(\vec \beta \vec  \beta)$,
it is straightforward to see that $\vec \beta$ is the unit vector 
perpendicular to the $\pi^{-}$
3-momentum. The angle $\alpha_{(\pi^{-}p)(\pi^{+}p')}$ coincides with the 
angle between the vectors $\vec \gamma$ and
$\vec \beta$.
So, the scalar product $( \vec \gamma
\vec \beta )$ allows determination of the angle
$\alpha_{(\pi^{-}p)(\pi^{+}p')}$ in eq.~(\ref{anglealpha}). 

The kinematic
variables for other hadron assignments for the
first, second and third final state particle
described above, were evaluated in a similar
way.

\clearpage
\newpage

\end{document}